\documentclass[12pt, draftclsnofoot, onecolumn]{IEEEtran}
\IEEEoverridecommandlockouts
\usepackage{cite}
\usepackage[T1]{fontenc}
\usepackage{aecompl}
\usepackage{eqparbox}
\usepackage[numbers,sort&compress]{natbib}
\usepackage{amsmath,amssymb,amsfonts}
\usepackage{amsthm}
\usepackage{amsmath}
\usepackage{float} 
\usepackage{booktabs}
\makeatletter
\renewcommand{\maketag@@@}[1]{\hbox{\m@th\normalsize\normalfont#1}}%
\makeatother
\usepackage{graphicx}
\usepackage{epstopdf}
\usepackage{verbatim}
\usepackage{subfigure}
\usepackage{textcomp}
\usepackage{xcolor}
\usepackage{stfloats}
\usepackage{tabularx}
\usepackage{array}
\usepackage{makecell}
\usepackage{url}
\makeatother
\usepackage{textcomp}
\usepackage[marginal]{footmisc}
\usepackage{setspace}

\allowdisplaybreaks[4]
\usepackage{algpseudocode}
 \usepackage[linesnumbered,ruled]{algorithm2e}                
\def\d{\,\mathrm{d}}

\def\BibTeX{{\rm B\kern-.05em{\sc i\kern-.025em b}\kern-.08em
    T\kern-.1667em\lower.7ex\hbox{E}\kern-.125emX}}
\usepackage{balance}
\newtheorem{myDef}{Definition}
\newtheorem{myTheo}{Theorem}

\newtheorem{myProp}{Proposition}
\newcommand*{\circled}[1]{\lower.7ex\hbox{\tikz\draw (0pt, 0pt)%
    circle (.4em) node {\makebox[1em][c]{\small #1}};}}
\begin{document}
\title{Streaming 360$^{\circ}$ VR Video with Statistical QoS Provisioning in mmWave Networks from Delay and Rate Perspectives}
\vspace{-0.5em}
\author{Yuang Chen, \emph{Student Member, IEEE}, Hancheng Lu, \emph{Senior Member, IEEE}, Langtian Qin, Chang Wu, and Chang Wen Chen, \emph{Fellow, IEEE}
\thanks{\setlength{\baselineskip}{2\baselineskip}This work was supported by Hong Kong Research Grants Council (GRF-15213322) and National Science Foundation of China (No. U21A20452, No. U19B2044). Yuang Chen, Hancheng Lu, Langtian Qin, and Chang Wu are with the CAS Key Laboratory of Wireless-Optical Communications, School of Information Science and Technology, University of Science and Technology of China, Hefei 230027, China (email: yuangchen21@mail.ustc.edu.cn; hclu@ustc.edu.cn; qlt315@mail.ustc.edu.cn; changwu@mail.ustc.edu.cn). Chang Wen Chen is with the Department of Computing, The Hong Kong Polytechnic University, Hong Kong (e-mail: changwen.chen@polyu.edu.hk).}}

\maketitle

\vspace{-3em}

\begin{abstract}
\vspace{-0.5em}
\par Millimeter-wave $\!$(mmWave) technology has emerged as a promising enabler for unleashing the full potential of 360$^{\circ}$ virtual reality (VR). However, the explosive growth of VR services, coupled with the reliability issues of mmWave communications, poses enormous challenges in terms of wireless resource and quality-of-service (QoS) provisioning for mmWave-enabled 360$^{\circ}$ VR. In this paper, we propose an innovative 360$^{\circ}$ VR streaming architecture that addresses three under-exploited issues: overlapping field-of-views (FoVs), statistical QoS provisioning (SQP), and loss-tolerant active data discarding.
Specifically, an overlapping FoV-based optimal joint unicast and multicast (JUM) task assignment scheme is designed to implement the non-redundant task assignments, thereby conserving wireless resources remarkably. Furthermore, leveraging stochastic network calculus, we develop a comprehensive SQP theoretical framework that encompasses two SQP schemes from delay and rate perspectives. Additionally, a corresponding optimal adaptive joint time-slot allocation and active-discarding (ADAPT-JTAAT) transmission scheme is proposed to minimize resource consumption while guaranteeing diverse statistical QoS requirements under loss-intolerant and loss-tolerant scenarios from delay and rate perspectives, respectively. Extensive simulations demonstrate the effectiveness of the designed overlapping FoV-based JUM optimal task assignment scheme. Comparisons with six baseline schemes validate that the proposed optimal ADAPT-JTAAT transmission scheme can achieve superior SQP performance in resource utilization, flexible rate control, and robust queue behaviors.
\end{abstract}

\vspace{-1.5em}

\begin{IEEEkeywords}
\vspace{-1em}
Virtual reality (VR), millimeter wave (mmWave), field of view (FoV), quality of service(QoS), stochastic network calculus (SNC).
\end{IEEEkeywords}

\vspace{-1.5em}

\section{Introduction}
\par The development of fifth-generation mobile wireless networks and beyond (5G/B5G) has created an unprecedented demand for more realistic human-digital interaction \cite{xiong2021augmented}. In particular, wireless immersive 360$^{\circ}$ virtual reality (VR), as the dominant content supply paradigm in future mobile networks, has enormous potential in various fields such as education, healthcare, industry, and entertainment \cite{yaqoob2020survey,wang2023road}. However, realizing these visions entails overcoming numerous highly intertwined challenges arising from the exceptionally rigorous and diverse QoS requirements for immersive VR \cite{hu2020cellular,xiong2021augmented,wang2023road,yaqoob2020survey}.

\par MmWave communications have the potential to provide multi-gigabits-per-second (Gbps) rates, which are anticipated to substantially gratify the resource demand for bandwidth-hungry wireless VR services \cite{li2022mobility,struye2022covrage}. However, the proliferation of wireless VR services has exponentially increased traffic, putting immense pressure on wireless networks \cite{yaqoob2020survey,xiao2022transcoding}. Delivering an immersive 360$^{\circ}$ VR experience necessitates exceptionally rigorous and diverse QoS provisioning, making it an essential prerequisite \cite{elbamby2018toward,zink2019scalable,yaqoob2020survey}. To this end, numerous research efforts have attempted to enhance wireless VR delivery performance. For instance, coordinated multi-point (CoMP) is integrated into mmWave communications to improve the immersive experience and resource utilization \cite{yang2022feeling}. Broadcasting is a more efficient way to improve resource utilization \cite{gui2020robust}. Researchers have developed a hybrid transmission mode selection scheme that achieves a good balance between resource utilization and QoS provisioning performance for 360$^{\circ}$ VR broadcasting \cite{feng2020smart}. However, the complex mode selection and limited application scenarios cannot guarantee reliable wireless VR delivery. To overcome the poor delivery reliability caused by transmission rate bottleneck, a dual-connectivity network architecture has been investigated that combines sub-6 GHz and mmWave networks with mobile edge computing to significantly enhance the reliability \cite{gu2021reliability}. Additionally, a transcoding-enabled tiled 360$^{\circ}$ VR streaming framework has also been exploited to enable flexible compromises between video bitrate and resource utilization \cite{xiao2022transcoding,long2020optimal}. Nevertheless, in the context of 5G/B5G, the diverse QoS requirements for 360$^{\circ}$ VR include not only bitrate but also latency, reliability, tolerable loss rate and so on \cite{zink2019scalable,yaqoob2020survey,xiong2021augmented,wang2023road,hu2020cellular}. Although previous studies have yielded valuable insights into QoS provisioning for wireless 360$^{\circ}$ VR, the underlying issues that result in low resource utilization and unsatisfactory QoS provisioning have not been thoroughly investigated, which are listed as follows:

\par The issue of redundant resource consumption resulting from overlapping field-of-views (FoVs). The potential downsides of overlapping FoV generated by frequent interactions of users in immersive virtual environments seems to have been overlooked. Specifically, users in identical virtual environment would request the same VR content, resulting in partial or complete overlap of corresponding FoVs. If these overlapping FoVs are not processed  properly, repeated streaming of the same VR content will lead to unnecessary resource consumption. However, research on this aspect has been inadequate.

\par Statistical QoS provisioning schemes for supporting 360$^{\circ}$ VR streaming. Aforementioned studies have primarily employed deterministic QoS provisioning (DQP) \cite{xiao2022transcoding,elbamby2018toward,yang2022feeling,gui2020robust,feng2020smart,gu2021reliability,long2020optimal}. Nevertheless, due to resource limitations and highly-varying mmWave channels, DQP performance is typically hard to guarantee \cite{li2020qos,wang2019unified,zhang2018heterogeneous}. Stochastic network calculus (SNC) is a potent methodology that has the potential to provide dependable theoretical insights into statistical QoS provisioning (SQP) for characterizing the QoS requirements of latency-sensitive services \cite{fidler2014guide,jiang2008stochastic}. SNC-based SQP methods can be generally categorized from the perspectives of delay and rate. From delay perspective, SQP focuses on analyzing non-asymptotic \emph{statistical delay violation probability} (SDVP) \cite{yang2018low,chen2023statistical,zhang2021aoi}, which is typically formulated as $\mathbb{P}\!\left[\rm{metric} \!>\! \rm{budget}\right]\! \leq \!\varepsilon_{th}$, where $\varepsilon_{th}$ represents the violation probability when the actual queuing delay $\rm{metric}$ exceeds the target delay $\rm{budget}$. From rate perspective, SQP focuses on analyzing the system's maximum asymptotic service capacity, commonly known as \emph{effective capacity} (EC) \cite{wu2003effective,chang2000performance,amjad2019effective}, which characterizes the maximum constant arrival data rate that can be maintained under statistical QoS requirements. Unfortunately, the effective SQP schemes for multi-layer tiled 360$^{\circ}$ VR streaming have not received sufficient attention and thoroughly investigated.

\par Loss-tolerant data discarding schemes for flexible rate control and robust queue behaviors. Studies have demonstrated that when motion-to-photon latency is relatively long (approximately $20\!\!\sim\!\!30$ ms, depending on the individual), the vestibulo-ocular reflex can produce conflicting signals, resulting in motion sickness and severe physiological discomfort for users \cite{elbamby2018toward}. Accordingly, video buffering caused by the rapid growth of queue length is a crucial factor in deteriorating the immersive experience for users \cite{zhang2021buffer,jedari2020video}. Therefore, it is essential to develop an optimal active data discarding scheme that can achieve flexible rate control and robust queue behaviors to effectively suppress video buffering \cite{yahia2019http,hong2019continuous}. Note that imposing exclusive tolerable loss rates on video quality layers with diverse QoS requirements is necessary and reasonable \cite{zhang2021buffer,jedari2020video,yahia2019http,hong2019continuous}. On the one hand, actively discarding low importance from FoV edges is favourable to achieve smoother video playback and improve QoE during poor channel conditions. On the other hand, 360$^{\circ}$ VR is a latency-sensitive service, where exceeding the target delay renders the video data obsolete for users. Even if HMDs successfully receive these obsolete video data, they would still be discarded.

\par To overcome the aforementioned issues, we propose an innovative wireless multi-layer tiled 360$^{\circ}$ VR streaming architecture with SQP in mmWave networks. Specifically, to deal with overlapping FoVs, we investigate a joint unicast and multicast (JUM) transmission scheme to implement non-redundant task assignments for tiles streaming at the base station (BS). Then, we develop an SNC-based comprehensive SQP theoretical architecture that encompasses two SQP schemes from SDVP and EC perspectives, respectively. Additionally, a corresponding optimal adaptive joint time-slot and active-discarding (ADAPT-JTAAT) transmission schemes is proposed to minimize resource consumption while guaranteeing diverse QoS requirements, flexible rate control, and robust queue behaviors, ultimately enabling seamless 360$^{\circ}$ VR. The contributions of this paper are summarized as follows:

\begin{itemize}
  \item We address the issue of redundant resource consumption resulting from overlapping FoVs and design an overlapping FoV-based optimal JUM task assignment scheme to implement non-redundant task assignments for tiles streaming through two processes: \emph{user grouping} and \emph{FoV clustering}. This scheme has been demonstrated to significantly conserve wireless resources.

  \item By leveraging SNC theory, we establish a comprehensive SQP theoretical framework that encompasses two SQP schemes from delay and rate perspectives. This theoretical framework provides dependable theoretical insights for 360$^{\circ}$ VR's SQP and valuable theoretical guidance for the development of resource optimization schemes.

  \item Based on the established theoretical framework, we propose an optimal ADAPT-JTAAT transmission scheme that encompasses delay and rate perspectives. From delay perspective, the proposed optimal ADAPT-JTAAT transmission scheme formulates the problems of minimizing resource consumption with non-asymptotic SDVP. Furthermore, two novel algorithms, namely the nested-shrinkage optimization algorithm and the stepwise-approximation optimization algorithm, is proposed to effectively address the resource optimization problem under loss-intolerant and loss-tolerant scenarios.

  \item From rate perspective, the proposed optimal ADAPT-JTAAT transmission scheme formulates the resource consumption minimization problem with non-asymptotic EC constraints. The expressions of the optimal time-slot allocation strategy and the optimal active data discarding strategy is derived, respectively. Additionally, a low-complexity subgradient-based optimization algorithm is proposed to address this resource optimization problem under both loss-intolerant and loss-tolerant scenarios.
\end{itemize}

\par Extensive simulations are carried out to demonstrate the effectiveness of the designed overlapping FoV-based optimal JUM task assignment scheme. Comparisons with six baseline schemes validate that the proposed optimal ADAPT-JTAAT transmission scheme can minimize resource consumption while achieving superior SQP performance, flexible rate control, and robust queue behaviors, from delay and rate perspectives, respectively.

\par The remainder of this paper is organized as follows. In Sec. II, the multi-layer tiled 360$^{\circ}$ VR streaming architecture with SQP is introduced. In Sec. III, a comprehensive SQP theoretical architecture is developed. The innovative optimal ADAPT-JTAAT transmission scheme and its solutions are presented in Sec. IV. In Sec. V, extensive performance evaluations and thorough analysis are presented. Finally, Sec. VI concludes the paper.

\vspace{-1.2em}

\section{Multi-layer Tiled 360$^{\circ}$ VR Streaming Architecture with SQP}
\vspace{-0.3em}
\par As illustrated in Fig. 1, we propose a wireless multi-layer tiled 360$^{\circ}$ VR streaming architecture with statistical QoS provisioning (SQP) over mmWave networks. The set of users is denoted by the subscript $\mathcal{N} \triangleq \left\{1,2,\cdots,N\right\}$. This architecture involves a mmWave BS equipped with a single antenna, which can provide wireless 360$^{\circ}$ services with diverse QoS requirements for these $N$ users simultaneously. Each user wears a single-antenna head-mounted display (HMD) and requests VR video services from the BS, specifying their expected QoS requirements. The video content within any rectangular region of the 360$^{\circ}$ VR video that a user may watch is referred to as FoV \cite{yang2022feeling,xiao2022transcoding}, with the central limit known as the viewing direction \cite{long2020optimal}. At any time, users can freely switch their current FoVs to another one that they find more engaging. Next, we explain the proposed architecture in three parts as follows.

\vspace{-1.0em}

\begin{figure*}[htbp]
\centering
\includegraphics[scale=0.40]{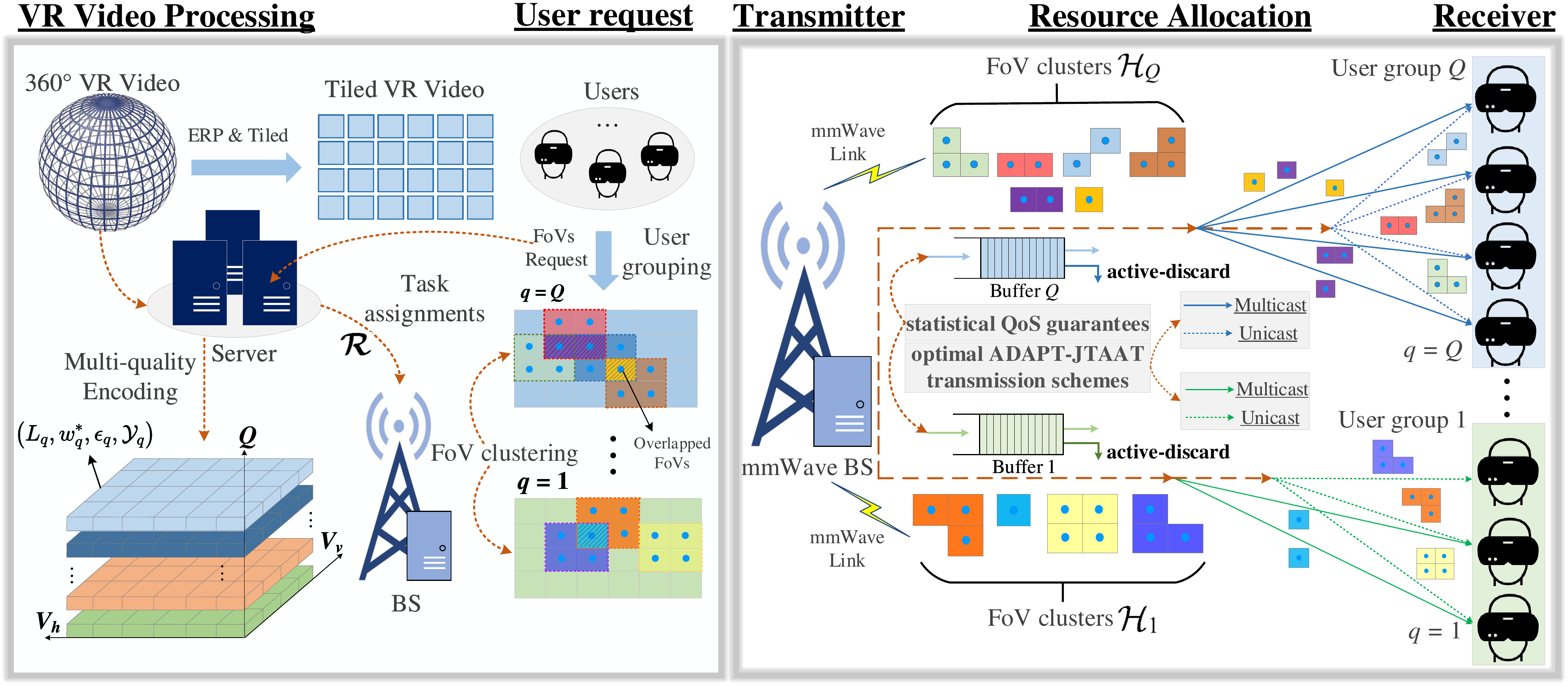}
\vspace{-1.0em}
\caption{The multi-layer tiled 360$^{\circ}$ VR streaming architecture with statistical QoS provisioning in mmWave
networks}
\label{fig:label}
\vspace{-2.0em}
\end{figure*}

\subsection{Statistical QoS Requirements}
\par The 360$^{\circ}$ VR video is first projected onto a two-dimensional tiled VR by using equirectangular projection (ERP) \cite{xiao2022transcoding,perfecto2020taming,yaqoob2020survey}. Then, the latest video encoding technologies, such as H.264 and HEVC \cite{ye2019omnidirectional,xiao2022transcoding}, are adopted to pre-encode each VR tile into $Q$ video quality layers with diverse encoding rates. The set of video quality layers is denoted by the subscript $\mathcal{Q}\triangleq\left\{1,2,\cdots,Q\right\}$. In contrast to existing state-of-the-art tiled 360$^{\circ}$ VR video streaming systems \cite{nguyen2019optimal,xiao2022transcoding}, our proposed streaming architecture features multi-layer tiled 360$^{\circ}$ VR video streams, each with multiple specific statistical QoS requirements, denoted by a quaternion $\left(L_{q},w_{q}^{\ast},\epsilon_{q},\mathcal{Y}_{q}\right)$, where $L_{q}$ and $w_{q}^{\ast}$ represent the encoding rate and target delay of the $q$-th layer, respectively, while $\epsilon_{q}$ and $\mathcal{Y}_{q}$ indicate the SDVP and the tolerable loss-rate of the $q$-th video layer, respectively. More precisely, SDVP characterizes the tail probability that the actual delay exceeds the target delay $w_{q}^{\ast}$, which aims to describe the delivery reliability of 360$^{\circ}$ VR video streaming \cite{chen2023statistical}.

\vspace{-1.0em}
\subsection{Overlapping FoV-based Optimal Joint Unicast and Multicast Task Assignment Scheme}
\vspace{-0.2em}
\par The enormous data volume of 360$^{\circ}$ VR leads to significant challenges in delivery latency and reliability due to overlapping FoVs resulting from frequent user interactions. To this end, we design an overlapping FoV-based optimal JUM task assignment scheme to avoid a waste of wireless resources by providing non-redundant task assignments for the BS. This scheme comprises two parts: \emph{user grouping} and \emph{FoV clustering}, which are further explained below.

\par \emph{\textbf{User grouping:}} Users are grouped based on the video quality layer $L_{q}, q \in \mathcal{Q}$ they requested. All users requesting the same video quality layer form a group, and all groups can be represented by the set $\boldsymbol{\mathcal{N}} \triangleq \left\{\mathcal{N}_{1},\mathcal{N}_{2},\cdots,\mathcal{N}_{Q}\right\}$, where $\mathcal{N}_{q}$ denotes the group formed by all users requesting the video quality layer $L_{q}$.

\par \emph{\textbf{FoV clustering:}} Assume that each FoV contains $a\!\times \!b$ tiles of the same size. For each user group $\!\mathcal{N}_{q}$, we first calculate the union of the tiles in it. The result can be denoted as $\mathcal{F}_{q} \!\triangleq \! \bigcup_{n \in \mathcal{N}_{q}} \! \mathcal{F}_{n}$, where $\!\mathcal{F}_{n}\!$ denotes the set of tiles corresponding to the FoV of user $n$. Then, we calculate the union of tiles in the set $\mathcal{N}_{q}\backslash \mathcal{M}_{q}$, and the result is represented as $\bigcup_{n \in \mathcal{N}_{q}\backslash \mathcal{M}_{q}} \mathcal{F}_{n}$, where $\mathcal{M}_{q}$ is one of the non-empty subsets of user cluster $\mathcal{N}_{q}$, and the set $\mathcal{N}_{q} \backslash \mathcal{M}_{q}$ denotes the users who are in user cluster $\mathcal{N}_{q}$, but not in set $\mathcal{M}_{q}$. Note that all non-empty subsets $\mathcal{M}_{q}$ of user cluster $\mathcal{N}_{q}$ will constitute set $\mathcal{H}_{q}$. Then, all tiles in subset $\mathcal{M}_{q}$ can be denoted as $\mathcal{F}_{q} - \bigcup_{n \in \mathcal{N}_{q}\backslash \mathcal{M}_{q}} \! \mathcal{F}_{n}$, and the overlapped tiles of user subset $\mathcal{M}_{q}$ can be denoted as $\bigcap_{n \in \mathcal{M}_{q}}\! \mathcal{F}_{n}$. Finally, the tiles should be streamed of user subset $\mathcal{M}_{q}$ can be given as follows:

\vspace{-1.3em}
\begin{equation}\label{e1}
  \mathcal{R}_{\!\mathcal{M}_{q}} \! \triangleq \! \bigg(\!\mathcal{F}_{q}-\!\!\!\!\!\! \bigcup_{n\in \mathcal{N}_{q}\backslash\mathcal{M}_{q}}\!\!\!\!\! \mathcal{F}_{n}\!\!\bigg) \! \bigcap \! \bigg(\bigcap_{n\in \mathcal{M}_{q}}\!\!\!\mathcal{F}_{n}\!\!\bigg).
\end{equation}

\vspace{-0.8em}

\par According to (\ref{e1}), we observe that if $\!\!\mathcal{M}_{q}\!\!$ is a single-user subset, $\!\mathcal{R}_{\mathcal{M}_{q}}\!\!$ denotes the tiles corresponding to the non-overlapping part of the user's FoV. If $\!\mathcal{M}_{q}$ is a multi-user subset, $\!\mathcal{R}_{\mathcal{M}_{q}}\!\!$ denotes the tiles that correspond to the FoVs of the overlapping parts of these users; an example of this is illustrated in Fig. 2 for ease of comprehension. The user group $\mathcal{N}_{q}=\{1,2,3\}$ has seven non-empty subsets $\mathcal{M}_{q}$, which make up the set $\mathcal{H}_{q}$. According to (\ref{e1}), the VR tiles contained in each non-empty subset $\mathcal{M}_{q}$ are shown in $\mathcal{R}_{\mathcal{M}_{q}}$ in Fig. 2.

\vspace{-1.5em}
\begin{figure}[htbp]
\centering
\includegraphics[scale=0.12]{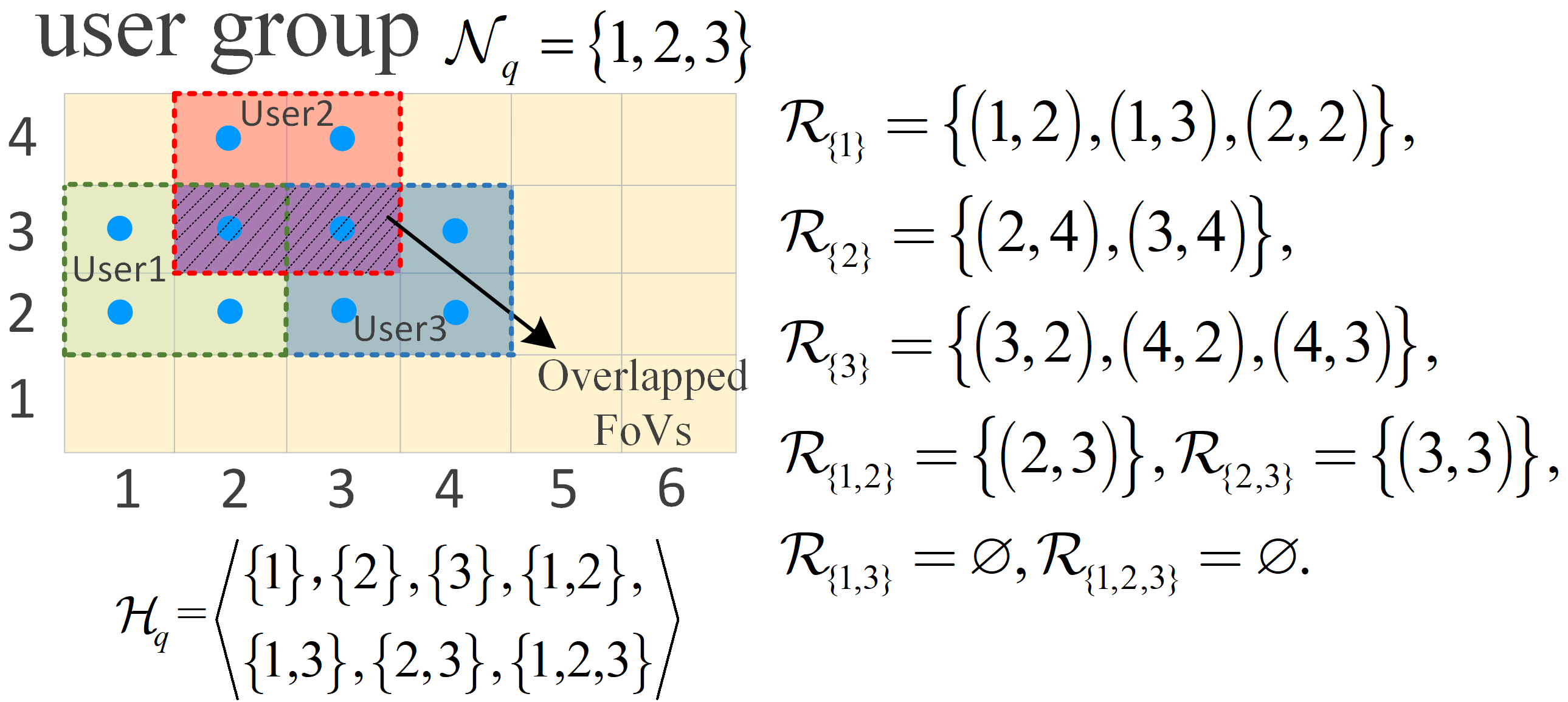}
\vspace{-0.8em}
\caption{An example of FoV clustering.}
\label{fig:label}
\vspace{-1.3em}
\end{figure}

\par Through \emph{user grouping} and \emph{FoV clustering}, we can obtain non-redundant task assignments, denoted by $\boldsymbol{\mathcal{R}} \triangleq \left\{\mathcal{R}_{\mathcal{M}_{q}}\right\}_{q \in \mathcal{Q}, \mathcal{M}_{q} \in \mathcal{H}_{q}}$. The resulting non-redundant task assignments can then be supplied to the BS. To achieve the goal of conserving wireless resources, we adopt the JUM transmission scheme to stream the obtained non-redundant $\boldsymbol{\mathcal{R}}$. Specifically, when $\mathcal{M}_{q}$ is a multi-user subset with overlapping FoVs, we select \textbf{multicast mode} to stream the tiles in $\mathcal{R}_{\mathcal{M}_{q}}$ to the corresponding users. All tiles in $\mathcal{R}_{\mathcal{M}_{q}}$ are first aggregated together, and then served in a single multicast session. On the other hand, if $\mathcal{M}_{q}$ is a single-user subset, we select \textbf{unicast mode} to stream the tiles in $\mathcal{R}_{\mathcal{M}_{q}}$ to the user.

\vspace{-1.5em}
\subsection{Channel Model and Active-discarding Scheme}
\vspace{-0.3em}
\par Similar to many literatures \cite{yang2018low,yu2017coverage,bai2014coverage}, the channel coefficients of mmWave are commonly modeled as random variables following Nakagami-m distributions for the ease of tractability. We assume that the small-scale fading channel with a Nakagami-m distribution is independent and identically distributed (i.i.d.) within each fading period. Then the capacity of mmWave channel can be expressed as follows:

\vspace{-2em}

\begin{equation}\label{e2}
  R = \log_{2}\left(1+l^{-\alpha} \xi \zeta\right),
  \vspace{-1.0em}
\end{equation}
where $l$ denotes the distance between users and the BS, $\alpha$ denotes the path loss exponent. $\xi$ denotes the transmit power, which is normalized with respect to the background noise \cite{yang2018low}, while the random variable $\zeta$ denotes the channel gain, which follows a Nakagami-m distribution. The probability density function (PDF) of channel gain $\zeta$ is given as

\vspace{-0.8em}

\begin{equation}\label{e3}
  f\left(\zeta,M\right)= \frac{\zeta^{M-1}}{\Gamma(M)}\!\left(\frac{M}{\bar{\zeta}}\right)^{M}\!\!\! \cdot e^{-M\frac{\zeta}{\bar{\zeta}}}, \zeta \ge 0,
  \vspace{-0.3em}
\end{equation}
where $\Gamma(M)=\int_{0}^{\infty}M^{t-1}e^{-t}dt$ is Gamma function, $\bar{\zeta}$ denotes the average SNR, and $M$ represents the fading parameter.

\par Due to the low-latency and ultra-reliability requirements of immersive 360$^{\circ}$ VR services, video buffering caused by the rapid growth of queue length is the culprit that deteriorates the QoE of users. Therefore, it is essential to develop an effective active data discarding scheme that achieves flexible rate control and robust queue behaviors for streaming 360$^\circ$ VR video with enormous data volume. Let $J_{\!\mathcal{M}_{q}}$ (in bits) denote the amount of discarded video data in the subset $\mathcal{M}_{q}$. Then, the normalized active-discarding rate during a frame can be expressed as $\mathcal{J}_{\!\mathcal{M}_{q}} \triangleq J_{\!\mathcal{M}_{q}}/B$ (in bits/Hz). Thus, the service rate provided by the BS for streaming the task assignment $\mathcal{R}_{\!\mathcal{M}_{q}}$ can be reformulated as follows:

\vspace{-1.5em}

\begin{equation}\label{e4}
  r_{\!\mathcal{M}_{q}} = B\left(t_{\!\mathcal{M}_{q}}\log_{2}(1 + \zeta^{'}) + \mathcal{J}_{\!\mathcal{M}_{q}}\right),
  \vspace{-1.0em}
\end{equation}
where $\zeta^{'} = l^{-\alpha}\xi\zeta$ denotes the normalized average SNR, $B$ represents the bandwidth, and $t_{\!\mathcal{M}_{q}}$ denotes the time-slot allocated to stream the task assignment $\mathcal{R}_{\mathcal{M}_{q}}$. Apparently, the integration of an active-discarding scheme holds promise in effectively curbing video buffering by actively discarding the data with low importance from the FoV edges. However, a robust transmission can only be effectuated through a good balance between the active-discarding rates and the QoE of users, as excessive high active-discarding rates lead to unnecessary data loss, thereby reducing the user's QoE. Conversely, excessively low active-discarding rates cannot ensure smooth video playback under poor channel conditions.

\vspace{-1.0em}

\section{SQP Theoretical Framework: A Comprehensive Approach From Delay and Rate Perspectives}
\vspace{-0.2em}
\par Compared to sub-6 GHz networks, mmWave networks exhibit higher propagation loss, link variability, and susceptibility blockages \cite{wang2021intelligent,yang2018low,yu2017coverage,bai2014coverage}. As a result, guaranteeing DQP performance is challenging due to resource limitations and highly-unreliable channels \cite{li2020qos,wang2019unified,zhang2018heterogeneous}. Moreover, flexible rate control and robust queueing behaviors rely on effective active-discarding schemes. The consistency assumptions imposed by classical queuing theory on the arrival and service processes are inadequate for analyzing such networks \cite{miao2019stochastic,fidler2014guide,jiang2008stochastic,yang2018low,chang2000performance}. In this section, we leverage SNC theory to develop a comprehensive SQP theoretical framework that encompasses two SQP schemes from the delay and rate perspectives.

\vspace{-1em}
\subsection{Statistical QoS Provisioning from Delay Perspective}
\par \par For each task assignment $\mathcal{R}_{\mathcal{M}_{q}}$ with the specific statistical QoS requirements $\left(L_{q},w_{q}^{\ast},\epsilon_{q},\mathcal{Y}_{q}\right)$, $q \!\!\in\!\! \mathcal{Q}, \mathcal{M}_{q} \!\!\in\!\! \mathcal{H}_{q}$, the cumulative arrival, departure, and service processes from time slot $s$ to $(t\!-\!1)$ are defined as bivariate processes $A_{\!\mathcal{M}_{q}}\!(s,t)\!\!=\!\!\sum_{i = s}^{t}a_{\!\mathcal{M}_{q}}\!(i)$, $D_{\!\mathcal{M}_{q}}\!(s,t) \!\!=\!\! \sum_{i = s}^{t} \! d_{\!\mathcal{M}_{q}}\!(i)$, and $S_{\!\mathcal{M}_{q}}\!(s,t) \!\!=\!\! \sum_{i = s}^{t}r_{\!\mathcal{M}_{q}}\!(i)$, respectively. Here, $a_{\mathcal{M}_{q}}(i)$, $d_{\mathcal{M}_{q}}(i)$, and $r_{\mathcal{M}_{q}}(i)$ represent the instantaneous video data arrival rate, corresponding departure rate, and achievable service rate of the task assignment $\!\!\mathcal{R}_{\mathcal{M}_{q}}\!\!$ at time slot $i$ ($s \!\leq \!i \!\leq \!t-1$), respectively.  Assume that all of the queues are work-conserving first-come-first-served queues. From SNC theory \cite{fidler2014guide,jiang2008stochastic,yang2018low,chen2023statistical}, the queueing delay $w_{\mathcal{M}_{q}}(t)$ of task assignment $\mathcal{R}_{\mathcal{M}_{q}}$ at time slot $t$ can be expressed as follows:

\vspace{-1.7em}

\begin{equation}\label{e5}
   w_{\!\mathcal{M}_{q}}\!(t) = \inf\!\left\{u\geq0: A_{\!\mathcal{M}_{q}}\!(0,t) \!\leq \! D_{\!\mathcal{M}_{q}}\!(0,t+u)\right\}.
\end{equation}

\vspace{-1.0em}

\par The cumulative processes $A_{\mathcal{M}_{q}}(s,t)$, $D_{\mathcal{M}_{q}}(s,t)$, and $S_{\mathcal{M}_{q}}(s,t)$ are all defined in the bit domain, which means that these processes of video data are measured in terms of the number of bits. To facilitate the modeling, the $(min,\times)$-algebra is introduced to convert these cumulative processes from the bit domain to the SNR domain \cite{al2014network} \footnote{Given a random process $\mathcal{U}(s,t)$ in the bit-domain, its counterpart in the SNR-domain can be expressed as $\mathcal{U}(s,t) = e^{U(s,t)}$.}. The counterparts of $A_{\mathcal{M}_{q}}(s,t)$, $D_{\mathcal{M}_{q}}(s,t)$, and $S_{\mathcal{M}_{q}}(s,t)$ of the task assignment $\mathcal{R}_{\mathcal{M}_{q}}$ in the SNR-domain can be expressed as $\mathcal{A}_{\mathcal{M}_{q}}(s,t) = e^{A_{\mathcal{M}_{q}}(s,t)}$, $\mathcal{D}_{\mathcal{M}_{q}}(s,t) = e^{D_{\mathcal{M}_{q}}(s,t)}$, and $\mathcal{S}_{\mathcal{M}_{q}}(s,t) = e^{S_{\mathcal{M}_{q}}(s,t)}$, respectively. Given any nonnegative random process $U(s,t)$, its Mellin transform can be expressed as $\mathcal{M}_{U}(\theta,s,t) = \mathbb{E}\left[\right(X(s,t)\left)^{\theta-1}\right]$ for any free parameter $\theta$, whenever the expectation exists \cite{al2014network}. Then, the steady-state kernel between $\mathcal{A}_{\mathcal{M}_{q}}(s,t) = e^{A_{\mathcal{M}_{q}}(s,t)}$ and $\mathcal{S}_{\mathcal{M}_{q}}(s,t) = e^{S_{\mathcal{M}_{q}}(s,t)}$ can be given as the following expression \cite{al2014network,chen2023statistical}:

\vspace{-1.8em}

\begin{equation}\label{e6}
       \mathcal{K}_{\!\mathcal{M}_{q}}\!(\theta_{\!\mathcal{M}_{q}},\!L_{q},w_{q}^{\ast}) \!\!=\!\! \lim_{t \rightarrow \infty}\! \sum_{v = 0}^{t}\! \mathcal{M}_{\!\mathcal{A}_{\!\mathcal{M}_{q}}}\!(1\!+\!\theta_{\!\mathcal{M}_{q}},\!v,\!t)\cdot\mathcal{M}_{\!\mathcal{S}_{\!\mathcal{M}_{q}}}(1\!-\!\theta_{\!\mathcal{M}_{q}},\!v,t\!+\!w_{q}^{\ast}).
\end{equation}

\vspace{-0.8em}

\par $A_{\!\mathcal{M}_{q}}\!(0,t)\!=\!\sum_{i = 0}^{t}a_{\!\mathcal{M}_{q}}\!(i)$ and $S_{\!\mathcal{M}_{q}}\!(0,t) \!=\! \sum_{i = 0}^{t}r_{\!\mathcal{M}_{q}}\!(i)$ can be rewritten as two incremental processes over $t$ consecutive time slots, where $a_{\!\mathcal{M}_{q}}\!(i)$ and $r_{\!\mathcal{M}_{q}}\!(i)$ denote the increments of arrivals and services for the task assignment $\!\mathcal{R}_{\!\mathcal{M}_{q}}\!$ at time slot $i$, respectively. We consider that $a_{\!\mathcal{M}_{q}}\!(i) \! \equiv \! a_{\!\mathcal{M}_{q}}$ is constant among different time slots. Consequently, the constant arrival rate of the task assignment $\mathcal{R}_{\!\mathcal{M}_{q}}$ can be denoted as $\lambda_{\!\mathcal{M}_{q}}\!\equiv\! \frac{a_{\!\mathcal{M}_{q}}}{T}$. Assume that the channel of each user subset $\mathcal{M}_{\!q}$ is no-dispersive block fading, and $r_{\!\mathcal{M}_{q}}\!(i)$ is i.i.d. at different time slots. Then, the achievable service rate of streaming the task assignment $\!\mathcal{R}_{\!\mathcal{M}_{q}}\!$ at different time slots can be denoted as a time independent random variable $r_{\!\mathcal{M}_{q}}$. In this case, the steady-state kernel formulated by (\ref{e6}) can be rewritten as follows:

\vspace{-2.0em}

\begin{equation}\label{e7}
      \mathcal{K}_{\!\mathcal{M}_{q}}\!(\theta_{\!\mathcal{M}_{q}},\!L_{q},\!w_{q}^{\ast}) \!=\! \frac{\mathcal{M}_{\!\delta_{\!\mathcal{M}_{q}}}^{w_{q}^{\ast}}\!\!(1\!-\!\theta_{\!\mathcal{M}_{q}})}{1 \!-\! \mathcal{M}_{\alpha_{\!\mathcal{M}_{q}}}\!(1\!+\!\theta_{\!\mathcal{M}_{q}}\!)\mathcal{M}_{\!\delta_{\!\mathcal{M}_{q}}}\!(1\!-\!\theta_{\!\mathcal{M}_{q}}\!)}= \frac{\left(\mathbb{E}\!\!\left[(\delta_{\!\mathcal{M}_{q}}\!)^{-\theta_{\!\mathcal{M}_{q}}}\!\right]\right)^{\!w_{q}^{\ast}}}{1 \!-\! \mathbb{E}\!\!\left[\!(\!\alpha_{\!\mathcal{M}_{q}}\!)^{\theta_{\!\mathcal{M}_{q}}}\!\right]\!\mathbb{E}\!\!\left[\!(\delta_{\!\mathcal{M}_{q}}\!)^{-\theta_{\!\mathcal{M}_{q}}}\!\right]},
\vspace{-0.6em}
\end{equation}
where $\!\alpha_{\!\mathcal{M}_{q}} \!\!=\!\! e^{a_{\!\mathcal{M}_{q}}}\!$, $\delta_{\!\mathcal{M}_{q}} \!=\! e^{r_{\!\mathcal{M}_{q}}}$, and $\mathcal{M}_{\!\delta_{\mathcal{M}_{q}}}^{w_{q}^{\ast}}\!\!(1\!-\!\theta_{\!\mathcal{M}_{q}}\!)$ denotes the $w_{q}^{\ast}$-th power of $\mathcal{M}_{\!\delta_{\!\mathcal{M}_{q}}}\!(1\!-\!\theta_{\mathcal{M}_{q}}\!)$. Notably, when the ``stability condition'', denoted by $S(\theta_{\!\mathcal{M}_{q}}\!)\!\! = \!\! \mathcal{M}_{\delta_{\!\mathcal{M}_{q}}}\!\!(1\!+\!\theta_{\!\mathcal{M}_{q}}\!)\mathcal{M}_{\alpha_{\!\mathcal{M}_{q}}}\!(1\! -\! \theta_{\mathcal{M}_{q}}\!) \!< \! 1$ holds, the (\ref{e7}) is meaningful; otherwise the summation in (\ref{e6}) would be unbounded\footnote{According to SNC \cite{fidler2014guide,jiang2008stochastic,yang2018low,chen2023statistical,al2014network}, the value of the free parameter $\theta_{\!\mathcal{M}_{q}} > 0$ corresponding to the task assignment $\mathcal{R}_{\!\mathcal{M}_{q}}$ portrays the exponential decaying exponent of the violation probability of its statistical QoS requirement and reveals the decay rate of the queue length. A larger value of $\theta_{\!\mathcal{M}_{q}}$ (e.g., $\theta_{\!\mathcal{M}_{q}} \rightarrow \infty$) indicates more stringent statistical QoS requirements of streaming $\mathcal{R}_{\!\mathcal{M}_{q}}$. Conversely, a smaller value of $\theta_{\!\mathcal{M}_{q}}$ (e.g., $\theta_{\!\mathcal{M}_{q}}\rightarrow 0$) implies looser statistical QoS requirements.}. From (\ref{e7}), it can be obtained that $\mathbb{E}[e^{\theta_{\mathcal{M}_{q}} a_{\mathcal{M}_{q}}}] \mathbb{E}[e^{-\theta_{\mathcal{M}_{q}} r_{\mathcal{M}_{q}}}] > 0$ for $\theta_{\mathcal{M}_{q}} > 0$. And it is necessary to determine the supremum of $\theta_{\!\mathcal{M}_{q}}$ to make $\mathbb{E}[e^{\theta_{\mathcal{M}_{q}} a_{\mathcal{M}_{q}}}] \mathbb{E}[e^{-\theta_{\mathcal{M}_{q}} r_{\mathcal{M}_{q}}}] < 1$ holds, thereby making (\ref{e6}) bounded and (\ref{e7}) meaningful. We set $\theta_{\mathcal{M}_{q}}^{max} = \sup\!\left\{\theta_{\!\mathcal{M}_{q}}\!\!: \! \mathbb{E}[e^{\theta_{\!\mathcal{M}_{q}} a_{\!\mathcal{M}_{q}}}] \mathbb{E}[e^{-\theta_{\!\mathcal{M}_{q}} r_{\!\mathcal{M}_{q}}}] < 1\right\}$, and the UB-SDVP of task assignment $\mathcal{M}_{q}$ can be written as \cite{chen2023statistical}

\vspace{-1.8em}

\begin{equation}\label{e8}
   \mathbb{P}\!\left\{w_{\!\mathcal{M}_{q}} \!>\! w_{q}^{\ast}\right\} \leq \inf\limits_{0 \leq \theta_{\mathcal{M}_{q}} < \theta_{\mathcal{M}_{q}}^{max}}\left\{\mathcal{K}_{\!\mathcal{M}_{q}}\!(\theta_{\!\mathcal{M}_{q}},\!L_{q},\!w_{q}^{\ast})\right\}= \mathcal{V}(w_{q}^{\ast},\boldsymbol{t},\boldsymbol{\mathcal{J}}),
   \vspace{-0.8em}
\end{equation}
where $\!\boldsymbol{t}\!\triangleq \!\left\{\!t_{\!\mathcal{M}_{q}}\!\right\}_{\!q\in\mathcal{Q},\mathcal{M}_{q}\in\mathcal{H}_{q}}\!$ and $\!\boldsymbol{\mathcal{J}}\!\!\triangleq\!\! \left\{\!\mathcal{J}_{\!\mathcal{M}_{q}}\!\right\}_{q\in\mathcal{Q},\mathcal{M}_{q}\in\mathcal{H}_{q}}\!\!$ denote the time-slot allocation strategy and active-discarding strategy for streaming the task assignments $\!\boldsymbol{\mathcal{R}}\!\triangleq\! \left\{\mathcal{R}_{\!\mathcal{M}_{q}}\right\}_{q\in\mathcal{Q},\mathcal{M}_{q}\in\mathcal{H}_{q}}$, respectively.

\vspace{-1.5em}
\subsection{Statistical QoS Provisioning from Rate Perspective}
\vspace{-0.5em}
\par Ensuring the stability condition in (\ref{e7}) requires the service capacity to surpass the arrival rate of video data, which represents the maximum processing capacity for video data without causing video buffering. Thus, it is crucial to investigate the maximum asymptotic service capacity of the proposed multi-layer tiled 360$^{\circ}$ VR streaming architecture under statistical QoS provisioning. Assume that the arrival process $A_{\!\mathcal{M}_{q}}\!(s,t)$ and service process $S_{\!\mathcal{M}_{q}}\!(s,t)$ of the task assignment $\!\mathcal{R}_{\!\mathcal{M}_{q}}\!$ are ergodic stochastic processes satisfying  $\mathbb{E}\!\left[a_{\mathcal{M}_{q}}\!(i)\right] \!<\! \mathbb{E}\!\left[r_{\mathcal{M}_{q}}\!(i)\right]$ at any time slot $i$. According to the Mellin transform of $\delta_{\!\mathcal{M}_{q}}$, the effective capacity can be defined as follows \cite{chen2023statistical}:
\vspace{-0.4em}
\begin{equation}\label{e9}
   \mathcal{EC}_{\!\mathcal{M}_{q}}\!(\theta_{\!\mathcal{M}_{q}}) = -\frac{1}{\theta_{\mathcal{M}_{q}}TB} \ln \mathcal{M}_{\delta_{\!\mathcal{M}_{q}}}\!(1-\theta_{\!\mathcal{M}_{q}}).
   \vspace{-1em}
\end{equation}

\vspace{-0.4em}

\par From the rate perspective of SQP \cite{yang2018low,chen2023statistical,al2014network}, given a certain statistical QoS exponent $\theta_{\!\mathcal{M}_{q}}$ that enables

\vspace{-2.4em}

\begin{equation}\label{e10}
  \lim_{w_{q}^{\ast} \rightarrow \infty}\!\frac{\log\!\left(\pi^{-1}\mathbb{P}(w_{\mathcal{M}_{q}} \! \geq \! w_{q}^{\ast})\right)}{w_{q}^{\ast}} = -\theta_{\!\mathcal{M}_{q}}\mathcal{EC}_{\!\mathcal{M}_{q}}(\theta_{\!\mathcal{M}_{q}}),
  \vspace{-0.4em}
\end{equation}
where $\!\pi\!$ is the non-empty probability of the queue. If $\mathbb{E}\!\left[a_{\mathcal{M}_{q}}(i)\right] \!<\! \mathbb{E}\!\left[r_{\mathcal{M}_{q}}(i)\right]$ is satisfied, then \emph{G$\ddot{a}$rtner-Ellis Theorem} holds \cite{yang2018low,amjad2019effective}, and SDVP $\mathbb{P}\left(w_{\mathcal{M}_{q}} \geq w_{q}^{\ast}\right)$ is approximated as follows \cite{amjad2019effective}:

\vspace{-2.5em}

\begin{equation}\label{e11}
  \mathbb{P}\left\{w_{\mathcal{M}_{q}} \geq w_{q}^{\ast}\right\} \approx \pi \exp\left(-\theta_{\mathcal{M}_{q}}\mathcal{EC}_{\mathcal{M}_{q}}(\theta_{\mathcal{M}_{q}})w_{q}^{\ast}\right),
  \vspace{-0.9em}
\end{equation}
where the arrival rate $a_{\mathcal{M}_{q}}$ and service rate $r_{\mathcal{M}_{q}}$ should satisfy the condition as follows:
\vspace{-0.4em}
\begin{equation}\label{e12}
   \mathcal{EB}_{\!\mathcal{M}_{q}}\!(\theta_{\!\mathcal{M}_{q}}) \! \triangleq \! \frac{\ln \mathcal{M}_{\!\alpha_{\!\mathcal{M}_{q}}}\!(1\!+\!\theta_{\!\mathcal{M}_{q}})}{\theta_{\!\mathcal{M}_{q}}TB} \leq \mathcal{EC}_{\!\mathcal{M}_{q}}\!(\theta_{\!\mathcal{M}_{q}}\!),
   \vspace{-0.4em}
\end{equation}
where $\mathcal{EB}_{\!\mathcal{M}_{q}}\!(\theta_{\!\mathcal{M}_{q}})$ denotes the effective bandwidth (EB) of the task assignment $\mathcal{R}_{\mathcal{M}_{q}}$. Note that EC and EB are dual concepts \cite{wu2003effective,yang2018low,amjad2019effective}, commonly employed to characterize the SQP performance of networks \cite{chen2023statistical,li2020qos,wang2019unified,zhang2018heterogeneous}. Specifically, EC denotes the maximum arrival rate that guarantees the statistical QoS requirements at a given service rate, while EB represents the minimum service rate that guarantees the statistical QoS requirements at a given arrival rate.

\vspace{-1em}
\section{Adaptive joint time-slot and active-discarding transmission scheme}
\vspace{-0.4em}
\par The theoretical framework developed in Sec. III provides dependable theoretical insights and guidance for the development of resource optimization problems with SQP. In this section, an optimal ADAPT-JTAAT transmission scheme is proposed for the multi-layer tiled 360$^{\circ}$ VR streaming architecture from SDVP and EC perspectives, respectively. Additionally, two scenarios: 1) loss-intolerant (w/o, loss) scenarios; and 2) loss-tolerant (w, loss) scenarios are considered.

\vspace{-1.5em}
\subsection{Optimal ADAPT-JTAAT Transmission Scheme from SDVP Perspective}
\vspace{-0.4em}
\subsubsection{Under (w/o,loss) scenarios} Setting $\!J_{\!\mathcal{M}_{q}} \!\!=\! 0, q \!\in \! \mathcal{Q}, \mathcal{M}_{q} \! \in \! \mathcal{H}_{q}$. Following the overlapping FoV-based optimal JUM task assignment scheme and SQP theoretical framework described in
Sec. II and Sec. III, the resource optimization problem with SDVP constraint can be formulated as follows:
\vspace{-1.5em}
\begin{subequations}\label{e13}
	\begin{align}
		\mathrm{\mathcal{P}1}:
		&\quad \min_{\left\{\boldsymbol{t}\right\}} \mathbb{E}_{\zeta}\bigg\{\sum_{q\in\mathcal{Q}} \sum_{\mathcal{M}_{q} \in \mathcal{H}_{q}} t_{\mathcal{M}_{q}} \bigg\}, \nonumber \\
		s.t.\;
        &\quad \mathbb{P}\left\{w_{\mathcal{M}_{q}} \geq w_{q}^{\ast}\right\} < \varepsilon_{q}, \ \forall q, \mathcal{M}_{q},\\
        &\quad \sum_{q\in\mathcal{Q}} \sum_{\mathcal{M}_{q} \in \mathcal{H}_{q}} t_{\mathcal{M},q} \leq T, \quad \forall \zeta,\\
        &\quad t_{\mathcal{M}_{q}} \ge 0, \ \forall q, \mathcal{M}_{q}.
	\end{align}
\end{subequations}

\par Problem $\mathcal{P}1$ is intractable to address since the analytical expression of SDVP $\mathbb{P}\!\left\{w_{\!\mathcal{M}_{q}} \! \geq w_{q}^{\ast}\right\}$ is typically unavailable. For this reason, we adopt the manageable UB-SDVP from (\ref{e8}) to replace the SDVP in constraint (\ref{e13}a). This allows us to reformulate problem $\mathcal{P}1$ as follows:
\vspace{-0.8em}
\begin{subequations}\label{e14}
	\begin{align}
		\mathrm{\mathcal{P}2}:
		&\ \min_{\left\{\boldsymbol{t}\right\}} \mathbb{E}_{\zeta}\bigg\{\sum_{q\in\mathcal{Q}} \sum_{\mathcal{M}_{q} \in \mathcal{H}_{q}} t_{\mathcal{M}_{q}} \bigg\}\nonumber,\\
		s.t.\;
        &\inf\limits_{0 \leq \theta_{\!\mathcal{M}_{q}} \! < \theta_{\!\mathcal{M}_{q}}^{max}} \!\! \left\{\!\!\frac{\mathcal{M}_{\!\delta_{\mathcal{M}_{q}}}^{w_{q}^{\ast}}\!\!(1\!-\!\theta_{\!\mathcal{M}_{q}}\!)}{1 \!-\! \mathcal{M}_{\alpha_{\!\mathcal{M}_{q}}}\!\!(1\!+\!\theta_{\!\mathcal{M}_{q}}\!)\mathcal{M}_{\!\delta_{\mathcal{M}_{q}}}\!(1\!-\!\theta_{\!\mathcal{M}_{q}}\!)}\!\!\right\}\! \leq \! \varepsilon_{q},\\
        & \mathcal{M}_{\alpha_{\!\mathcal{M}_{q}}}\!\!(1\!+\!\theta_{\!\mathcal{M}_{q}}\!)\mathcal{M}_{\delta_{\!\mathcal{M}_{q}}}\!\!(1\!-\!\theta_{\!\mathcal{M}_{q}}) \! < \! 1, \ \forall \zeta, q, \mathcal{M}_{q},\\
        & (\ref{e13}b) \ \rm{and}\  (\ref{e13}c),
        \vspace{-3.0em}
	\end{align}
\end{subequations}
where constraints (\ref{e13}b) and (\ref{e13}c) ensure that the time-slot allocated to the task assignment $\!\mathcal{R}_{\!\mathcal{M}_{q}}\!\!$ is non-negative and the cumulative time-slot consumption cannot surpass the total time-slot resource. Constraint (\ref{e14}a) specifies that the UB-SDVP for streaming the task assignment $\mathcal{R}_{\!\mathcal{M}_{q}}$ should not surpass the SDVP threshold of the $q$-th video quality layer. Constraint (\ref{e14}b) represents the corresponding stability condition for UB-SDVP\footnote{Streaming 360$^{\circ}$ VR over mmWave networks is plagued by resource allocation issues. The established facts show that unreasonable time-slot resource allocation in mmWave networks tends to result in poor time-slot utilization \cite{ford2017achieving}. This is because highly directional phased array is required to realize the desired directional gain. This results in the entire bandwidth being allocated to one user at a time. However, mmWave can transmit Mbytes of data even during a short slot (e.g., 0.1 ms) \cite{dutta2016mac}. This ultimately makes it prone to underutilization of resources.}. Regrettably, the complex form of constraint (\ref{e14}a) for UB-SDVP makes solving the non-convex problem $\mathcal{P}2$ challenging. Consequently, we deeply investigate the intrinsic properties of $\mathcal{P}2$ from a structure perspective.

\vspace{-1.0em}

\begin{myTheo}\label{theo1}
  For a given time-slot allocation strategy $\left\{\boldsymbol{t}\right\}$, the stability condition $S(\theta_{\!\mathcal{M}_{q}}\!)$ is a convex function with respect to $\theta_{\!\mathcal{M}_{q}}$, and has a value of 1 when with $\theta_{\!\mathcal{M}_{q}}\rightarrow 0$.
\end{myTheo}

\vspace{-1.2em}

\begin{proof}
  The proof of \emph{Theorem \ref{theo1}} is given in Appendix A.
\end{proof}

\vspace{-0.6em}

\par From \emph{Theorem \ref{theo1}}, for a given time-slot allocation strategy $\left\{\boldsymbol{t}\right\}$, there exists a unique $\theta_{\mathcal{M}_{q}}^{max}(\boldsymbol{t}) \!>\! 0$, such that $\forall \theta_{\!\mathcal{M}_{q}} \! \in \! (0,\theta_{\!\mathcal{M}_{q}}^{max}(\boldsymbol{t}))$, the stability condition $S(\theta_{\!\mathcal{M}_{q}}) \! < \! 1$ holds, and for any $\triangle \! > \! 0$, we have $S(\theta_{\mathcal{M}_{q}}^{max}(\boldsymbol{t}) \! + \! \triangle) \! \geq \! 1$. Inspired by \emph{Theorem \ref{theo1}}, we further derive \emph{Theorem \ref{theo2}}.

\vspace{-0.8em}
\begin{myTheo}\label{theo2}
  Within the feasible domain $(0,\theta_{\mathcal{M}_{q}}^{max}(\boldsymbol{t}))$, the steady-state function $\mathcal{K}_{\mathcal{M}_{q}}(\theta_{\mathcal{M}_{q}},-w_{q}^{\ast})$ is also a convex function with respect to $\theta_{\mathcal{M}_{q}}$.
\end{myTheo}

\vspace{-1.2em}

\begin{proof}
  The proof of \emph{Theorem \ref{theo2}} is given in Appendix B.
\end{proof}
\vspace{-0.5em}

\par According to \emph{Theorem \ref{theo2}}, a unique $\theta_{\!\mathcal{M}_{q}}^{\ast}\!(\boldsymbol{t})$ exists in the feasible domain $(0,\theta_{\!\mathcal{M}_{q}}^{max}(\boldsymbol{t}))$, such that the kernel function $\mathcal{K}(\theta_{\!\mathcal{M}_{q}}\!(\boldsymbol{t}),\!L_{q},\!w_{q}^{\ast})$ is minimized. Moreover, owing to the monotonicity of the infimum function $\inf\left\{\cdot\right\}$, the value of $\mathcal{K}(\theta_{\!\mathcal{M}_{q}}\!(\boldsymbol{t}),L_{q},\!w_{q}^{\ast})$ attains the UB-SDVP at the point $\theta_{\!\mathcal{M}_{q}}\!(\boldsymbol{t}) \!=\! \theta_{\!\mathcal{M}_{q}}^{\ast}\!(\boldsymbol{t})$. Additionally, it can be easily proven that $\mathcal{V}(w_{q}^{\ast},\boldsymbol{t},\!\boldsymbol{\mathcal{J}})$ decreases monotonically with the increase of time-slot $t_{\!\mathcal{M}_{q}}$ (where $\boldsymbol{\mathcal{J}} \!=\! 0$). This decrease in $\mathcal{V}(w_{q}^{\ast},\boldsymbol{t},\boldsymbol{\mathcal{J}})$ being a monotonically decreasing function with respect to $t_{\!\mathcal{M}_{q}}$, as defined by the Mellin transform. Thus, if $\mathcal{V}(w_{q}^{\ast},\boldsymbol{t},\!\boldsymbol{\mathcal{J}})\!<\! \varepsilon_{q}$, the allocated time-slot is excessively large and should be reduced accordingly; otherwise, it should be increased. Based on these arguments, we propose a novel nested-shrinkage optimization algorithm for effectively solving $\mathcal{P}2$. The solution steps for $\mathcal{P}2$ are outlined in detail in \textbf{Algorithm 1}.

\subsubsection{Under (w,loss) scenarios} Next, we consider the active-discarding rate $J_{\mathcal{M}_{q}} \!\!>\!\! 0, q \!\in\! \mathcal{Q},\mathcal{M}_{q}\!\! \in \!\!\mathcal{H}_{q}$. The integration of the active-discarding strategy is indispensable since the throughput for streaming the task assignment $\!\mathcal{R}_{\!\mathcal{M}_{q}}\!\!$ of the user subset $\!\!\mathcal{M}_{q}\!\!$ is determined by the user with the worst CSI. Therefore, for the designed overlapping FoV-based optimal JUM task assignment scheme, data loss-tolerable is crucial for improving video buffering of 360$^{\circ}$ VR with huge data volumes. Given a tolerable loss-rate constraint $\mathcal{Y}_{q} > 0$ of the $q$-th layer, the loss rate of the task assignment $\!\mathcal{R}_{\!\mathcal{M}_{q}}\!\!$ can be defined as follows:

\vspace{-1.0em}
\begin{equation}\label{e15}
      \mathcal{Y}_{\!\mathcal{M}_{q}} = \frac{\mathbb{E}_{\zeta}\!\left[C_{\!\mathcal{M}_{q}}\right] - B\mathbb{E}_{\zeta}\!\left[t_{\!\mathcal{M}_{q}}\log_{2}\!\big(1 \!+\! \zeta^{'}\!\big)\right]}{\mathbb{E}_{\zeta}\!\left[C_{\!\mathcal{M}_{q}}\right]}= 1 \!-\! \frac{\mathbb{E}_{\zeta}\!\left[t_{\mathcal{M}_{q}}\!\log_{2}\!\big(1 \!+\! \zeta^{'}\big)\right]}{\mathbb{E}_{\zeta}\!\left[t_{\!\mathcal{M}_{q}}\!\log_{2}\!\left(1 \!+\! \zeta^{'}\right) \!+\! \mathcal{J}_{\!\mathcal{M}_{q}}\right]}.
      \vspace{-1.2em}
\end{equation}
\vspace{-0.8em}
\par To effectively suppress video buffering, it is critical to develop an optimal ADAPT-JTAAT transmission scheme that achieves flexible rate control and robust queue behaviors, striking a good balance between the active-discarding rate and the QoE of users. This can be accomplished by formulating and addressing the resource optimization problem as follows:
\vspace{-0.6em}
\begin{subequations}\label{e16}
	\begin{align}
		\mathrm{\mathcal{P}3}:
		&\ \min_{\left\{\boldsymbol{t},\boldsymbol{\mathcal{J}}\right\}} \mathbb{E}_{\zeta}\bigg\{\sum_{q\in\mathcal{Q}} \sum_{\mathcal{M}_{q} \in \mathcal{H}_{q}} t_{\mathcal{M}_{q}} \bigg\} ,\nonumber\\
		s.t.\;
        &\inf\limits_{0 \leq \theta_{\!\mathcal{M}_{q}} \! < \theta_{\!\mathcal{M}_{q}}^{max}} \!\! \left\{\!\!\frac{\mathcal{M}_{\!\delta_{\mathcal{M}_{q}}}^{w_{q}^{\ast}}\!\!(1\!-\!\theta_{\!\mathcal{M}_{q}}\!)}{1 \!-\! \mathcal{M}_{\alpha_{\!\mathcal{M}_{q}}}\!\!(1\!+\!\theta_{\!\mathcal{M}_{q}}\!)\mathcal{M}_{\!\delta_{\mathcal{M}_{q}}}\!(1\!-\!\theta_{\!\mathcal{M}_{q}}\!)}\!\!\right\}\! \leq \! \varepsilon_{q},\\
        & \mathcal{M}_{\alpha_{\!\mathcal{M}_{q}}}\!\!(1\!+\!\theta_{\!\mathcal{M}_{q}}\!)\mathcal{M}_{\delta_{\!\mathcal{M}_{q}}}\!\!(1\!-\!\theta_{\!\mathcal{M}_{q}}) \! < \! 1, \ \forall \zeta, q, \mathcal{M}_{q},\\
        & 1\!-\!\frac{\mathbb{E}_{\zeta}\!\left[t_{\!\mathcal{M}_{q}}\!\log_{2}\!\left(\!1 \!+\! \zeta^{'}\!\right)\!\right]}{\mathbb{E}_{\zeta}\!\left[t_{\!\mathcal{M}_{q}}\!\log_{2}\!\left(1 \!+\! \zeta^{'}\right) \!+\! \mathcal{J}_{\!\mathcal{M}_{q}}\right]} \!\leq\! \mathcal{Y}_{q}, \forall \zeta, q, \mathcal{M}_{q},\\
        &\mathcal{J}_{\mathcal{M}_{q}} > 0, \ \forall q, \mathcal{M}_{q},\\
        &(\ref{e13}b)-(\ref{e13}c),
	\end{align}
\end{subequations}
where constraint $(\ref{e16}c)$ denotes the loss rate of streaming the task assignment $\mathcal{R}_{\!\mathcal{M}_{q}}$ cannot exceed a loss rate constraint $\mathcal{Y}_{q}$, and constraint (\ref{e16}d) ensures that the active-discarding rate of streaming $\mathcal{R}_{\!\mathcal{M}_{q}}$ is greater than zero.

\vspace{-0.8em}

\begin{myProp}
   Given a fixed active-discarding rate $\tilde{\mathcal{J}}_{\!\mathcal{M}_{q}} > 0$, the optimal time-slot allocation $t_{\mathcal{M}_{q}}^{\star}$ of $\mathcal{R}_{\!\mathcal{M}_{q}}$ depends on which of the more time-slot resource is required to satisfy the UB-SDVP constraint (\ref{e16}a) and the tolerable loss rate constraint (\ref{e16}c).
\end{myProp}

\vspace{-1em}

\par \emph{Proposition 1} is easy to prove. Given a fixed active-discarding rate $\tilde{\mathcal{J}}_{\mathcal{M}_{q}} > 0$, $\mathcal{P}3$ degenerates into the subproblem $\mathcal{P}3^{\prime}$, as follows:
\vspace{-0.5em}
\begin{subequations}\label{e17}
	\begin{align}
		\mathrm{\mathcal{P}3^{\prime}}\!:\!
		&\ \min_{\left\{\boldsymbol{t}\right\}} \mathbb{E}_{\zeta}\bigg\{\sum_{q\in\mathcal{Q}} \sum_{\mathcal{M}_{q} \in \mathcal{H}_{q}} t_{\mathcal{M}_{q}} \bigg\},\nonumber\\
		\!\!\!\!\! s.t.\;
        &\ \mathbb{E}_{\zeta}\!\left[t_{\mathcal{M}_{q}}\!\log_{2}\!\!\left(\!1\!+\!\zeta^{'}\!\right)\!\right] \! \geq \! \frac{1\!-\!\mathcal{Y}_{q}}{\mathcal{Y}_{q}}\mathbb{E}_{\zeta}\!\left[\tilde{\mathcal{J}}_{\!\mathcal{M}_{q}}\!\right],\\
        &\ (\ref{e13}b)-(\ref{e13}c), (\ref{e16}a), (\ref{e16}b),\ \rm{and}\ (\ref{e16}d).
	\end{align}
\end{subequations}

\vspace{-1.5em}

\par It can be observed that constraints (\ref{e17}a) and (\ref{e17}b) are only coupled with respect to $t_{\!\mathcal{M}_{q}}$, thus problem $\mathcal{P}3^{\prime}$ can be further decomposed into two subproblems, as follows:

\vspace{-2.5em}

\begin{subequations}\label{P3-1-1}
	\begin{align}
		\mathrm{\mathcal{P}3^{\prime}\text{-}1}\!:\!
		&\ \min_{\left\{\boldsymbol{t}\right\}} \mathbb{E}_{\zeta}\bigg\{\sum_{q\in\mathcal{Q}} \sum_{\mathcal{M}_{q} \in \mathcal{H}_{q}} t_{\mathcal{M}_{q}} \bigg\},\quad \quad \quad \quad  \mathrm{\mathcal{P}3^{\prime}\text{-}2}\!:\!\min_{\left\{\boldsymbol{t}\right\}} \mathbb{E}_{\zeta}\bigg\{\sum_{q\in\mathcal{Q}} \sum_{\mathcal{M}_{q} \in \mathcal{H}_{q}} t_{\mathcal{M}_{q}} \bigg\},\nonumber\\
		\!\!\!\!\! s.t.\;
        &\ (\ref{e17}a).  \quad \quad \quad \quad \quad \quad \quad \quad \quad \quad \quad \quad \quad  s.t.\; \ (\ref{e17}b).\nonumber
	\end{align}
\end{subequations}

\vspace{-1.0em}

\par $\mathcal{P}3^{\prime}\text{-}1$ is a standard convex problem, whose optimal solution can be represented by $\boldsymbol{\breve{t}}^{\star}_{1}(\boldsymbol{\tilde{\mathcal{J}}})\!\!=\!\!\big\{\breve{t}_{\!\mathcal{M}_{q}}(\tilde{\mathcal{J}}_{\!\mathcal{M}_{q}}\!)\big\}$. And $\mathcal{P}3^{\prime}\text{-}2$ is equivalent to $\mathcal{P}2$, whose optimal solution $\boldsymbol{\tilde{t}}^{\star}_{1}(\boldsymbol{\tilde{\mathcal{J}}}) \!\!=\!\! \big\{\tilde{t}_{\!\mathcal{M}_{q}}(\tilde{\mathcal{J}}_{\!\mathcal{M}_{q}})\big\}$ can be obtained by executing \textbf{Algorithm 1}. To make the constraints (\ref{e17}a) and (\ref{e17}b) satisfied simultaneously, the optimal time-slot allocation should be selected as $\boldsymbol{t}^{\star}(\boldsymbol{\tilde{\mathcal{J}}}) \!\!=\!\! \left\{\max\!\left\{\tilde{t}_{\mathcal{M}_{q}}(\tilde{\mathcal{J}}_{\mathcal{M}_{q}}),\breve{t}_{\mathcal{M}_{q}}(\tilde{\mathcal{J}}_{\mathcal{M}_{q}})\!\right\}\!\!\right\}$, where $q\in\mathcal{Q},\mathcal{M}_{q}\in\mathcal{H}_{q}$. On the other hand, we can prove that $\mathcal{V}(w_{q}^{\ast},\!\boldsymbol{t},\!\boldsymbol{\mathcal{J}})$ is a monotonically decreasing function with respect to the active-discarding rate $\!\! \mathcal{J}_{\!\mathcal{M}_{q}}\!$, since $\mathcal{V}(w_{q}^{\ast},\!\boldsymbol{t},\!\boldsymbol{\mathcal{J}})$ is a monotonically increasing function of $\mathcal{M}_{\delta_{\!\mathcal{M}_{q}}}\!(1\!-\!\theta_{\!\mathcal{M}_{q}}\!)$, and  $\mathcal{M}_{\delta_{\mathcal{M}_{q}}}\!(1\!-\!\theta_{\!\mathcal{M}_{q}}\!)$ decreases monotonically with respect to the active-discarding rate $\mathcal{J}_{\!\mathcal{M}_{q}}$. Given a fixed time-slot allocation $\left\{\boldsymbol{\tilde{t}}\right\}$, $\mathcal{P}3$ can be equivalently simplified as determining the optimal active-discarding rate, which should satisfy both the UB-SDVP constraint and the loss rate constraint, as follows:

\vspace{-2.8em}

\begin{subequations}\label{e20}
         \begin{align}
         & \mathcal{V}(w_{q}^{\ast},\!\boldsymbol{\tilde{t}},\!\boldsymbol{\mathcal{J}})\leq \varepsilon_{q}, \\
         & 0 < \left(1-\mathcal{Y}_{q}\right)\mathbb{E}_{\zeta}[\mathcal{J}_{\!\mathcal{M}_{q}}] \!\leq \!\mathcal{Y}_{q}\mathbb{E}\!\left[\tilde{t}_{\!\mathcal{M}_{q}}\!\log_{2}\!\left(1 \!+\! \zeta^{'}\right)\right].
         \end{align}
\end{subequations}
\vspace{-2.5em}

\par For constraint (\ref{e20}a), by exploiting the similar process of \emph{Algorithm 1}, a minimum active-discarding rate $\!\!\mathcal{J}_{\!\mathcal{M}_{q}}^{min}\!\!\left(\boldsymbol{\tilde{t}}\right)\!\!$ can be determined. For constraint (\ref{e20}b), a maximum active-discarding rate $\mathcal{J}_{\!\mathcal{M}_{q}}^{max}\!\!\left(\boldsymbol{\tilde{t}}\right)$ that satisfies $\mathbb{E}_{\zeta}\!\big[\mathcal{J}_{\mathcal{M}_{q}}^{max}\big] \!=\! \frac{\mathcal{Y}_{th}}{1 - \mathcal{Y}_{th}} \mathbb{E}_{\zeta}\!\big[\tilde{t}_{\mathcal{M}_{q}}\!\log_{2}\big(1 \!+\! \zeta^{'}\big)\big]\!\!$ can be easily searched. If $\mathcal{J}_{\!\mathcal{M}_{q}}^{min}\!\!\left(\boldsymbol{\tilde{t}}\right) \! \leq \! \mathcal{J}_{\!\mathcal{M}_{q}}^{max}\!\!\left(\boldsymbol{\tilde{t}}\right)$, the effective range of $\mathcal{J}_{\!\mathcal{M}_{q}}\!\!\left(\boldsymbol{\tilde{t}}\right)$ is denoted as $\mathcal{J}_{\!\mathcal{M}_{q}}^{min}\!\!\left(\boldsymbol{\tilde{t}}\right)\leq \mathcal{J}_{\!\mathcal{M}_{q}}\!\!\left(\boldsymbol{\tilde{t}}\right) \leq \mathcal{J}_{\!\mathcal{M}_{q}}^{max}\!\!\left(\boldsymbol{\tilde{t}}\right)$; otherwise, if $\!\mathcal{J}_{\!\mathcal{M}_{q}}^{min}\!\!\left(\boldsymbol{\tilde{t}}\right) \!>\!  \mathcal{J}_{\!\mathcal{M}_{q}}^{max}\!\!\left(\boldsymbol{\tilde{t}}\right)$, it means that the limited wireless resources cannot satisfy the statistical QoS provisioning guarantees with respect to SDVP and tolerable loss rate of streaming 360$^{\circ}$ VR video data simultaneously. Based on the aforementioned arguments, a novel stepwise-approximation optimization algorithm is proposed to address $\mathcal{P}3$, whose optimal solution can be denoted as $\!\left\{\boldsymbol{t}^{\star},\boldsymbol{\mathcal{J}}^{\star}\right\}$. The details are listed in \textbf{Algorithm 2}, whose execution sketch is depicted in Fig. 3.

\vspace{-1em}

\begin{algorithm}[h]
\small
\setstretch{0.7}
        \caption{Nested-Shrinkage Optimization Algorithm}
        \KwIn{ $\left(w_{q}^{\ast},\varepsilon_{q},L_{q}\right)$; $\Psi_{th}$; $\Phi_{th}$; $\Delta_{s}$; $N_{max}$; $\kappa_{th}$;
}
        Set the lower bound $\underline{t}_{\mathcal{M}_{q}}(n) = 0$, the upper bound $\overline{t}_{\mathcal{M}_{q}}(n) = T$, iteration index $n = 1$, and the UB-SDVP $\tilde{\varepsilon}_{\mathcal{M}_{q}} = 0$;\\
        \While{$n \leq N_{max}$ \rm{\textbf{and}} $\left|\frac{\tilde{\varepsilon}_{\mathcal{M}_{q}}}{\varepsilon_{q}}-1\right| > \kappa_{th}$}{
           Set $t_{\mathcal{M}_{q}} = (\underline{t}_{\mathcal{M}_{q}}(n) + \overline{t}_{\mathcal{M}_{q}}(n))/2$;\\
           \!\!\!\!\!\tcc{\!\!Step1: \!\!Determine feasible domain\!\!\!}  
           Set $\theta_{\mathcal{M}_{q}}^{max}(t_{\mathcal{M}_{q}}) = 0$, step length $\Phi_{s} = 1$;\\
           \While{$\Phi_{s} > \Phi_{th}$}{
               \eIf{$S(\theta_{\mathcal{M}_{q}}^{max}(t_{\mathcal{M}_{q}} + \Phi_{s})) < 1$}
               {$\theta_{\mathcal{M}_{q}}^{max}(t_{\mathcal{M}_{q}}) = \theta_{\mathcal{M}_{q}}^{max}(t_{\mathcal{M}_{q}}) + \Phi_{s}$;}
               {$\Phi_{s} = \Phi_{s}/2$;}
            }
           \!\!\!\!\!\tcc{\!\!Step2: \!\!Determine \!\!$\theta_{\mathcal{M}_{q}}^{\ast}(t_{\mathcal{M}_{q}}\!) \!\in\! (0,\theta_{\mathcal{M}_{q}}^{max})$\!\!\!\!\!\!\!\!}
           Set $\theta_{\mathcal{M}_{q}}^{\ast} = \frac{1}{2}\theta_{\mathcal{M}_{q}}^{max}(t_{\mathcal{M},q})$;\\
           Calculate the gradient of $\mathcal{K}_{\mathcal{M}_{q}}(\theta_{\mathcal{M}_{q}},L_{q},w_{q}^{\ast})$ at $\theta_{\mathcal{M}_{q}} = \theta_{\mathcal{M}_{q}}^{\ast}$:
           $\nabla\mathcal{K}_{\mathcal{M}_{q}}(\theta_{\mathcal{M}_{q}}^{\ast},L_{q},w_{q}^{\ast}) = (\nabla\mathcal{K}_{\mathcal{M}_{q}}(\theta_{\mathcal{M}_{q}}^{\ast}+\Psi_{th},L_{q},w_{q}^{\ast})-\nabla\mathcal{K}_{\mathcal{M}_{q}}(\theta_{\mathcal{M}_{q}}^{\ast},L_{q},w_{q}^{\ast}))/\Psi_{th}$;\\
           \While{$\left|\nabla\mathcal{K}_{\mathcal{M}_{q}}(\theta_{\mathcal{M}_{q}}^{\ast},L_{q},w_{q}^{\ast})\triangle_{s}\right| > \Psi_{th}$}{
               $\theta_{\mathcal{M}_{q}}^{\ast} = \theta_{\mathcal{M}_{q}}^{\ast} - \nabla\mathcal{K}_{\mathcal{M}_{q}}(\theta_{\mathcal{M}_{q}}^{\ast},L_{q},w_{q}^{\ast})\triangle_{s}$;\\
               Update $\!\nabla\!\mathcal{K}_{\mathcal{M}_{q}}\!(\!\theta_{\mathcal{M}_{q}}^{\ast},L_{q},w_{q}^{\ast}\!)\triangle_{s}$ according to line $13$;\\
            }
           Update the UB-SDVP $\tilde{\varepsilon}_{\mathcal{M}_{q}} = \mathcal{K}_{\mathcal{M}_{q}}\left(\theta_{q}^{\ast},L_{q},w_{q}^{\ast}\right)$;\\
           \!\!\!\!\!\tcc{\!\!Step3: \!\!Nested-Sectioning for $t_{\mathcal{M}_{q}}$\!\!\!\!\!}
           \eIf{$\tilde{\varepsilon}_{\mathcal{M}_{q}} > \varepsilon_{q}$}
           {$\underline{t}_{\mathcal{M}_{q}}(n) = (\underline{t}_{\mathcal{M}_{q}}(n) + \overline{t}_{\mathcal{M}_{q}}(n))/2$;}
           {$\overline{t}_{\mathcal{M}_{q}}(n) = (\underline{t}_{\mathcal{M}_{q}}(n) + \overline{t}_{\mathcal{M}_{q}}(n))/2$;}
           $n = n +1$;\\
        }
       \KwOut{$\left\{\boldsymbol{t}^{\star}\right\} = \left\{t_{\mathcal{M}_{q}}\right\}_{q\in\mathcal{Q},\mathcal{M}_{q}\in\mathcal{H}_{q}}$.}
\end{algorithm}

\vspace{-2em}

\subsection{Optimal ADAPT-JTAAT Transmission Scheme from EC Perspective}
\par The service rate of the task assignment $\mathcal{R}_{\!\mathcal{M}_{q}}$ can be denoted as $r_{\!\mathcal{M}_{q}}/T, q \!\in\! \mathcal{Q},\mathcal{M}_{q} \!\in\! \mathcal{H}_{q}$ (in bits/s). According to (\ref{e9}), the EC can be reformulated as follows:
\begin{equation}\label{e21}
   \mathcal{EC}_{\!\mathcal{M}_{q}}\!\left(\theta_{\!\mathcal{M}_{q}}\right) = -\frac{1}{\theta_{\mathcal{M}_{q}}^{\ast}BT}\ln\!\left(\!\mathbb{E}_{\zeta}\!\left[e^{-\frac{\theta_{\mathcal{M}_{q}}^{\ast}r_{\mathcal{M}_{q}}}{T} }\right]\right).
\end{equation}

\vspace{-0.5em}

\par Based on (\ref{e12}), in order to effectively guarantee the statistical QoS requirements $\left(L_{q},w_{q}^{\ast},\varepsilon_{q},\mathcal{Y}_{q}\right)$, the EC of $\mathcal{R}_{\!\mathcal{M}_{q}}$ should not less than corresponding EB \cite{yang2018low,amjad2019effective,chen2023statistical}, as follows:

\vspace{-1.5em}

\begin{equation}\label{e22}
   \mathcal{EC}_{\!\mathcal{M}_{q}}\!\left(\theta_{\!\mathcal{M}_{q}}^{\ast}\right) \geq |\mathcal{R}_{\!\mathcal{M}_{q}}| \mathcal{EB}_{\!\mathcal{M}_{q}}\!\left(\theta_{\mathcal{M}_{q}}^{\ast}|L_{q}\right).
   \vspace{-1em}
\end{equation}
where $|\mathcal{R}_{\mathcal{M}_{q}}|$ represents the number of tiles of the subset $\mathcal{M}_{q}$.

\vspace{-1.5em}

\begin{figure}[h] 
\centering
\includegraphics[scale=0.1]{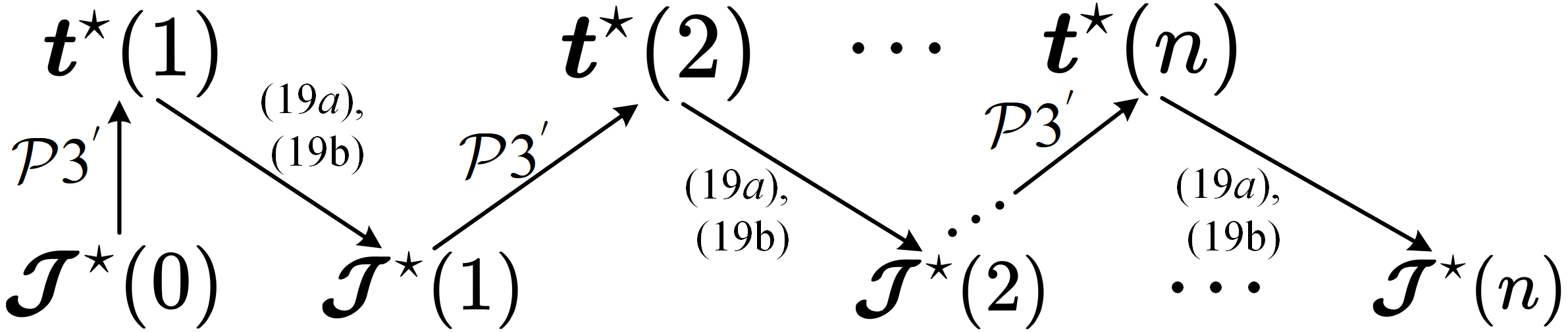}
\vspace{-1em}
\caption{The execution sketch of Algorithm 2.}
\label{fig:label}
\end{figure}

\vspace{-1.5em}

\subsubsection{Under (w/o,loss) scenarios} Setting $\!\mathcal{J}_{\mathcal{M}_{q}} \!=\! 0, q\!\in\!\mathcal{Q},\mathcal{M}_{q}\!\in\!\mathcal{H}_{q}$. The resource optimization problem with EC constraint can be formulated as follows:
\vspace{-0.6em}
\begin{subequations}\label{e23}
	\begin{align}
		\mathrm{\mathcal{P}4}:
		&\quad \min_{\left\{\boldsymbol{t}\right\}} \mathbb{E}_{\zeta}\bigg\{\sum_{q\in\mathcal{Q}} \sum_{\mathcal{M}_{q} \in \mathcal{H}_{q}} t_{\mathcal{M}_{q}} \bigg\}, \nonumber \\
		s.t.\;
        & \quad (\ref{e22}), (\ref{e13}b),\ \rm{and}\ (\ref{e13}c).
	\end{align}
\end{subequations}

\vspace{-1.5em}

\par It can be proven that $\mathcal{P}4$ is convex, and the detailed proof is described in \emph{Theorem \ref{theo3}} below:

\vspace{-0.8em}

\begin{myTheo}\label{theo3}
  The problem $\mathcal{P}4$ is a standard convex optimization problem.
\end{myTheo}

\vspace{-1.5em}

\begin{proof}
  The constraint (\ref{e22}) can be equivalently reformulated as follows:
\vspace{-0.3em}
  \begin{equation}\label{e24}
    \!\! \mathbb{E}_{\zeta}\!\left[e^{-\Theta_{\mathcal{M}_{q}}^{\ast}t_{\mathcal{M}_{q}}\ln\left(1+\zeta^{'}\right)}\right] \leq e^{-\theta_{\mathcal{M}_{q}}^{\ast}\left|\mathcal{R}_{\mathcal{M}_{q}}\right|\overline{\mathcal{EB}}_{\mathcal{M}_{q}}},
    \vspace{-1em}
  \end{equation}
where $\overline{\mathcal{EB}}_{\!\mathcal{M}_{q}} \!=\! BT\!\cdot\! \mathcal{EB}_{\!\mathcal{M}_{q}}\!\left(\theta_{\mathcal{M}_{q}}^{\ast}|L_{q}\right)$, and $\Theta_{\mathcal{M}_{q}}^{\ast} \triangleq \frac{\theta_{\mathcal{M}_{q}}^{\ast}B}{T\ln 2}$ denotes the normalized statistical QoS exponent.
 \par The right side of inequality (\ref{e24}) is constant. Let function $F\!\left(t_{\!\mathcal{M}_{q}}\!\right)$ represent the left part of inequality (\ref{e24}), and taking the first- and second-order partial derivatives of $F\!\left(t_{\!\mathcal{M}_{q}}\!\right)$ with respect to $t_{\!\mathcal{M}_{q}}$, yields:
\begin{small}
\begin{equation*}\label{e}
   \begin{aligned}
     & \!\!\!\frac{\partial F\!\left(t_{\!\mathcal{M}_{q}}\!\right)}{\partial t_{\!\mathcal{M}_{q}}}
      = -\Theta_{\mathcal{M}_{q}}^{\ast}\mathbb{E}_{\zeta}\!\bigg[e^{-\Theta_{\mathcal{M}_{q}}^{\ast}t_{\mathcal{M}_{q}}\!\ln\left(1+\zeta^{'}\right)}\cdot\ln\!\big(1\!+\!\zeta^{'}\big)\bigg] < 0,\\
    & \!\!\!\frac{\partial^{2} F\!\left(t_{\mathcal{M}_{q}}\!\right)}{\partial t_{\!\mathcal{M}_{q}}^{2}} = \mathbb{E}_{\zeta}\!\left\{\!e^{-\Theta_{\mathcal{M}_{q}}^{\ast}t_{\!\mathcal{M}_{q}}\!\ln\big(1+\zeta^{'}\big)}\!\cdot\!\left(\!\Theta_{\mathcal{M}_{q}}^{\ast} \ln\!\big(1+\zeta^{'}\big)\!\right)^{2}\!\right\}>0.
   \end{aligned}
   \vspace{-0.5em}
  \end{equation*}
\end{small}
\vspace{-1em}
\par Therefore, the function $F\left(t_{\mathcal{M}_{q}}\right)$ is a strictly decreasing convex function of $t_{\mathcal{M}_{q}}$. So the proof of \emph{Theorem \ref{theo3}} is concluded.
\end{proof}

\vspace{-1em}

\par From \emph{Theorem \ref{theo3}}, $\mathcal{P}4$ can be equivalently reformulated as follows:
\vspace{-1.0em}
\begin{subequations}\label{e25}
	\begin{align}
		\mathrm{\mathcal{P}4^{'}}:
		&\quad \min_{\left\{\boldsymbol{t}\right\}} \mathbb{E}_{\zeta}\bigg\{\sum_{q \in \mathcal{Q}} \sum_{\mathcal{M}_{q} \in \mathcal{H}_{q}} t_{\mathcal{M}_{q}} \bigg\}, \nonumber \\
		s.t.\;
		&\quad F\left(t_{\mathcal{M}_{q}}\right) \leq e^{-\theta_{\mathcal{M}_{q}}^{\ast}\left|\mathcal{R}_{\mathcal{M}_{q}}\right|\overline{\mathcal{EB}}_{\mathcal{M}_{q}}},\forall q, \mathcal{M}_{q}, \\
        &\quad (\ref{e13}b),\ \rm{and}\ (\ref{e13}c).
	\end{align}
\end{subequations}

\vspace{-1.5em}

\par Based on the convex optimization theory \cite{boyd2004convex}, $\mathcal{P}4^{'}$ can be solved by exploiting Karush-Kuhn-Tucker (KKT) method, which is summarized in \emph{Theorem \ref{theo4}}:

\vspace{-1.0em}

\begin{myTheo}\label{theo4}
  If the optimal solution $\boldsymbol{t}^{\star}$ of $\mathcal{P}4^{'}$ exists, it can be determined by
\vspace{-0.5em}
\begin{small}
  \begin{equation}\label{e}
     \boldsymbol{t}^{\star} = \left[\!\frac{\ln\!\left(\!\boldsymbol{\psi}^{\star}\!g\big(\theta_{\mathcal{M}_{q}}^{\ast},\zeta\big)\!\right)\!-\!\ln\left(\!1\!+\!\mu_{\zeta}^{\star}\!\right)}{g\!\!\left(\!\theta_{\mathcal{M}_{q}}^{\ast},\zeta\!\right)}\!\right]^{+},\forall q, \mathcal{M}_{q},\forall \zeta,
     \vspace{-0.5em}
  \end{equation}
\end{small}
where $[x]^{+}\! \triangleq \!\max\{x,0\}$, and $g\big(\theta_{\mathcal{M}_{q}}^{\ast},\zeta\big) = \Theta_{\mathcal{M}_{q}}^{\ast}\!\ln\!\big(1\!+\!\zeta^{'}\big)$. $\mu_{\zeta}^{\star}$ and $\boldsymbol{\psi} \!\! \triangleq \!\! \big\{\psi_{\mathcal{M}_{q}}^{\star}\!\big\}_{q \in \mathcal{Q},\mathcal{M}_{q}\in\mathcal{H}_{q}}\!\!$ are optimal Lagrange Multipliers associated with (\ref{e25}a) and (\ref{e25}b), respectively.
\end{myTheo}

\vspace{-1.7em}

\begin{proof}
  The proof of \emph{Theorem \ref{theo4}} is given in Appendix C.
\end{proof}

\vspace{-1.7em}

\begin{algorithm}[h]
\small
\setstretch{0.80}
        \caption{Stepwise-Approximation optimization Algorithm}
        \KwIn{$\left(w_{q}^{\ast},\varepsilon_{q},L_{q}\right)$; $\mathcal{Y}_{th}$; Maximum iteration number $I_{max}$; Convergence accuracy $\Omega_{th}$;}
        Initialize $\boldsymbol{\mathcal{J}}^{\star}(0)$;\\
        Set iteration index $n = 0$;\\
        \tcc{For all $q \in \mathcal{Q}$, and $\mathcal{M}_{q}\in\mathcal{H}_{q}$, execute the following program:} %
        \While{$\!\!\big|\!(\!1\!-\!\mathcal{Y}_{q}\!)\mathbb{E}_{\zeta}\!\big[\!\mathcal{J}_{\!\mathcal{M}_{q}}^{\star}\!\big] \!-\! \mathcal{Y}_{q}\mathbb{E}_{\zeta}\!\big[t_{\!\mathcal{M}_{q}}^{\star}\!\!\log_{2}(1\!+\!\zeta^{'}\!)\!\big]\!\big| \!<\! \Omega_{th}$}
        {
           Use $\mathcal{J}^{\star}_{\mathcal{M}_{q}}(n-1)$ as input;\\
           Execute \textbf{Algorithm 2} to solve $\mathcal{P}3^{\prime}$ and obtain $t^{\star}_{\mathcal{M}_{q}}(n)$;\\
           Use $t^{\star}_{\mathcal{M}_{q}}(n)$ as input;\\
           Determine $\mathcal{J}_{\mathcal{M}_{q}}^{max}$ according to (\ref{e20}a);\\
           Determine $\mathcal{J}_{\mathcal{M}_{q}}^{min}$ according to (\ref{e20}b);\\
           \eIf{$\mathcal{J}_{\mathcal{M}_{q}}^{min} \leq \mathcal{J}_{\mathcal{M}_{q}}^{max}$}
           {
             Set $\mathcal{J}^{\star}_{\mathcal{M}_{q}}(n) = \mathcal{J}_{\mathcal{M}_{q}}^{max}$;\\
           }
           {
             \rm{\textbf{return:}} \texttt{QoS is too stringent!};\\
           }
           Set $n = n+1$;\\
        }
       \KwOut{$\left\{\boldsymbol{t}^{\star},\boldsymbol{\mathcal{J}}^{\star}\right\} = \left\{t^{\star}_{\mathcal{M}_{q}}(n),\mathcal{J}^{\star}_{\mathcal{M}_{q}}(n)\right\}_{q\in\mathcal{Q},\mathcal{M}_{q}\in\mathcal{H}_{q}}$.}
\end{algorithm}

\vspace{-1.7em}

\par According to \emph{Theorem \ref{theo4}}, to determine the optimal solution $\boldsymbol{t}^{\star}$, we need first to obtain the value of Lagrange multipliers $\boldsymbol{\psi}^{\star}$ and $\mu_{\zeta}^{\star}$. From \emph{Theorem \ref{theo4}}, we can obtain that $\boldsymbol{\psi}^{\star} \succeq \textbf{0}$ since $\boldsymbol{\psi} \rightarrow 0^{+}$ implies $\boldsymbol{t}^{\star} \rightarrow 0$, which violate the EC-based SQP constraint (\ref{e25}a). Then, given $\mu_{\zeta}$, we can construct the dual problem $\mathcal{P}5$ by utilizing convex theory \cite{boyd2007notes}, as follows:


\vspace{-2.3em}

\begin{subequations}\label{e27}
	\begin{align}
		\mathrm{\mathcal{P}5}:
		&\quad \max_{\left\{\boldsymbol{\psi}\right\}} D\!\left(\boldsymbol{\psi},\mu_{\zeta}(\boldsymbol{\psi})\right), \\
		s.t.\;
		&\quad \psi_{\mathcal{M}_{q}} > 0, \quad \forall q \in \mathcal{Q}, \mathcal{M}_{q} \in \mathcal{H}_{q},
	\end{align}
\vspace{-1.0em}
\end{subequations}
where
\begin{small}
\begin{equation}\label{e}
  \begin{aligned}
      & D(\boldsymbol{\psi},\mu_{\zeta}(\boldsymbol{\psi}))\triangleq \min_{\boldsymbol{t}} \sum\limits_{q\in\mathcal{Q}}\sum_{\mathcal{M}_{q}\in\mathcal{H}_{q}}\!\!\! t_{\!\mathcal{M}_{q}} + \sum\limits_{q\in\mathcal{Q}}\sum_{\mathcal{M}_{q}\in\mathcal{H}_{q}}\!\!\! \psi_{\mathcal{M}_{q}}\times \left(\!F\!\left(t_{\mathcal{M}_{q}}\!\right)\!-\!e^{-\theta_{\mathcal{M}_{q}}^{\ast}\!\left|\mathcal{R}_{\!\mathcal{M}_{q}}\!\right|\overline{\mathcal{EB}}_{\!\mathcal{M}_{q}}}\!\!\right),\\
      & s.t.\quad (\ref{e13}a) \ \mathrm{and} \ (\ref{e13}b),
  \end{aligned}
  \vspace{-1.0em}
\end{equation}
\end{small}
where $\mu_{\zeta}$ is regarded as the function of $\boldsymbol{\psi}$ in this dual problem.

\par Note that the Slater condition of problem $\mathcal{P}5$ is satisfied due to the convexity of problem $\mathcal{P}4$, and consequently, the dual gap distance between $\mathcal{P}4$ and $\mathcal{P}5$ is zero. Thus, $\mathcal{P}5$ shares the same optimal solution with $\mathcal{P}4$ \cite{boyd2007notes}. By solving the dual problem $\mathcal{P}5$, we can obtain the value of the optimal $\boldsymbol{\psi}^{\star}$. Subsequently, we obtain the value of the Lagrange multiplier $\mu_{\zeta}^{\star}$ according to the following theorem:

\vspace{-0.8em}

\begin{myTheo}\label{theo5}
   Given $\zeta$ and $\psi_{\mathcal{M}_{q}}^{\star}$, there is either a unique solution $\mu_{\zeta}^{\star} > 0$ satisfies the following equation
\vspace{-0.5em}
  \begin{equation}\label{e}
    \sum_{q\in\mathcal{Q}}\sum_{\mathcal{M}_{q}\in \mathcal{H}_{q}} t_{\mathcal{M}_{q}}(\zeta,\psi_{\mathcal{M}_{q}}^{\star},\mu_{\zeta}^{\star}) = T,
  \end{equation}
\vspace{-0.5em}
  if and only if
\vspace{-0.5em}
  \begin{equation}\label{e}
  \lim_{\mu_{\zeta}\rightarrow 0}\sum\limits_{q\in\mathcal{Q}}\sum\limits_{\mathcal{M}_{q}\in\mathcal{H}_{q}}t_{\mathcal{M}_{q}}\left(\zeta,\psi_{\mathcal{M}_{q}}^{\star},\mu_{\zeta}\right) \geq T,
\end{equation}
\vspace{-0.5em}
 holds; otherwise $\mu_{\zeta}^{\star} = 0$.
\end{myTheo}

\vspace{-1.2em}

\begin{proof}
  The proof of \emph{Theorem \ref{theo5}} is given in Appendix D.
\end{proof}

\vspace{-0.5em}

\par According to \emph{Theorem \ref{theo5}}, the solution of $\mu_{\zeta}^{\star}$ is unique, and can be easily obtained since  $t_{\mathcal{M}_{q}}(\zeta,\psi_{\mathcal{M}_{q}}^{\star},\mu_{\zeta}^{\star})$ is a decreasing function w.r.t. $\mu_{\zeta}^{\star}$. In order to obtain $\boldsymbol{\psi^{\star}}$, we propose a subgradient-based optimization algorithm to effectively determine it. We select the decay step size sequence $\eta_{\mathcal{M}_{q}}, q\in \mathcal{Q}, \mathcal{M}_{q} \in \mathcal{H}_{q}$ that ensures the step size gradually decays to zero without changing greatly. The detailed algorithm is described in \textbf{Algorithm 3}.

\subsubsection{Under (w,loss) scenarios} We consider the active-discarding rate $\!\mathcal{J}_{\mathcal{M}_{q}} \!>\! 0, q\!\in\!\mathcal{Q},\mathcal{M}_{q}\in\mathcal{H}_{q}$. The EC-based statistical QoS constraint in (\ref{e22}) can be reformulated as follows:
\vspace{-0.6em}
\begin{equation}\label{e31}
  \mathbb{E}_{\zeta}\!\bigg[e^{-\Theta_{\mathcal{M}_{q}}^{\ast}\!\left(\!t_{\!\mathcal{M}_{q}} \! \log_{2}(1 + \zeta^{'}) + \mathcal{J}_{\!\mathcal{M}_{q}}\!\right)}\!\bigg]\! \leq \!e^{-\theta_{\mathcal{M}_{q}}^{\ast}\!|\mathcal{R}_{\!\mathcal{M}_{q}}\!|\overline{\mathcal{EB}}\!_{\mathcal{M}_{q}}}.
\end{equation}

\vspace{-1.0em}

\par Then, the resource optimization problem with EC-based SQP under (w,loss) scenarios can be formulated as follows:

\vspace{-2.0em}

\begin{subequations}\label{e32}
	\begin{align}
		\mathrm{\mathcal{P}6}:
		&\quad \min_{\left\{\boldsymbol{t}, \boldsymbol{\mathcal{J}}\right\}} \mathbb{E}_{\zeta}\bigg\{\sum_{q\in\mathcal{Q}} \sum_{\mathcal{M}_{q} \in \mathcal{H}_{q}} t_{\mathcal{M}_{q}} \bigg\}, \nonumber \\
		s.t.\;
        &\ (\ref{e31}), (\ref{e16}c), (\ref{e16}d), (\ref{e13}b), \rm{and}\ (\ref{e13}c).
	\end{align}
\end{subequations}

\vspace{-1.7em}

\par The constraint (\ref{e16}c) can be rewritten as follows:

\vspace{-1.5em}

\begin{equation}\label{e33}
     \big(\!1\!-\!\mathcal{Y}_{q}\!\big)\mathbb{E}_{\zeta}\!\left[\mathcal{J}_{\!\mathcal{M}_{q}}\right]\!-\!\mathcal{Y}_{q}\mathbb{E}_{\zeta}\!\left[t_{\!\mathcal{M}_{q}}\!\log_{2}\!\big(1 + \zeta^{'}\big)\right]\leq 0, \forall \zeta, q \in \mathcal{Q}, \mathcal{M}_{q} \in \mathcal{H}_{q}.
\end{equation}

\vspace{-1.0em}

\par We can prove that the problem $\mathcal{P}6$ is still a standard convex problem \cite{boyd2007notes}. Moreover, the Lagrange multiplier method can still be used to determine the optimal solution, which inspires \emph{Theorem \ref{theo6}}, as follows:

\vspace{-0.8em}

\begin{myTheo}\label{theo6}
   The problem $\mathcal{P}6$ is a standard convex problem, and if $\mathcal{P}6$ exists the optimal solution $\left(\boldsymbol{t}^{\star},\boldsymbol{\mathcal{J}}^{\star}\right)$, which can be given as follows:
   \vspace{-0.5em}
   \begin{small}
   \begin{equation}\label{e34}
     \begin{aligned}
      \!\!& \mathcal{J}_{\mathcal{M}_{q}}^{\star}\!\!\left(\zeta,\mu_{\zeta}^{\star},\psi_{\mathcal{M}_{q}}^{\star},\phi_{\mathcal{M}_{q}}^{\star}\!\right) \!\triangleq \!\left\{\!\!\!
                   \begin{array}{ll}
                     0, & \hbox{Case 1;} \\
                     \left[-\frac{\log\left(\!\!\frac{\phi_{\mathcal{M}_{q}}^{\star}\!\left(\!1-\mathcal{Y}_{\mathcal{M}_{q}}^{th}\!\right)}{\Theta_{\mathcal{M}_{q}}^{\ast}\psi_{\mathcal{M}_{q}}^{\star}}   \!\!\right)}{\Theta_{\mathcal{M}_{q}}^{\ast}}\right]^{+}\!, & \hbox{Case 2;} \\
                     0, & \hbox{Case 3.}
                   \end{array}
                 \right.
     \end{aligned}
   \end{equation}
   \end{small}
\vspace{-0.5em}
and
\vspace{-0.5em}
   \begin{small}
   \begin{equation}\label{e35}
     \begin{aligned}
      \!\!& t_{\mathcal{M}_{q}}^{\star}\!\!\left(\zeta,\mu_{\zeta}^{\star},\psi_{\mathcal{M}_{q}}^{\star},\phi_{\mathcal{M}_{q}}^{\star}\!\right) \!\triangleq \! \left\{\!\!\!
                   \begin{array}{ll}
                     \infty, \ \ \ \ \ \ \ \ \ \ \ \ \ \ \ \ \ \ \ \ \ \ \ \ \ \ \hbox{Case 1;} \\
                     0, \ \ \ \ \ \ \ \ \ \ \ \ \ \ \ \ \ \ \ \ \ \ \ \ \ \ \ \ \hbox{Case 2;} \\
                     \left[-\frac{\log\left(\frac{1+\mu_{\zeta}^{\star} - \mathcal{Y}_{\mathcal{M}_{q}}^{th}\phi_{\mathcal{M}_{q}}^{\star}\log(1+\zeta)}{\Theta_{q}^{\star}\psi_{\mathcal{M}_{q}}^{\star}\log(1+\zeta)}\right)}{\Theta_{q}^{\star}\log(1+\zeta)}    \right]^{+}\!,\\
          \ \ \ \ \ \ \ \ \ \ \ \ \ \ \ \ \ \ \ \ \ \ \ \ \ \ \ \ \ \ \hbox{Case 3,}
                   \end{array}
                 \right.
     \end{aligned}
   \end{equation}
   \end{small}
where Case 1-3 can be defined as follows:
\begin{equation}\label{e36}
   \left\{\!\!\!\!\!\!\!\!\!
     \begin{array}{ll}
       & Case 1: \frac{1+\mu_{\zeta}^{\star}}{\log(1 + \zeta)} \leq \phi_{\mathcal{M}_{q}}^{\star}\mathcal{Y}_{\mathcal{M}_{q}}^{\star}; \\
       & Case 2: \frac{1+\mu_{\zeta}^{\star}}{\log(1 + \zeta)} \ge \phi_{\mathcal{M}_{q}}^{\star}; \\
       & Case 3: \phi_{\mathcal{M}_{q}}^{\star}\mathcal{Y}_{\mathcal{M}_{q}}^{\star} < \frac{1+\mu_{\zeta}^{\star}}{\log(1 + \zeta)} < \phi_{\mathcal{M}_{q}}^{\star}.
     \end{array}
   \right.
\end{equation}
\par Similar to \emph{Theorem \ref{theo5}}, given $\zeta$, $\{\psi_{\mathcal{M}_{q}}^{\star}\}_{q \in\mathcal{Q}, \mathcal{}_{q}\in \mathcal{H}_{q}}$, and $\{\phi_{\mathcal{M}_{q}}^{\star}\}_{q \in\mathcal{Q}, \mathcal{M}_{q}\in \mathcal{H}_{q}}$, if $\sum_{q\in \mathcal{Q}}\sum_{\mathcal{M}_{q}\in\mathcal{H}_{q}} t_{\mathcal{M}_{q}} \geq T$, then $\mu_{\zeta}^{\star}$ is selected such that $\sum_{q\in \mathcal{Q}}\sum_{\mathcal{M}_{q}\in\mathcal{H}_{q}} t_{\mathcal{M}_{q}} = T$ holds; otherwise $\mu_{\zeta}^{\star} = 0$. In addition, $\{\psi_{\mathcal{M}_{q}}^{\star}\}_{q \in\mathcal{Q}, \mathcal{M}_{q}\in \mathcal{H}_{q}}$ and $\{\phi_{\mathcal{M}_{q}}^{\star}\}_{q \in\mathcal{Q},\mathcal{M}_{q}\in \mathcal{H}_{q}}$ should be optimized jointly such that ``='' holds in both constraints (\ref{e31}) and (\ref{e16}c).
\end{myTheo}

\vspace{-1.0em}

\begin{proof}
  The proof of \emph{Theorem \ref{theo6}} is given in Appendix E.
\end{proof}

\vspace{-2.0em}

\subsection{Computational Complexity Analysis}
\vspace{-0.5em}
\par The complexity of \emph{Algorithm 1} can be represented by $\mathcal{O}\!\!\left(\!\tilde{N}_{max}\big(\mathcal{C}_{1} \!+\! \mathcal{C}_{2}\!\big) \!+\! \tilde{N}_{max}\!\right)$, where $\!\!\mathcal{O}(\mathcal{C}_{1})\!\!=\!\!\mathcal{O}\left(\log\Phi_{s}/\Phi_{th}\right)$ and $\!\!\mathcal{O}(\mathcal{C}_{2})\!\!=\!\!\mathcal{O}\!\left(\log\!\left|\nabla\mathcal{K}_{\mathcal{M}_{q}}\!\big(\theta_{\mathcal{M}_{q}}^{\ast},L_{q},w_{q}^{\ast}\big)\Delta_{s}\right|\!/\Psi_{th}\!\right)$ denotes the complexities of \emph{Step 1} and \emph{Step 2} in each iteration, respectively, $\tilde{N}_{max}$ denotes the actual number of iterations, and the complexity of \emph{Step 3} is $\tilde{N}_{max}$. The complexity of \emph{Algorithm 2} can be denoted by $\mathcal{O}\!\!\left(\tilde{I}_{max}\!\left(\!\tilde{N}_{max}\big(\mathcal{C}_{1} \!+\! \mathcal{C}_{2}\!\big) \!+\! \tilde{N}_{max}\!\right)\!\right)$, where $\tilde{I}_{max}$ is the actual number of iterations of \emph{Algorithm 2}. As shown in Fig. 3, $\boldsymbol{J}^{\star}(0)$ is first initialized. In each iteration, $\boldsymbol{J}^{\star}(n)$ gradually increases, and $\boldsymbol{t}^{\star}(n)$  gradually decreases. Thus, $t^{\star}_{\!\mathcal{M}_{q}}\!(n) \!\!\leq\!\! t^{\star}_{\!\mathcal{M}_{q}}\!(n\!+\!1)$ and $\mathcal{J}^{\star}_{\!\mathcal{M}_{q}}\!(n) \!\!\geq\!\! \mathcal{J}^{\star}_{\!\mathcal{M}_{q}}\!(n\!+\!1)$. Due to the resource limitations and rigorous QoS constraints, \emph{Algorithm 2} gradually converges and achieves the optimal solution in each iteration. \emph{Algorithm 3} has a convergence rate of $\mathcal{O}(\frac{1}{\sqrt{z}})$, where $z$ denotes the number of iterations \cite{boyd2003subgradient}.

\vspace{-1em}

\section{Performance Evaluation}
\par In this section, extensive simulations and discussions are presented to demonstrate the effectiveness of the designed overlapping FoV-based optimal JUM task assignment scheme, as well as the proposed optimal ADAPT-JTAAT transmission scheme. Here, we refer to the proposed ADAPT-JTAAT from SDVP and EC perspectives as \textbf{Proposed 1} and \textbf{Proposed 2}, respectively. Two video quality layers with different statistical QoS requirements are considered, namely \textbf{Layer 1}: $\left(L_{1},w_{1}^{\ast},\varepsilon_{1},\mathcal{Y}_{1}\right)$ and \textbf{Layer 2}: $\left(L_{2},w_{2}^{\ast},\varepsilon_{2},\mathcal{Y}_{2}\right)$, where $L_{1} \!\!\!=\!\!\! 13 \!\times\! 10^{6}$ bits/s, $L_{2} \!=\! 18\!\times\!10^{6}$ bits/s, $w_{1}^{\ast} \!\!=\!\! 20$ ms, $w_{2}^{\ast} \!\!=\!\! 15$ ms, $\varepsilon_{1} \!\!=\!\! 10^{-3}$, $\varepsilon_{2} \!\!=\!\! 10^{-4}$, $\mathcal{Y}_{1} \!=\! 0.01$, and $\mathcal{Y}_{2} \!=\! 0.001$. The architecture includes six users, with subscripts $\mathcal{N} \!\!=\!\! \{1,2,\cdots,6\}$, which is grouped into two user groups, namely $\mathcal{N}_{1}$, consisting of $\left\{\rm{user}\ 1, \rm{user}\ 2, \rm{user}\ 3\right\}$, and $\mathcal{N}_{2}$, consisting of $\left\{\rm{user}\ 4, \rm{user}\ 5, \rm{user}\ 6\right\}$. The non-redundant task assignments $\boldsymbol{\mathcal{R}}$ are obtained by the overlapping FoV-based optimal JUM task assignment scheme. Other simulation parameters include a bandwidth of $B \!\!=\!\! 500$ MHz, distances from users to the BS of $l \!=\! 50$ m, a path loss exponent of $\alpha \!=\! 2.45$, a time-slot length of $T \!=\! 10$ ms, a Nakagami-m parameter of $M \!=\! 3$, an ERP size of $V_{h}\! \times \!V_{v} = 6 \times 4$, an FoV size of $a \times b = 2 \times 2$.

\vspace{-1.0em}

\begin{algorithm}
\small
\setstretch{0.60}
        \caption{Subgradient-Based Optimization Algorithm} 
        \KwIn{Set $z = 1$; Select $\boldsymbol{\psi}^{\star}\left(0\right) \succeq 0$;  Select convergence criterion $\kappa_{th} > 0$.}
        \KwOut{$\mu_{\zeta}^{\star}$ and $\boldsymbol{\psi}^{\star} \triangleq \left\{\psi_{\mathcal{M}_{q}}^{\star}\right\}_{q\in \mathcal{Q},\mathcal{M}_{q}\in \mathcal{H}_{q}}$.}
        \While{$\mid \boldsymbol{\psi}^{\star}\left(z+1\right)-\boldsymbol{\psi}^{\star}\left(z\right)\mid > \kappa_{th}$}{
          Substituting $\boldsymbol{\psi}^{\star}\left(z\right)$ in to Eq.(28) to obtain the optimal $\mu_{\zeta}^{\star}\left(z\right)$;\\
          \For{\rm{\textbf{all}} $q \in \mathcal{Q}$ and $\mathcal{M}_{q} \in \mathcal{H}_{q}$}
          {
            Update ${t_{\mathcal{M}_{q}}^{\star}\left(\zeta,\psi_{\mathcal{M}_{q}}^{\star}(z),\mu_{\zeta}^{\star}(z)\right)}$ according to Eq.(25);\\
            Update $\psi_{\mathcal{M}_{q}}^{\star}\left(z\right)$ according to $\psi_{\mathcal{M}_{q}}^{\star}\left(z+1\right) = \max\left\{\psi_{\mathcal{M}_{q}}^{\star}\left(z\right)+\eta_{\mathcal{M}_{q}}\left(z\right)\mathcal{S}_{\mathcal{M}_{q}},0\right\}$, where $\mathcal{S}_{\mathcal{M}_{q}}\triangleq \left(F\left(t_{\mathcal{M}_{q}}^{\star}\right)-e^{-\theta_{\mathcal{M}_{q}}^{\ast}\left|\mathcal{R}_{\mathcal{M}_{q}}\right|\overline{\mathcal{EB}}_{\mathcal{M}_{q}}}\right)$, and the step size sequence $\eta_{\mathcal{M}_{q}}$ satisfies $\sum\limits_{z = 0}^{\infty}\eta_{\mathcal{M}_{q}}^{2}\left(z\right) < \infty$, $\sum\limits_{z = 1}^{\infty} \eta_{\mathcal{M}_{q}}(z) = \infty$.
            \\
            }
           Set $\boldsymbol{\psi}^{\star} = \boldsymbol{\psi}(z)$ and $\mu_{\zeta}^{\star} = \mu_{\zeta}^{\star}(z)$;\\
           $z = z+1$; \\
          }
\end{algorithm}

\vspace{-3em}

\subsection{Comparison Baseline Schemes}

\vspace{-0.7em}
\par To demonstrate the superiority of our proposed schemes, six baseline schemes are designed for comparison analysis from the perspectives of resource utilization, streaming mode, and video data discarding strategy.

\vspace{-0.18em}
\par \textbf{(1)} \emph{\textbf{Optimal-Unicast (w/o,loss)}:} This baseline scheme performs the optimal ADAPT-JTAAT transmission schemes under the unicast mode and (w/o,loss) scenarios.
\vspace{-0.18em}
\par \textbf{(2)} \emph{\textbf{Optimal-Unicast (w,loss)}:} This baseline scheme performs the optimal ADAPT-JTAAT transmission schemes under the unicast mode and (w,loss) scenarios.
\vspace{-0.18em}
\par \textbf{(3)} \emph{\textbf{Optimal-JUM Delay (w/o,loss)}:} This baseline scheme performs \textbf{Proposed 1} with the overlapping FoV-based optimal JUM task assignment scheme under (w/o,loss) scenarios.
\vspace{-0.18em}
\par \textbf{(4) }\emph{\textbf{Optimal-JUM Rate (w/o,loss)}:} This baseline scheme performs \textbf{Proposed 2} with the overlapping FoV-based Optimal JUM task assignment scheme under (w/o,loss) scenarios.
\vspace{-0.2em}
\par \textbf{(5)} \emph{\textbf{Fixed-Unicast (w/o,loss)}:} In comparison to \textbf{Proposed 1} and \textbf{Proposed 2}, the fixed slot $\bar{t}_{n,q}$ can be selected under (w/o,loss), such that
\vspace{-0.5em}
      \begin{equation}\label{e37}
        \!\!\!\!\!\left\{
         \begin{array}{ll}
           \!\!\!\mathcal{K}_{n,q}\left(w_{q}^{\ast},\bar{t}_{n,q}\right) = \varepsilon_{q}, & \hbox{for Proposed 1;}\\
           \!\!\!F\left(\bar{t}_{n,q}\right)=e^{-\theta_{n,q}^{\ast}\left|\mathcal{R}_{n,q}\right|\overline{\mathcal{EB}}_{n,q}}, & \hbox{for Proposed 2.}
         \end{array}
                  \right.
        \vspace{-0.3em}
      \end{equation}
where $\!\mathcal{M}_{q}\!$ is a single-user set in unicast mode, and can be rewritten as $(n,q), n \in \mathcal{N}, q\in\mathcal{Q}$. The values of $\bar{t}_{n,q}$ can be obtained by solving the equation set (\ref{e37}).
\vspace{-0.18em}
\par \textbf{(6)} \emph{\textbf{Fixed-Unicast (w,loss)}:} In comparison to \textbf{Proposed 1} and \textbf{Proposed 2}, the fixed slot $\bar{t}_{n,q}$ and fixed active-discarding rate $\bar{\mathcal{J}}_{n,q}$ are selected, respectively, under unicast mode and (w,loss) scenarios, such that the following two equation sets holds:
\vspace{-0.7em}
      \begin{small}
      \begin{equation}\label{e38}
          \begin{aligned}
               & \quad \hbox{\emph{For Proposed 1:}}  \quad \quad \quad \quad \quad \quad \quad \quad \quad \quad \quad \quad \quad \quad \quad \quad \quad \hbox{\emph{For Proposed 2:}}\\
               &\left\{
         \begin{array}{ll}
           \!\!\!\mathcal{K}_{n,q}\!\left(w_{q}^{\ast},\bar{t}_{n,q},\bar{\mathcal{J}}_{n,q}\right) \!=\! \varepsilon_{q},\\
           \!\!\!\mathbb{E}_{\zeta}\!\left[\big(1\!-\!\mathcal{Y}_{q}\big)\bar{\mathcal{J}}_{n,q} \!-\! \mathcal{Y}_{q}\bar{t}_{n,q}\log(1 \!+\! \zeta^{'})\right] = 0. \ (37)
         \end{array}
                  \right. \quad \quad \quad \left\{
         \begin{array}{ll}
           \!\!\!F\left(\bar{t}_{n,q}\right)=e^{-\theta_{n,q}^{\ast}\left|\!\mathcal{R}_{n,q}\!\right|\overline{\mathcal{EB}}_{n,q}},\\
           \!\!\!\mathbb{E}_{\zeta}\!\left[\big(1\!-\!\mathcal{Y}_{q}\big)\bar{\mathcal{J}}_{n,q} \!-\! \mathcal{Y}_{q}\bar{t}_{n,q}\log(1 \!+\! \zeta^{'})\right] = 0. \ (38)
         \end{array}
               \right. \nonumber \!\!\!
          \end{aligned}
          \vspace{-0.3em}
      \end{equation}
      \end{small}
where the values of $\bar{t}_{n,q}$ and $\bar{\mathcal{J}}_{n,q}$ can be obtained by solving equation set (37) for \textbf{Proposed 1} or equation set (38) for \textbf{Proposed 2}.

\par To verify the effectiveness of the proposed overlapping FoV-based optimal JUM task assignment scheme, the concept of FoV overlap ratio (FoR) is defined to characterize the overlapping degree of FoVs, which is given as follows:
\vspace{-0.7em}
\begin{myDef}
  $\forall \ q \in \mathcal{Q}$, the FoV Overlapped Ratio for user group $\mathcal{N}_{q}$ is define as
\vspace{-0.8em}
  \begin{equation}\label{e40}
    \rho_{q} = \big(\!\!\sum\limits_{n \in \mathcal{N}_{q}}\!\!\!\mathcal{F}_{n} \!\!-\!\!\!\!\sum\limits_{\mathcal{M}^{\dag}_{q} \in \mathcal{H}_{q}}\!\!\!\!\!\mathcal{R}_{\mathcal{M}^{\dag}_{q},q}\big)/\!\!\sum\limits_{n \in \mathcal{N}_{q}}\!\!\!\mathcal{F}_{n},
    \vspace{-0.5em}
  \end{equation}
where $0 \leq \rho_{q} \leq 1$, and $\mathcal{M}^{\dag}_{q}$ denotes the single-user subset in $\mathcal{H}_{q}$, and the tiles in $\mathcal{M}^{\dag}_{q}$ are non-overlapped. FoR $\rho_{q}$ measures the degree of FoV overlap for user group $\mathcal{N}_{q}, q\in\mathcal{Q}$. A larger $\rho_{q}$ implies that FoVs of $\mathcal{N}_{q}$ are overlapped seriously; while a smaller $\rho_{q}$ means the VR video data requested by users in $\mathcal{N}_{q}$ has lower repetition.
\end{myDef}

\vspace{-0.5em}

\begin{figure*}[t]
\centering
  \subfigure[]{
  \includegraphics[scale=0.238]{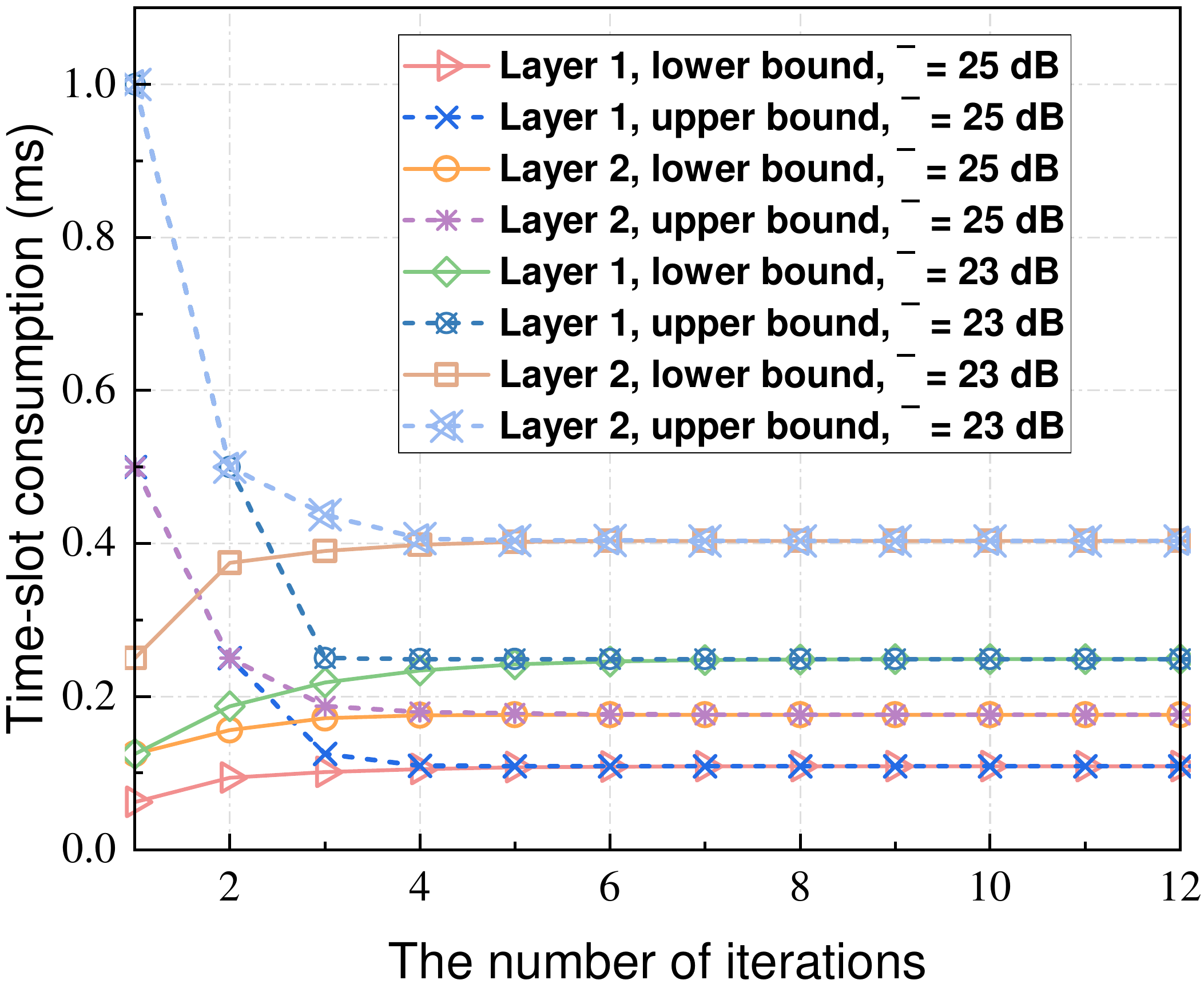}
  }
  \subfigure[]{
   \includegraphics[scale=0.238]{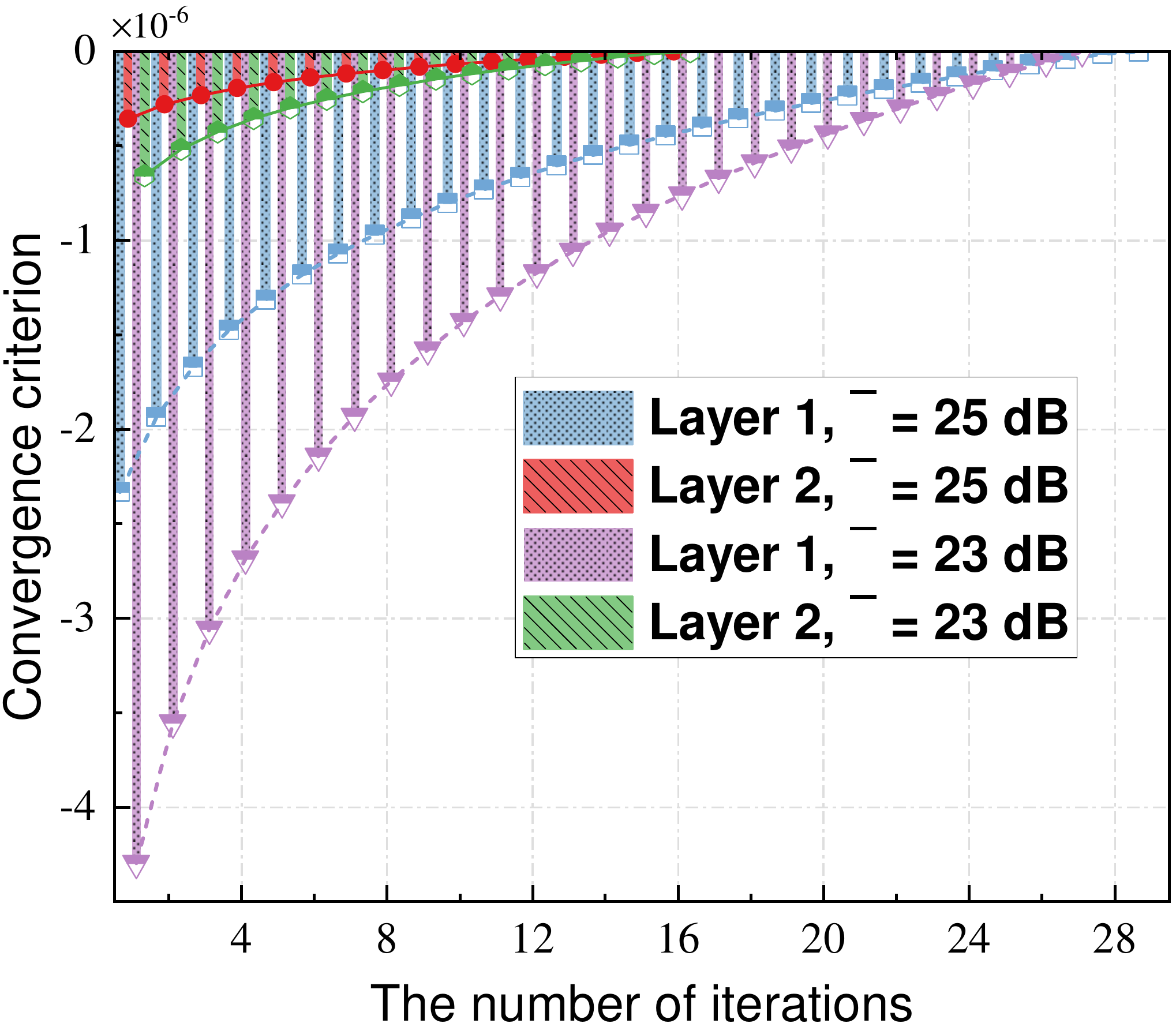}
  }
  \subfigure[]{
    \includegraphics[scale=0.238]{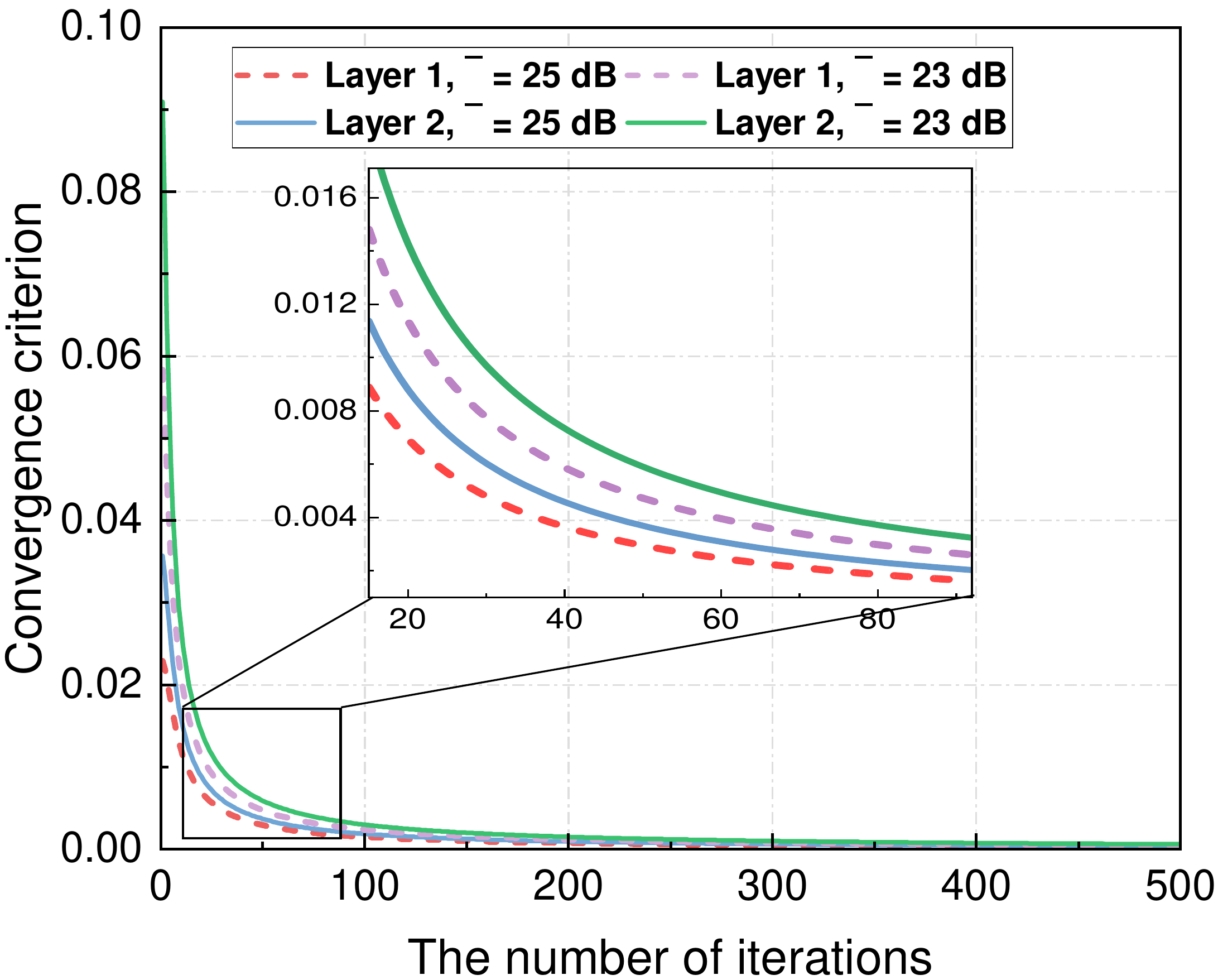}
  }
  \vspace{-1em}
  \caption{(a) Convergence analysis of Algorithm 1, where $k_{th} = 10^{-16}$, $N_{max} = 100$, and $\Phi_{th} = \Delta_{s} = \Psi_{th} = 10^{-5}$; (b) convergence analysis of Algorithm 2, where $\Omega_{th} = 10^{-16}$; (c) convergence analysis of Algorithm 3, where $\kappa_{th} = 10^{-16}$.}
\vspace{-2em}
\end{figure*}

\vspace{-2.0em}

\subsection{Effectiveness Discussion}
\vspace{-0.5em}
\par Fig. 4 (a), (b), and (c) elaborate the convergence behaviors of the algorithms proposed in this paper. In Fig. 4 (a), the convergence behaviors of the upper- and lower-search bounds for the proposed nested-shrinkage optimization algorithm are illustrated. Numerical results show that, irrespective of Layer 1 or Layer 2 and under different average SNRs, the upper-search bound drastically decreases while the lower-search bound increases with the increase of iteration number. Within six iterations, both bounds converge simultaneously to the optimal time-slot consumption, thus affirming the rapid convergence of \emph{Algorithm 1}. Fig. 4 (b) depicts the convergence behaviors of the stepwise-approximation algorithm. We initialize $\!\boldsymbol{\mathcal{J}}\!(0)\!=\!0$ and follow \emph{Algorithm 2}, where the active-discarding rate gradually increases with the increase of iteration number, resulting in a decrease in time-slot consumption, and the convergence criterion ultimately approaches zero. Moreover, the numerical results show that \emph{Algorithm 2} converges rapidly, with the convergence criterion significantly converging to zero for different average SNRs, as well as Layer 1 and Layer 2. In Fig. 4 (c), the convergence behaviors of the subgradient-based optimization algorithm are shown. According to the convergence and complexity analysis of the subgradient algorithm \cite{boyd2003subgradient}, if the optimal solution exists, $\mathcal{S}_{\mathcal{M}_{q}}$ will converge to zero as the iteration number increases. Numerical results indicate that $\mathcal{S}_{\mathcal{M}_{q}}$ converges rapidly to zero as the iteration number increases, thereby demonstrating the rapid convergence of \emph{Algorithm 3} for different average SNRs and for both Layer 1 and Layer 2.

\par Fig. 5 and Fig.6 illustrate the effectiveness of the designed overlapping FoV-based optimal JUM task assignment scheme for \textbf{Proposed 1} and \textbf{Proposed 2}, respectively. The numerical results make it abundantly clear that for both \textbf{Proposed 1} and \textbf{Proposed 2}, the designed task assignment scheme can significantly improve time-slot consumption with the increase of FoR, regardless of whether it is Layer 1 or Layer 2, as well as (w/o,loss) or (w,loss) scenarios. It can also be observed that even slight improvements of average SNR lead to a significant reduction of time-slot consumption, especially under relatively poor channel conditions (e.g., $21 \!\sim\! 24$ dB). These arguments unequivocally demonstrate the effectiveness of the designed overlapping FoV-based optimal JUM task assignment scheme for \textbf{Proposed 1} and \textbf{Proposed 2}.

\vspace{-1em}
\begin{figure*}[htbp]
\centering
  \subfigure[]{
  \includegraphics[scale=0.3]{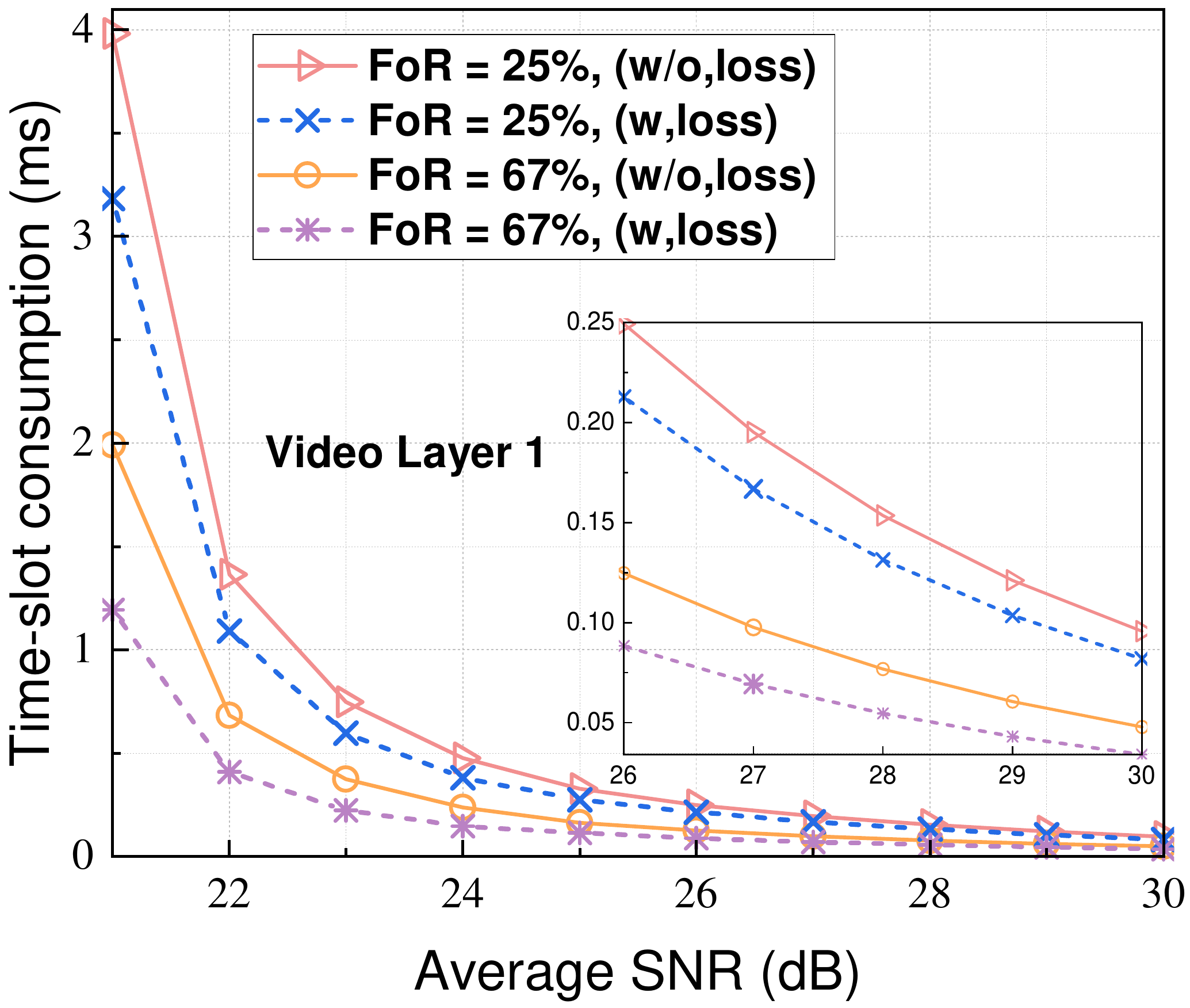}
  }
  \subfigure[]{
   \includegraphics[scale=0.3]{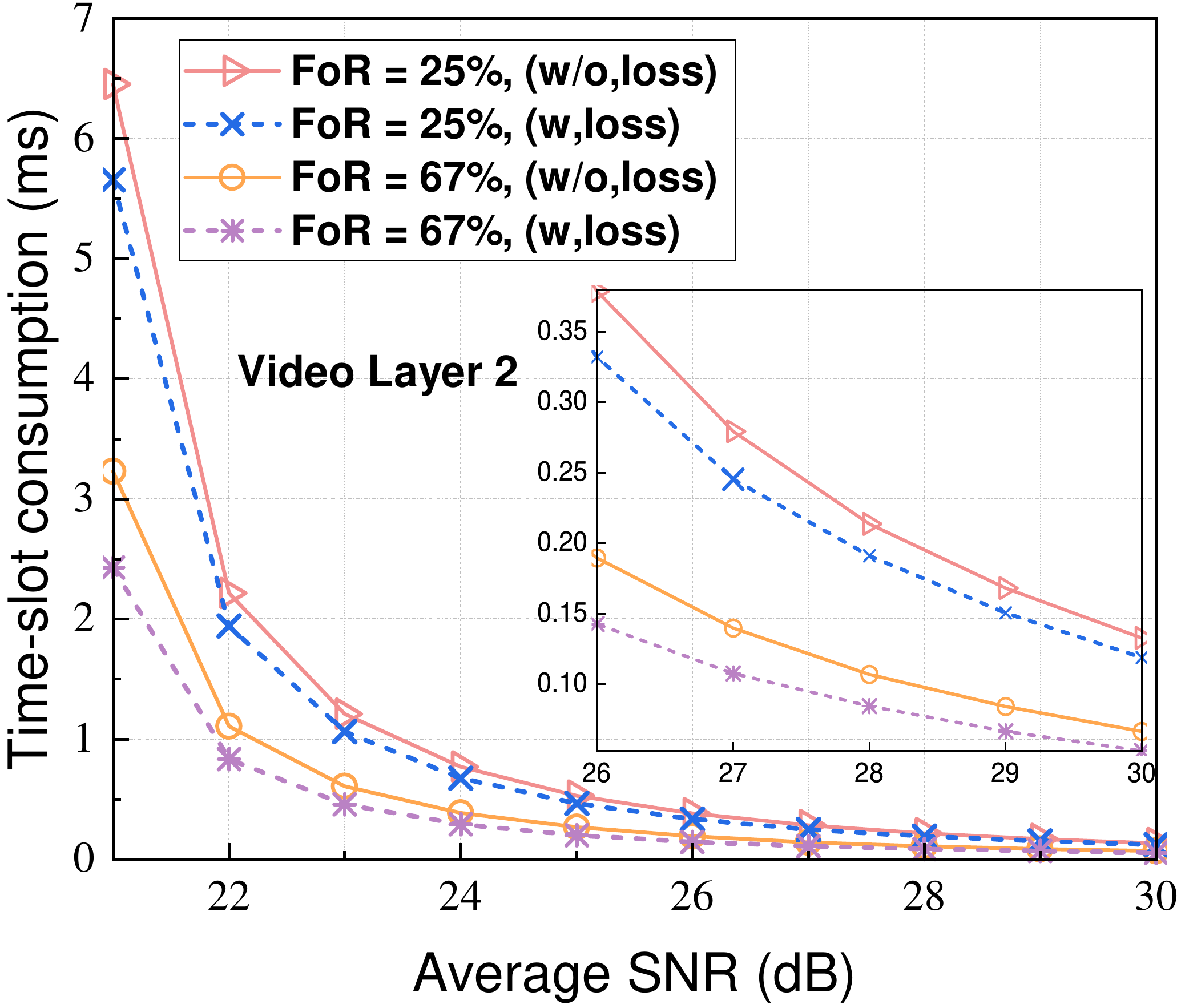}
  }
  \vspace{-0.8em}
  \caption{Performance of overlapping FoV-based optimal JUM task assignment for \textbf{Proposed 1}.}
\vspace{-0.5em}
\end{figure*}

\vspace{-2em}

\begin{figure*}[htbp]
\centering
  \subfigure[]{
  \includegraphics[scale=0.3]{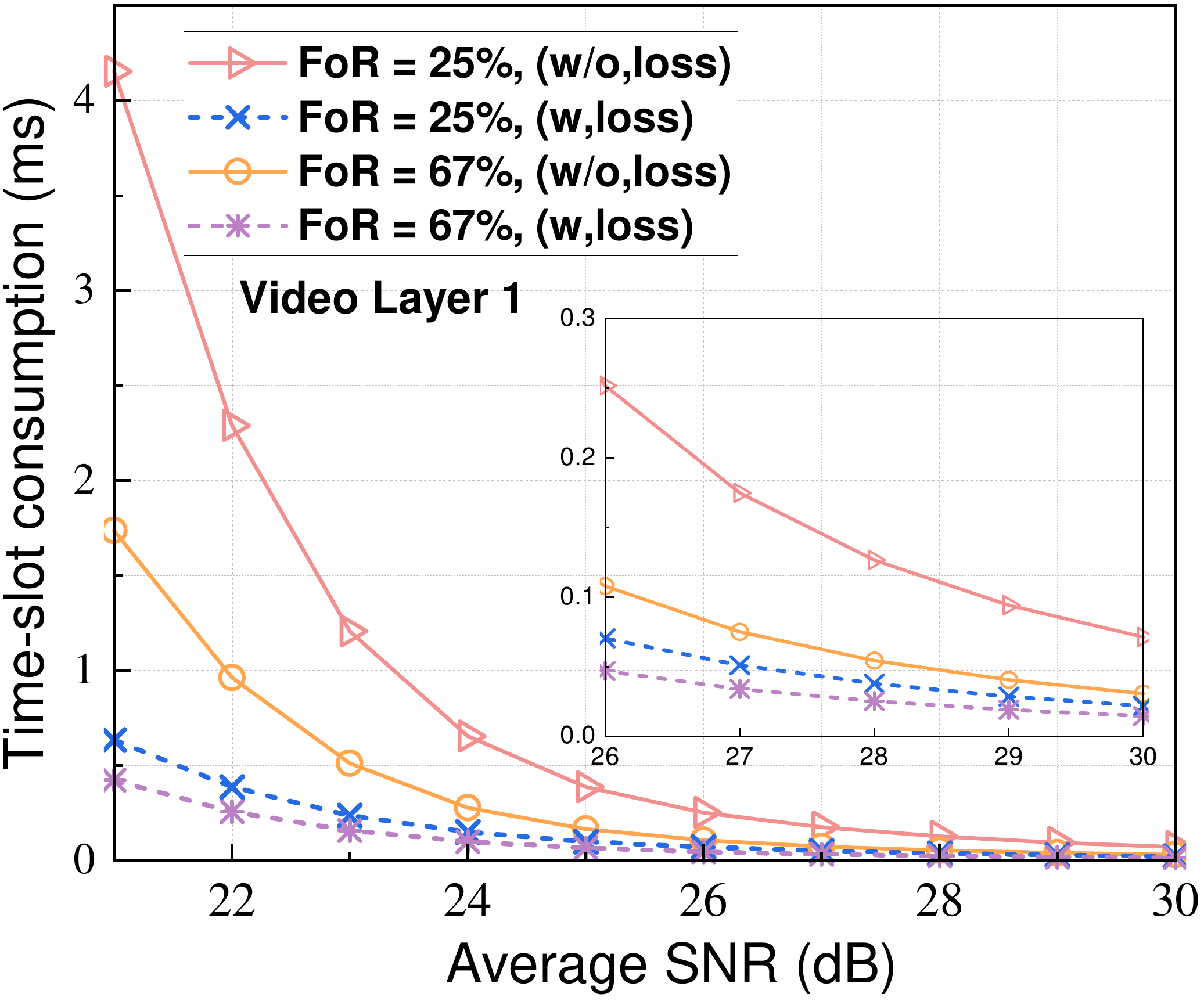}
  }
  \subfigure[]{
   \includegraphics[scale=0.3]{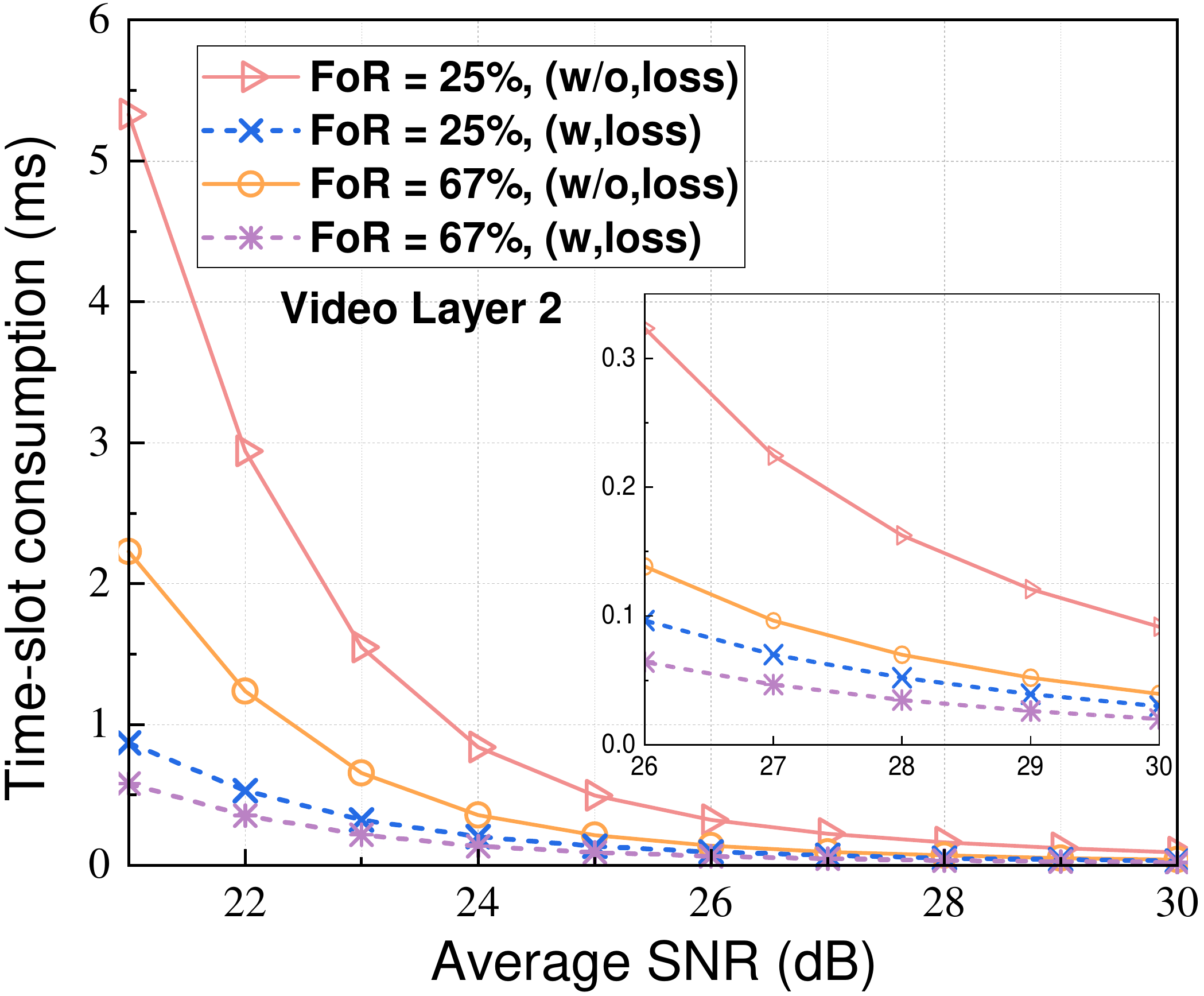}
  }
  \vspace{-0.8em}
  \caption{Performance of overlapping FoV-based optimal JUM task assignment for \textbf{Proposed 2}.}
\vspace{-1.5em}
\end{figure*}

\par Fig. 5 (a) and (b) reveal a substantial reduction in time-slot consumption when the FoR increases from 27$\%$ to 67$\%$ in both (w/o,loss) and (w,loss) scenarios. This improvement is primarily due to the fact that the increase of FoR implies more severe FoV overlapping, and our task assignment scheme can aggregate these overlapped VR tiles into a single multicast session for streaming, thus achieving significant conserving resources. A similar conclusion can also be drawn from Fig. 6 (a) and (b). Additionally, it is worth noting that \textbf{Proposed 1} and \textbf{Proposed 2} specifically focus on the optimal ADAPT-JTAAT transmission scheme. In \textbf{Proposed 1}, the improvements in time-slot consumption of Layer 1 and Layer 2 are less significant compared to \textbf{Proposed 2}. This is mainly due to the fact that \textbf{Proposed 1} provides SQP from the SDVP perspective, which requires a stringent guarantee that the UB-SDVP cannot be larger than the violation probability threshold. Thus, even though \textbf{Proposed 1} achieves lower time-slot consumption in (w,loss) scenarios, it still necessitates the consideration of statistical QoS requirements, along with a carefully designed adaptive active-discarding strategy to guarantee SQP performance at each layer. On the other hand, \textbf{Proposed 2} can achieve remarkable improvements in time-slot consumption under (w,loss) scenarios. The intuition behind this is that it focuses on the EC perspective, thus increasing the active-discarding rate as much as possible to enhance the maximum service capability while guaranteeing the statistical QoS requirements.

\begin{figure*}[t]
\centering
  \subfigure[]{
  \includegraphics[scale=0.238]{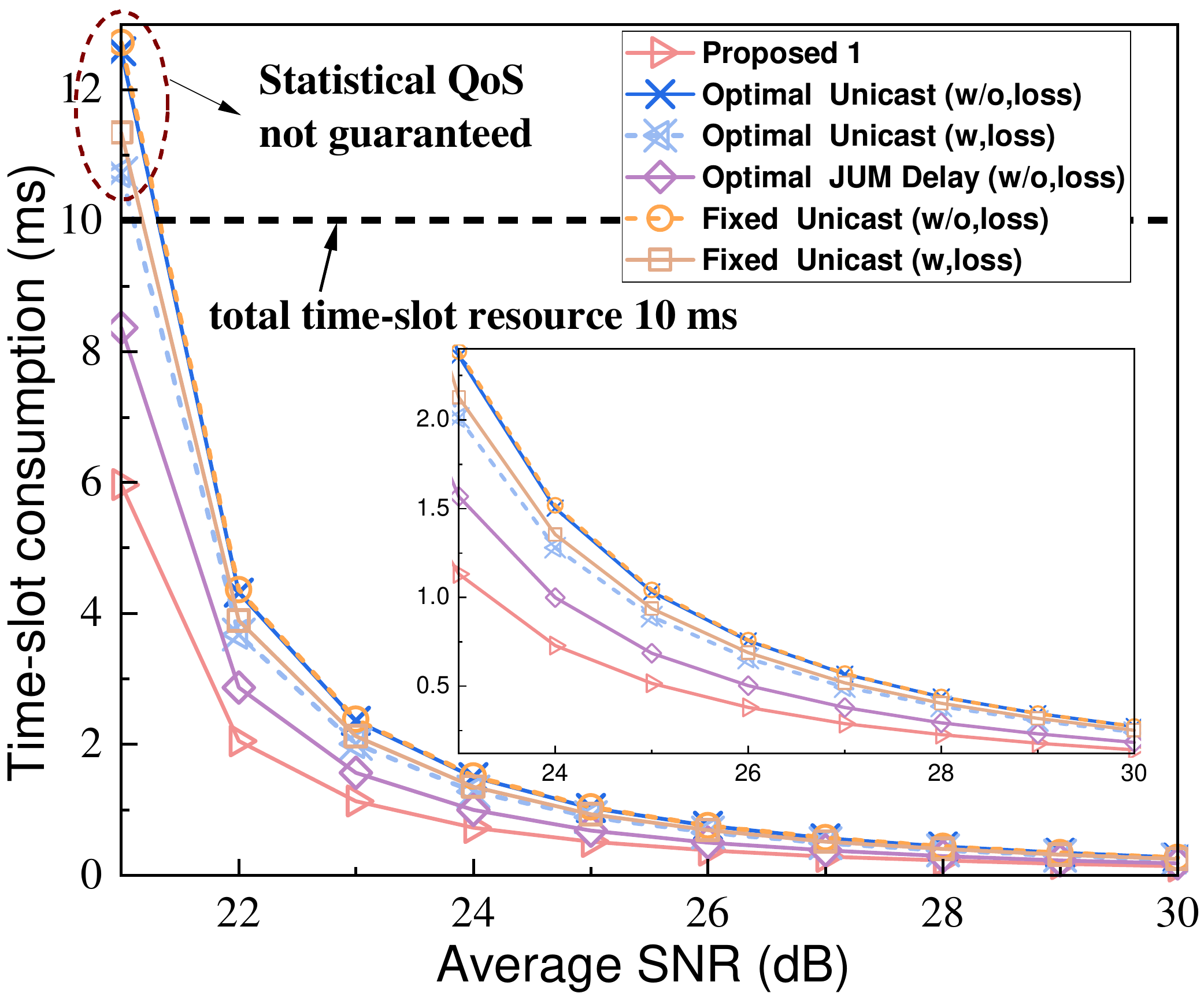}
  }
  \subfigure[]{
   \includegraphics[scale=0.238]{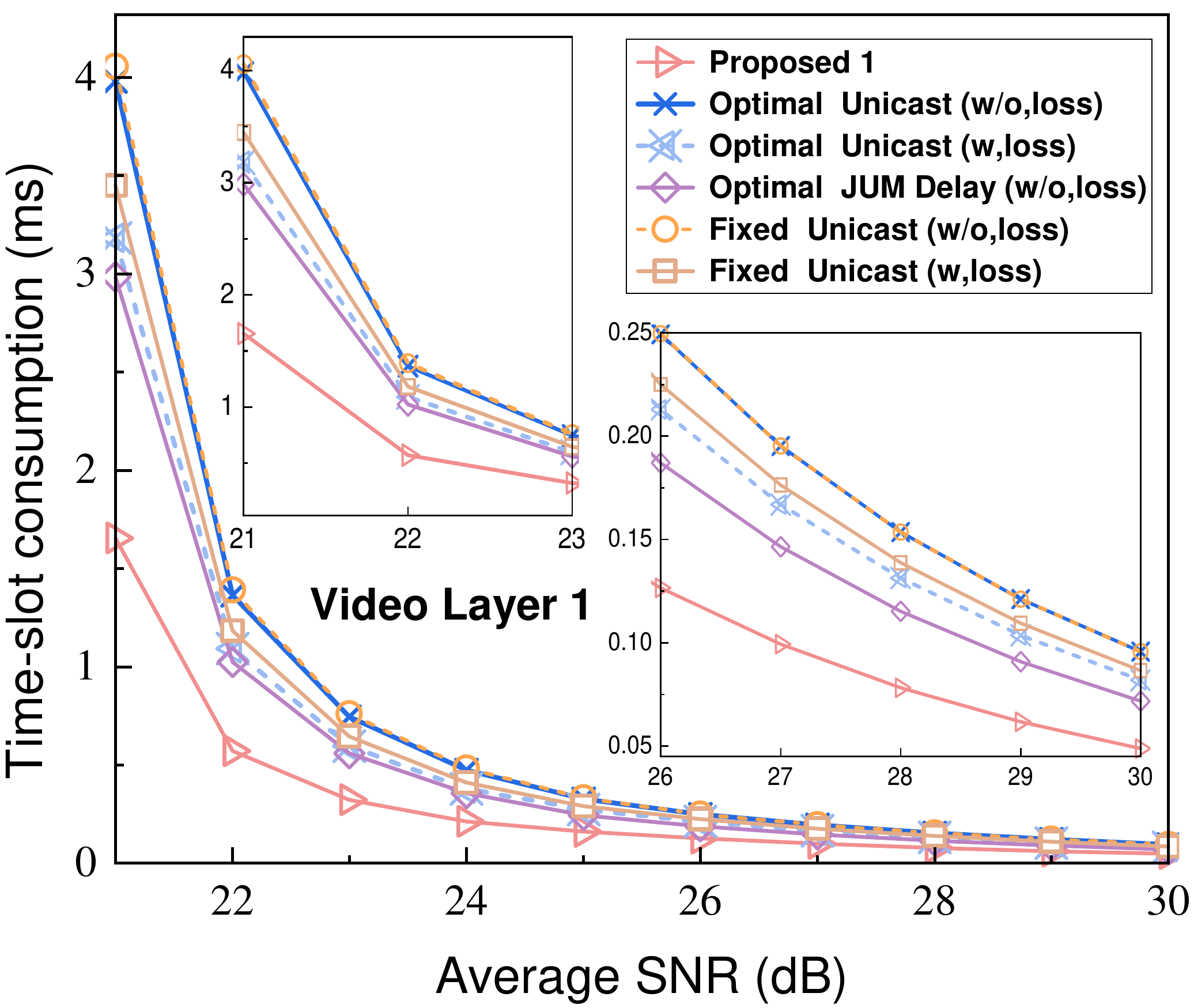}
  }
  \subfigure[]{
   \includegraphics[scale=0.238]{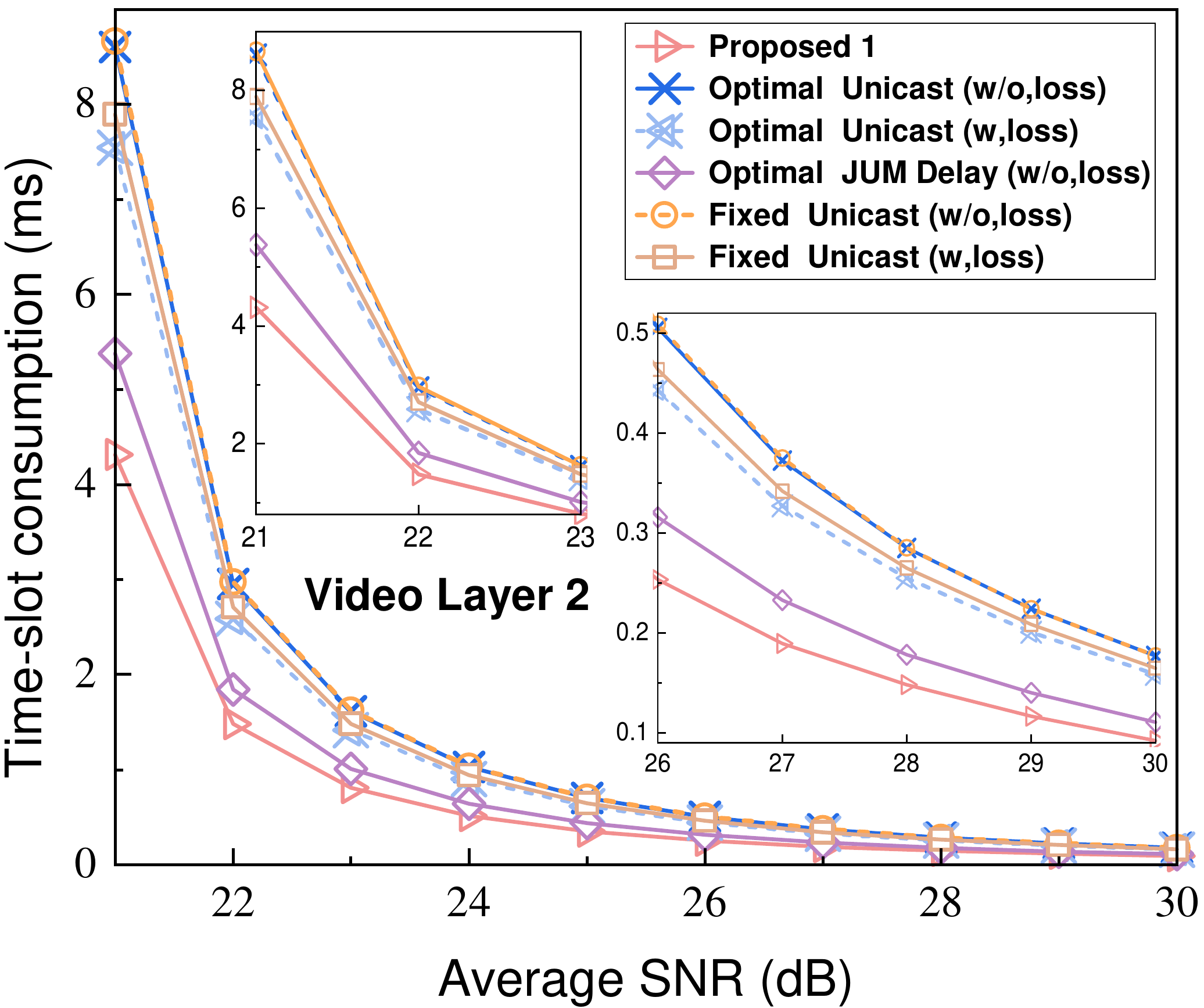}
  }
  \vspace{-0.8em}
  \caption{Performance comparison between \textbf{Proposed 1} and Optimal-Unicast (w/o,loss), Optimal-Unicast (w,loss), Optimal-JUM Delay (w/o,loss), Fixed-Unicast (w/o,loss), and Fixed-Unicast (w,loss).}
\vspace{-2em}
\end{figure*}


\vspace{-0.3em}
\par As depicted in Fig. 7 and 8, we compare the performance of \textbf{Proposed 1} and \textbf{Proposed 2} with their respective baseline schemes. It can be observed that the time-slot consumption of \textbf{Proposed 1}, and \textbf{Proposed 2}, as well as all baseline schemes, can be significantly improved as the channel conditions become better. From Fig. 7 (b) and (c), it can be seen that even when the average SNR is relatively low (around $21\!\!\!\sim\!\!\!22$ dB), \textbf{Proposed 1} still manages to achieve the expected SQP performance with minimal time-slot consumption, when compared to other baseline schemes. The comparison between \textbf{Proposed 1} and Optimal-JUM Delay (w/o,loss), Optimal-Unicast (w/o,loss) and Optimal- Unicast (w,loss), as well as Fixed-Unicast (w/o,loss) and Fixed-Unicast (w,loss), further highlights the significant time-slot resource savings that can be achieved in (w,loss) scenarios. This is mainly because \textbf{Proposed 1} integrates the optimal adaptive time-slot allocation strategy, as well as the adaptive active-discarding strategy, which thus enables the timely discarding of data with relatively low importance from FoV edges, and adaptively allocates the time-slot resources. In this way, smoother video playback can be ensured in relatively poor channel conditions while guaranteeing SQP performance.

\par From Fig. 7 (a), it can be observed that the SQP performance of the proposed multi-layer tiled 360$^{\circ}$ VR streaming architecture cannot be guaranteed when the channel conditions are poor (e.g., 21-22 dB) for all baseline schemes, even with the integration of adaptive active-discarding strategy and the exhaustion of all available time-slot resources. Moreover, enabling the overlapping FoV-based optimal JUM task assignment scheme, namely Optimal-JUM Delay (w/o,loss), leads to some improvement in time-slot consumption, which however still exceeds 8 ms. In contrast, \textbf{Proposed 1} consumes only 6 ms of time slot. A comparison among \textbf{Proposed 1}, Optimal-Unicast (w,loss), and Fixed-Unicast (w,loss) further supports the efficacy of the overlapping FoV-based optimal JUM task assignment scheme in conserving wireless resources. Additionally, a comparison between \textbf{Proposed 1} and Optimal-JUM Delay (w/o,loss) highlights the significant improvement in time-slot consumption achieved by \textbf{Proposed 1}. This improvement is primarily due to the integration of the adaptive active-discarding strategy, which leads more flexible rate and robust queuing behaviors, resulting in further enhancement of time-slot consumption.

\vspace{-1.3em}

\begin{figure*}[h]
\centering
  \subfigure[]{
  \includegraphics[scale=0.238]{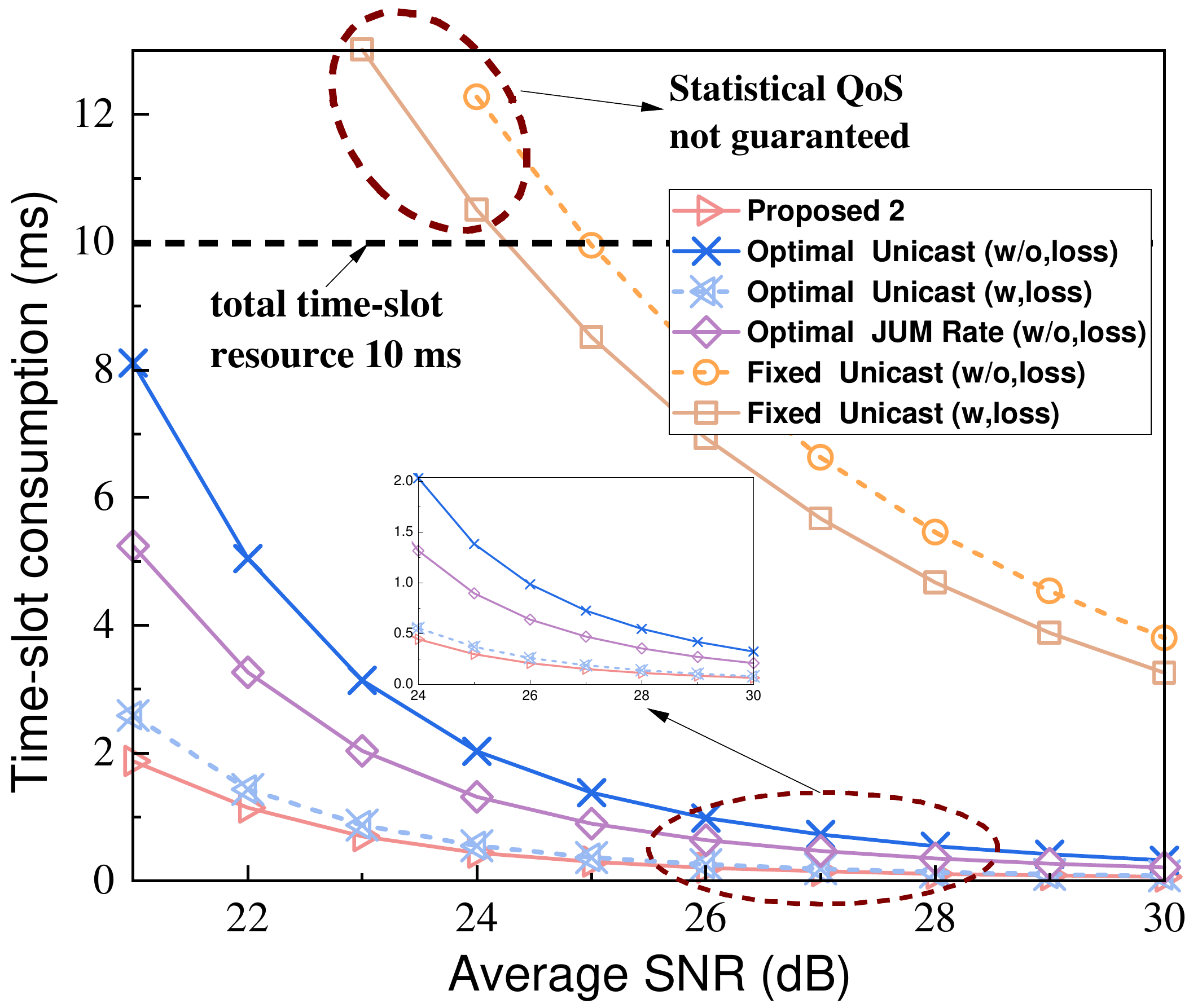}
  }
  \subfigure[]{
   \includegraphics[scale=0.238]{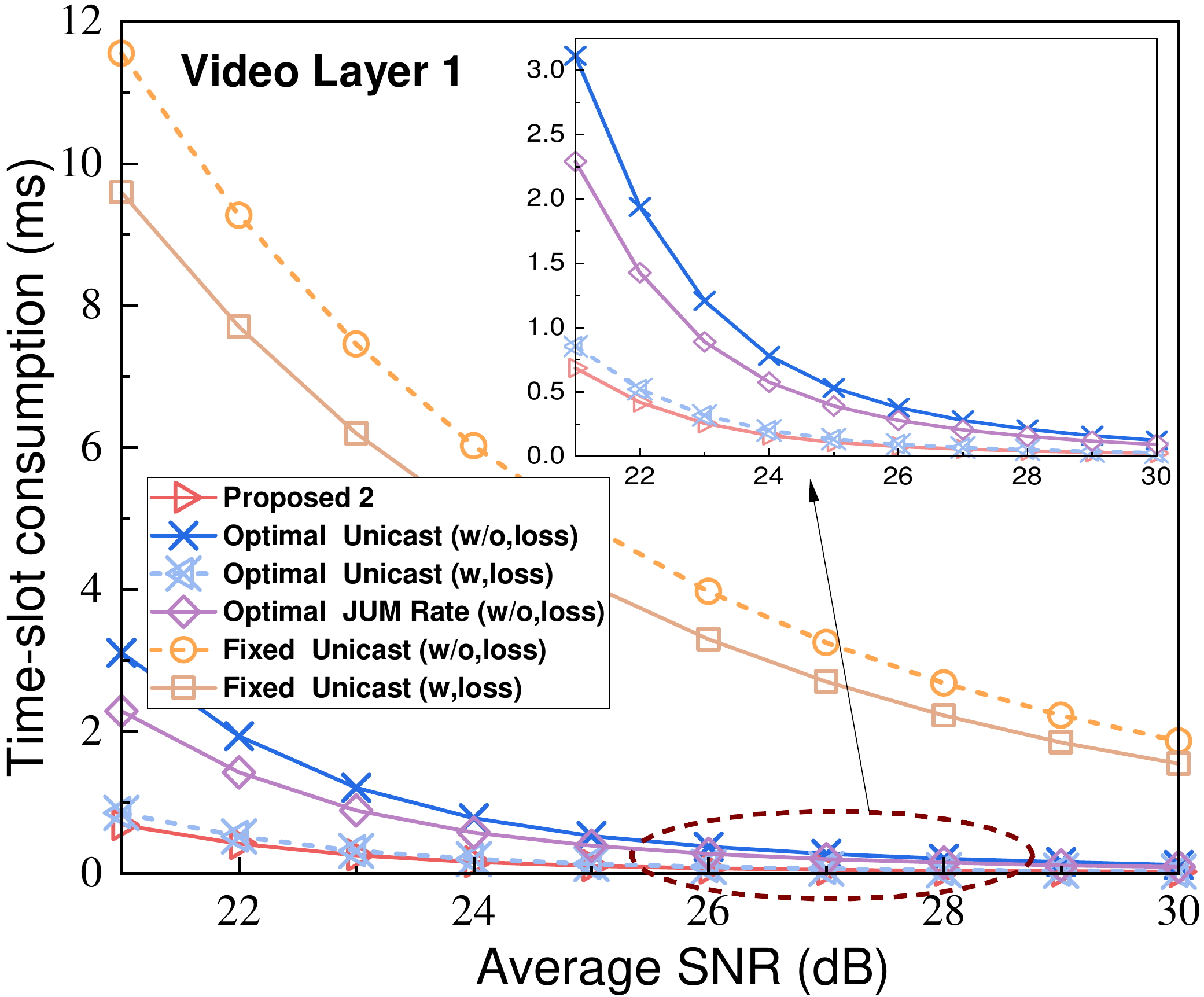}
  }
  \subfigure[]{
    \includegraphics[scale=0.238]{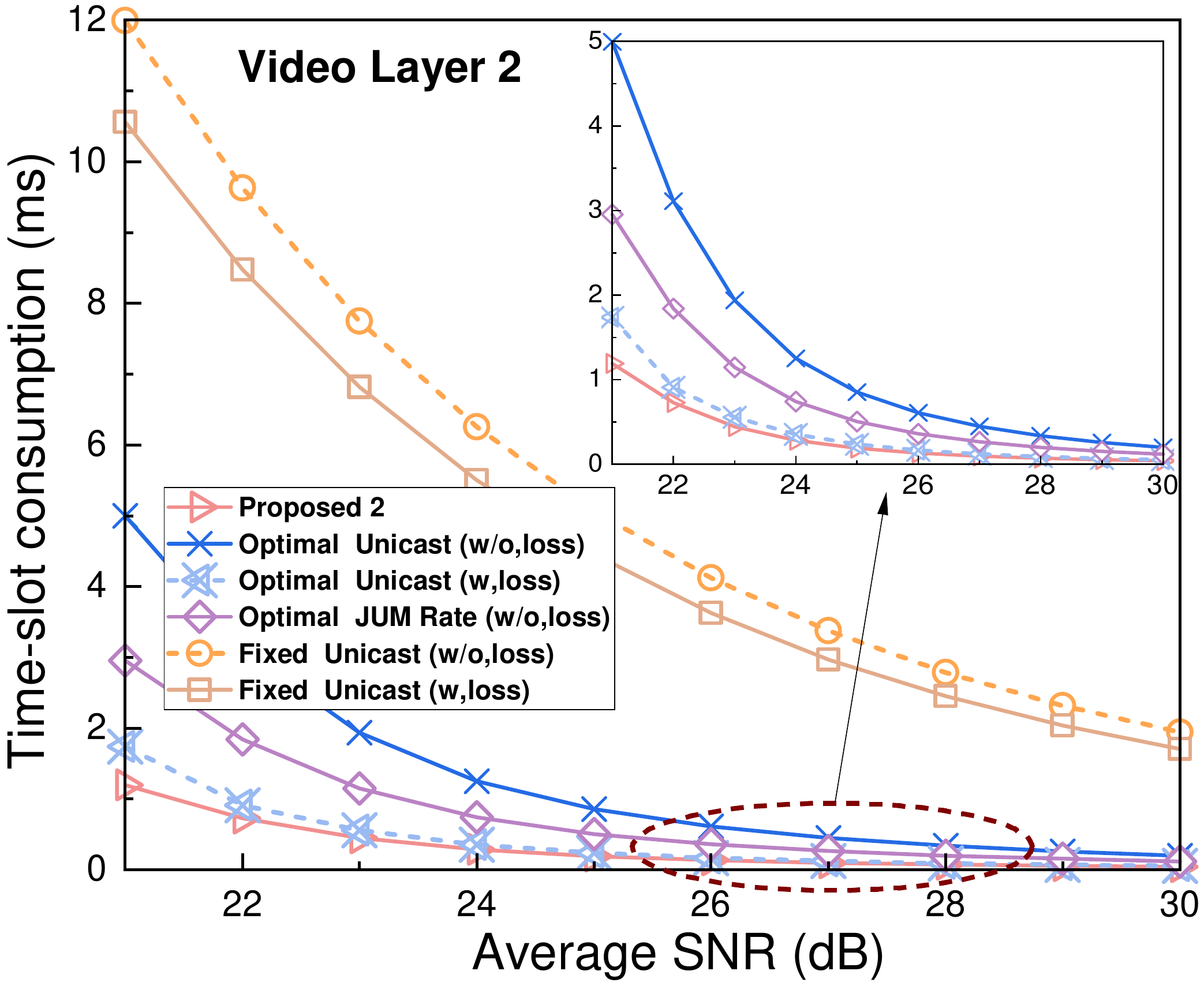}
  }
  \vspace{-0.8em}
  \caption{Performance comparison between \textbf{Proposed 2} and Optimal-Unicast (w/o,loss), Optimal-Unicast (w,loss), Optimal-JUM Rate (w/o,loss), Fixed-Unicast (w/o,loss), and Fixed-Unicast (w,loss).}
\vspace{-1em}
\end{figure*}
\vspace{-0.5em}

\par As illustrated in Fig. 8, the comparison among \textbf{Proposed 2}, Optimal-Unicast (w/o,loss), and Optimal-JUM Rate (w/o,loss) reveals that the optimal ADAPT-JTAAT transmission scheme from the EC perspective, can remarkably enhance wireless resource utilization, and achieve more significant improvement compared to \textbf{Proposed 1}. This numerical result can be drawn by comparing the performance of \textbf{Proposed 2}, Optimal-Unicast (w, loss), and Optimal-JUM Rate (w/o, loss) in Fig. 8 (a), (b), and (c). The underlying intuition behind this is that the adaptive active-discarding strategy can timely discard the data with relatively low importance from FoV edges to achieve flexible robust rate control and robust queuing behaviors, thereby obtaining greater service capacity, which naturally benefits the SQP performance from the EC perspective, since 360$^{\circ}$ VR is an enhanced mobile broadband (eMBB) service with huge data volume. Additionally, by comparing \textbf{Proposed 2} with Fixed-Unicast (w/o, loss) and Fixed-Unicast (w, loss), we observe that Fixed-Unicast (w/o,loss) and Fixed-Unicast (w,loss) fail to provide SQP performance for our 360$^{\circ}$ VR streaming architecture, even when the channel conditions are relatively ideal (e.g., 24-25 dB). This indicates that fixed slot allocation and fixed active-discarding rate are extremely detrimental for the SQP of 360$^{\circ}$ VR from the EC perspective, particularly for streaming 360$^{\circ}$ VR video with higher bitrates (e.g., referring Layer 2).

\vspace{-2.2em}

\section{Conclusion}
\par In this paper, we develop an innovative multi-layer tiled 360$^{\circ}$ VR architecture with SQP to resolve three underexploited aspects: overlapping FoVs, SQP, and loss-tolerant active data discarding. Considering the redundant resource consumption resulting from overlapping FoVs, we design an overlapping FoV-based optimal JUM task assignment scheme to implement non-redundant task assignments. Furthermore, we establish a comprehensive SQP theoretical framework by leveraging SNC theory, which encompasses two SQP schemes from SDVP and EC perspectives. Based on this theoretical framework, a corresponding optimal ADAPT-JTAAT transmission scheme is proposed to minimize resource consumption while guaranteeing diverse statistical QoS requirements under (w/o,loss) and (w,loss) scenarios from delay and rate perspectives, respectively. Extensive simulations and comparative analysis demonstrate that the proposed the multi-layer tiled 360$^{\circ}$ VR video streaming architecture can achieve superior SQP performance in resource utilization, flexible rate control, and robust queue behaviors.
\par For future work, we intend to further investigate the visual-haptic perception-enabled VR based on the SQP schemes we proposed in this research, aiming to provide users with a multimodal perception experience in virtual environments. Specifically, eMBB and URLLC traffic services need to be taken into account simultaneously.

\vspace{-1.3em}

\begin{appendices}
\section{Proof of Theorem 1}
\setcounter{equation}{0}
\renewcommand\theequation{A-\arabic{equation}}
\vspace{-0.5em}
\par The stability condition can be rewritten as follows:

\vspace{-3em}

\begin{equation}\label{A1}
    S(\theta\!_{\mathcal{M}_{q}}\!) \!=\! \mathcal{M}_{\alpha\!_{\mathcal{M}_{q}}}\!\!(\!1\!+\!\theta_{\mathcal{M}_{q}}\!)\mathcal{M}_{\delta\!_{\mathcal{M}_{q}}}\!\!(1\!-\!\theta\!_{\mathcal{M}_{q}}\!) = \mathbb{E}\!\!\left[e^{a_{\mathcal{M}_{q}}\theta_{\mathcal{M}_{q}}}\!\right]\!\cdot\!\mathbb{E}\!\!\left[e^{-\theta\!_{\mathcal{M}_{q}}r\!_{\mathcal{M}_{q}}}\right] \!=\! \mathbb{E}\!\!\left[e^{-\theta\!_{\mathcal{M}_{q}}\!\mathcal{X}\!_{\mathcal{M}_{q}}}\!\right]\! < \!1,
    \vspace{-1em}
\end{equation}
where the independence of $a\!_{\mathcal{M}_{q}}$ and $r\!_{\mathcal{M}_{q}}$ is exploited, and we substitute $\mathcal{X}\!_{\mathcal{M}_{q}}\! = \!r\!_{\mathcal{M}_{q}} - a\!_{\mathcal{M}_{q}}$. In order to prove the convexity of \ref{A1}, $\forall \theta_{\mathcal{M}_{q}}^{1} \!\neq\! \theta_{\mathcal{M}_{q}}^{2}$ and $\forall 0 \!\leq \!\varrho \!\leq \! 1$, where $\theta_{\mathcal{M}_{q}}^{1}, \theta_{\mathcal{M}_{q}}^{2} \in (0,\theta_{\mathcal{M}_{q}}^{max})$, the following inequality must hold:

\vspace{-1.2em}

\begin{equation}\label{A2}
    \mathbb{E}\!\!\left[\!e^{-\left(\varrho\theta_{\mathcal{M}_{q}}^{1} + (1-\varrho)\theta_{\mathcal{M}_{q}}^{2}\right)\mathcal{X}\!_{\mathcal{M}_{q}}}\!\right] \leq \varrho \mathbb{E}\!\!\left[e^{-\varrho\theta_{\mathcal{M}_{q}}^{1}\mathcal{X}\!_{\mathcal{M}_{q}}}\right] + (1-\varrho)\mathbb{E}\!\!\left[e^{-\varrho\theta_{\mathcal{M}_{q}}^{2}\mathcal{X}\!_{\mathcal{M}_{q}}}\right].
\end{equation}

\vspace{-0.5em}

\par By applying Hölder’s inequality, we can obtain that
\vspace{-0.5em}
\begin{equation}\label{A3}
   \begin{aligned}
     & \mathbb{E}\!\!\left[e^{-\left(\!\varrho\theta_{\mathcal{M}_{q}}^{1} + (1-\varrho)\theta_{\mathcal{M}_{q}}^{2}\!\right)\mathcal{X}\!_{\mathcal{M}_{q}}}\!\!\right] \!=\! \mathbb{E}\!\!\left[\left|e^{-\varrho\theta_{\mathcal{M}_{q}}^{1}\mathcal{X}\!_{\mathcal{M}_{q}}}\!\cdot\! e^{-(1-\varrho)\theta_{\mathcal{M}_{q}}^{2}\mathcal{X}\!_{\mathcal{M}_{q}}}\!\right|\right] \leq \! \left(\!\mathbb{E}\!\!\left[\left|e^{-\varrho\theta_{\mathcal{M}_{q}}^{1}\mathcal{X}\!_{\mathcal{M}_{q}}}\!\right|^{1/\varrho}\right]\right)^{\varrho}\!\!\times\\
     &\left(\!\mathbb{E}\!\!\left[\left|e^{-(1-\varrho)\theta_{\mathcal{M}_{q}}^{2}\mathcal{X}\!_{\mathcal{M}_{q}}}\right|^{1/1-\varrho}\right]\right)^{1-\varrho} = \left(\mathbb{E}\left[e^{-\theta_{\mathcal{M}_{q}}^{1}\mathcal{X}_{\mathcal{M}_{q}}}\right]\right)^{\varrho}\cdot \left(\mathbb{E}\left[e^{-\theta_{\mathcal{M}_{q}}^{2}\mathcal{X}_{\mathcal{M}_{q}}}\right]\right)^{1-\varrho} \\
     & \overset{(a)}{\leq} \varrho \mathbb{E}\!\!\left[e^{-\varrho\theta_{\mathcal{M}_{q}}^{1}\mathcal{X}_{\mathcal{M}_{q}}}\right] + (1-\varrho)\mathbb{E}\!\!\left[e^{-\varrho\theta_{\mathcal{M}_{q}}^{2}\mathcal{X}_{\mathcal{M}_{q}}}\right].
   \end{aligned}
\end{equation}

\vspace{-0.5em}

\par For the inequality (a) in (\ref{A3}) holds based on the fact that
\vspace{-1.5em}
\begin{equation}\label{A4}
   x_{1}^{\varrho}x_{2}^{1-\varrho} \leq \varrho x_{1} + (1-\varrho) x_{2}, \forall 0 \leq \varrho \leq 1.
   \vspace{-1em}
\end{equation}
\par We define $g(\varrho) \!=\! \varrho x_{1} \!+\! (1-\varrho) x_{2} \!-\! x_{1}^{\varrho}x_{2}^{1-\varrho}$. $\forall \varrho \in [0,1]$ and $x_{2},x_{2} < 1$, the second-order derivation $g^{''}(\varrho) = - x_{1}^{\varrho}x_{2}^{1-\varrho}\left(\log(x_{2}) - \log(x_{1})\right)^{2} \leq 0$. Since $g(0) = g(1) = 0$ and $g^{''}(\varrho) \leq 0$, it follows that $g(\varrho)$ reaches a local maximum for $\varrho \in [0,1]$. Thus, $\forall \varrho \in [0,1]$, $g(\varrho) \geq 0$ and (\ref{A4}) holds. Therefore, we show that the stability condition is convex. So the proof of Theorem \ref{theo1} is concluded.

\vspace{-1.5em}

\section{Proof of Theorem 2}
\setcounter{equation}{0}
\renewcommand\theequation{B-\arabic{equation}}
\par Based on Theorem \ref{theo1}, in the feasible domain $(0,\theta_{\mathcal{M}_{q}}^{max})$, it follows that $1 - S(\theta_{\mathcal{M}_{q}})$ is a concave and positive function. Hence, its reciprocal is convex, i.e., $\frac{1}{1-S(\theta_{\mathcal{M}_{q}})}$ \cite{boyd2004convex}. According to the definition of convex function, $\forall \theta_{\mathcal{M}_{q}}^{1}, \theta_{\mathcal{M}_{q}}^{2} \in (0,\theta_{\mathcal{M}_{q}}^{max})$ and $0 \leq \varrho \leq 1$, we have
\begin{small}
\begin{equation}\label{B1}
   \begin{aligned}
      \frac{1}{1 \!-\! \mathbb{E}\!\!\left[e^{\left(\!\varrho\theta_{\mathcal{M}_{q}}^{1} + (1-\varrho)\theta_{\mathcal{M}_{q}}^{2}\!\!\right)a\!_{\mathcal{M}_{q}}}\!\!\right]\!\!\cdot\! \mathbb{E}\!\!\left[\!e^{-\left(\!\varrho\theta_{\mathcal{M}_{q}}^{1} \!+ (1-\varrho)\theta_{\mathcal{M}_{q}}^{2}\!\right)r\!_{\mathcal{M}_{q}}}\!\!\right]} \!\!\! \leq \!\! \frac{\varrho}{1 \!-\!\mathbb{E}\!\!\left[e^{\theta_{\mathcal{M}_{q}}^{1}a\!_{\mathcal{M}_{q}}}\!\right]\! \cdot \! \mathbb{E}\!\!\left[e^{-\theta_{\mathcal{M}_{q}}^{1}r\!_{\mathcal{M}_{q}}}\!\right]} \!+\! \frac{1\!-\!\varrho}{1\!-\!\mathbb{E}\!\!\left[e^{\theta_{\mathcal{M}_{q}}^{2}a\!_{\mathcal{M}_{q}}}\!\right]\!\!\cdot\! \mathbb{E}\!\!\left[e^{-\theta_{\mathcal{M}_{q}}^{2}\!r\!_{\mathcal{M}_{q}}}\!\!\right]}.
   \end{aligned}
\end{equation}
\end{small}
\par By multiplying $\left(\!\mathbb{E}\!\!\left[e^{-(\varrho \theta_{\mathcal{M}_{q}}^{1} + (1-\varrho)\theta_{\mathcal{M}_{q}}^{2}\!)r_{\mathcal{M}_{q}}}\!\right]\!\right)^{w_{q}^{\ast}}$ to the left- and right-hand side of (\ref{B1}), we can obtain (\ref{B2}).

\begin{figure*}[b]
\vspace{-0.8em}
\centering
\hrulefill
\begin{small}
\begin{equation}\label{B2}
  \begin{aligned}
   \!\!\! \frac{\left(\mathbb{E}\!\!\left[\!e^{-(\!\varrho \theta_{\mathcal{M}_{q}}^{1}\! + (1-\varrho)\theta_{\mathcal{M}_{q}}^{2}\!)r\!_{\mathcal{M}_{q}}}\!\right]\!\right)\!\!^{w_{q}^{\ast}}}{1 \!\!-\! \mathbb{E}\!\!\left[\!e^{\left(\!\varrho\theta_{\mathcal{M}_{q}}^{1} \! + (1-\varrho)\theta_{\mathcal{M}_{q}}^{2}\!\right)a\!_{\mathcal{M}_{q}}}\!\!\right]\!\!\cdot\! \mathbb{E}\!\!\left[\!e^{-\left(\!\varrho\theta_{\mathcal{M}_{q}}^{1} \!+ (1-\varrho)\theta_{\mathcal{M}_{q}}^{2}\!\right)r\!_{\mathcal{M}_{q}}}\!\right]} \!\leq \!\frac{\varrho \! \left(\!\mathbb{E}\!\!\left[\!e^{-(\varrho \theta_{\mathcal{M}_{q}}^{1} \! + (1-\varrho)\theta_{\mathcal{M}_{q}}^{2}\!)r\!_{\mathcal{M}_{q}}}\!\right]\!\right)\!\!^{w_{q}^{\ast}}}{1 \!\!-\!\mathbb{E}\!\!\left[\!e^{\theta_{\mathcal{M}_{q}}^{1}a\!_{\mathcal{M}_{q}}}\!\right]\!\! \cdot \! \mathbb{E}\!\!\left[\!e^{-\theta_{\mathcal{M}_{q}}^{1}r\!_{\mathcal{M}_{q}}}\!\!\right]} \!+\! \frac{(1\!\!-\!\varrho)\!\!\left(\!\mathbb{E}\!\!\left[\! e^{-(\varrho \theta_{\mathcal{M}_{q}}^{1} \! + (\!1\!-\!\varrho)\theta_{\mathcal{M}_{q}}^{2}\!)r\!_{\mathcal{M}_{q}}}\!\!\right]\!\right)\!\!^{w_{q}^{\ast}}}{1\!-\!\mathbb{E}\!\left[e^{\theta_{\mathcal{M}_{q}}^{2}a_{\mathcal{M}_{q}}}\!\right]\!\cdot\! \mathbb{E}\!\left[e^{-\theta_{\mathcal{M}_{q}}^{2}r_{\mathcal{M}_{q}}}\!\right]}.
  \end{aligned}
\end{equation}
\end{small}
\hrulefill
\vspace{-0.8em}
\end{figure*}

\par By applying Hölder’s inequality, we can obtain that
\vspace{-0.4em}
\begin{small}
\begin{equation}\label{B3}
   \begin{aligned}
     \!\!\! & \left(\mathbb{E}\!\left[e^{-(\varrho \theta_{\mathcal{M}_{q}}^{1} + (1-\varrho)\theta_{\mathcal{M}_{q}}^{2})r_{\mathcal{M}_{q}}}\right]\!\right)\!^{w_{q}^{\ast}} = \left(\!\mathbb{E}\!\left[\left|e^{\!-\varrho \theta_{\mathcal{M}_{q}}^{1}r_{\mathcal{M}_{q}}}\!\right|\right] \!\!\cdot\! \mathbb{E}\!\!\left[\left|e^{\!-(1-\varrho) \theta_{\mathcal{M}_{q}}^{2}r_{\mathcal{M}_{q}}}\!\right|\right]\!\right)\!\!^{w_{q}^{\ast}} \leq \!\! \left(\!\mathbb{E}\!\!\left[\!\left|e^{\!-(1-\varrho) \theta_{\mathcal{M}_{q}}^{2}r_{\mathcal{M}_{q}}}\!\right|^{1/1-\varrho}\right]\!\right)\!\!^{w_{q}^{\ast}(1-\varrho)} \times \\
     \!\!\! & \left(\!\mathbb{E}\!\!\left[\!\left|e^{\!-\varrho \theta_{\mathcal{M}_{q}}^{1}r_{\mathcal{M}_{q}}}\!\right|^{1/\varrho}\right]\!\right)\!\!^{w_{q}^{\ast}\varrho} = \left(\mathbb{E}\!\!\left[\!\left|e^{\!-\theta_{\mathcal{M}_{q}}^{1}r_{\mathcal{M}_{q}}}\!\right|\right]\right)^{w_{q}^{\ast}\varrho} \!\! \cdot \!\! \left(\mathbb{E}\!\!\left[\!\left|e^{\!-\theta_{\mathcal{M}_{q}}^{2}r_{\mathcal{M}_{q}}}\!\right|\right]\right)^{w_{q}^{\ast}(1-\varrho)} \!\! \leq \!\! \left(\mathbb{E}\!\!\left[\!\left|e^{\!-\theta_{\mathcal{M}_{q}}^{1}r_{\mathcal{M}_{q}}}\!\right|\right] \cdot \mathbb{E}\!\!\left[\!\left|e^{\!-\theta_{\mathcal{M}_{q}}^{2}r_{\mathcal{M}_{q}}}\!\right|\right]\right)^{w_{q}^{\ast}},
   \end{aligned}
\end{equation}
\end{small}
since $\left(\mathbb{E}\!\!\left[\!\left|e^{\!-\theta_{\mathcal{M}_{q}}r_{\mathcal{M}_{q}}}\!\right|\right]\right)^{w_{q}^{\ast}} > 0$. Then, the following inequality holds for the right-side of (\ref{B2}), and we can obtain that
\begin{small}
\begin{equation}\label{B4}
\setstretch{0.95}
  \begin{aligned}
    & \frac{\varrho \! \left(\!\mathbb{E}\!\!\left[\!e^{\!-(\!\varrho \theta_{\mathcal{M}_{q}}^{1} \!\! + (\!1\!-\!\varrho)\theta_{\mathcal{M}_{q}}^{2}\!)r\!_{\mathcal{M}_{q}}}\!\!\right]\!\right)\!\!^{w_{q}^{\ast}}}{\!\!1 \! \!-\!\mathbb{E}\!\!\left[\!e^{\!\theta_{\mathcal{M}_{q}}^{1}\!a\!_{\mathcal{M}_{q}}}\!\right]\!\! \cdot \! \mathbb{E}\!\!\left[e^{\!-\theta_{\mathcal{M}_{q}}^{1}\!r\!_{\mathcal{M}_{q}}}\!\!\right]} \!+\! \frac{\!\!\!(\!1\!\!-\!\varrho\!)\!\!\left(\!\mathbb{E}\!\!\left[\!e^{-(\!\varrho \theta_{\mathcal{M}_{q}}^{1} \!+ (\!1\!-\!\varrho)\theta_{\mathcal{M}_{q}}^{2}\!)r\!_{\mathcal{M}_{q}}}\!\right]\!\right)\!\!^{w_{q}^{\ast}}}{\!\!\!1\!-\!\mathbb{E}\!\left[e^{\theta_{\mathcal{M}_{q}}^{2}\!a\!_{\mathcal{M}_{q}}}\!\right]\!\!\cdot\! \mathbb{E}\!\left[\!e^{-\theta_{\mathcal{M}_{q}}^{2}\!r\!_{\mathcal{M}_{q}}}\!\right]} \leq \! \Biggl(\frac{\!\varrho \! \left(\!\mathbb{E}\!\!\left[\!\big|e^{\!-\theta_{\mathcal{M}_{q}}^{1}\! r\!_{\mathcal{M}_{q}}}\!\big|\!\right] \!\!\cdot\! \mathbb{E}\!\!\left[\!\big|e^{\!-\theta_{\mathcal{M}_{q}}^{2}\!r\!_{\mathcal{M}_{q}}}\!\big|\!\right]\!\right)\!\!^{w_{q}^{\ast}}}{\!\!\!\!\!1 \!-\!\mathbb{E}\!\!\left[\!e^{\theta_{\mathcal{M}_{q}}^{1}\! a\!_{\mathcal{M}_{q}}}\!\right]\!\! \cdot \! \mathbb{E}\!\left[\! e^{-\theta_{\mathcal{M}_{q}}^{1}\! r\!_{\mathcal{M}_{q}}}\!\right]} + \\
    &\frac{(\!1\!-\!\varrho\!)\!\!\left(\!\mathbb{E}\!\!\left[\!\big|e^{\!-\theta_{\mathcal{M}_{q}}^{1}\!r\!_{\mathcal{M}_{q}}}\!\big|\!\right] \!\!\cdot\! \mathbb{E}\!\!\left[\!\big|e^{\!-\theta_{\mathcal{M}_{q}}^{2}r\!_{\mathcal{M}_{q}}}\!\big|\!\right]\!\right)\!\!^{w_{q}^{\ast}}}{\!1\!\!-\!\!\mathbb{E}\!\!\left[\!e^{\theta_{\mathcal{M}_{q}}^{2}\!a\!_{\mathcal{M}_{q}}}\!\right]\!\!\cdot\! \mathbb{E}\!\!\left[\!e^{\!-\theta_{\mathcal{M}_{q}}^{2}\!r\!_{\mathcal{M}_{q}}}\!\right]} \Biggl) \leq \! \frac{\varrho \! \left(\!\mathbb{E}\!\!\left[\!\big|e^{\!-\theta_{\mathcal{M}_{q}}^{1}\!r\!_{\mathcal{M}_{q}}}\!\big|\right]\!\right)\!\!^{w_{q}^{\ast}}}{1\! \!-\!\!\mathbb{E}\!\!\left[\!e^{\!\theta_{\mathcal{M}_{q}}^{1}\!a\!_{\mathcal{M}_{q}}}\!\right]\!\! \cdot \! \mathbb{E}\!\!\left[\!e^{\!-\theta_{\mathcal{M}_{q}}^{1}\!r\!_{\mathcal{M}_{q}}}\right]} \!\!+\!\! \frac{(\!1\!-\!\varrho\!)\!\!\left(\!\mathbb{E}\!\!\left[\!\big|e^ {\!-\theta_{\mathcal{M}_{q}}^{1}\!r\!_{\mathcal{M}_{q}}}\!\big|\!\right] \!\!\cdot\! \mathbb{E}\!\!\left[\!\big|e^{\!-\theta_{\mathcal{M}_{q}}^{2}\!r\!_{\mathcal{M}_{q}}}\!\big|\!\right]\!\right)\!\!^{w_{q}^{\ast}}.}{1\!-\!\mathbb{E}\!\!\left[\!e^{\theta_{\mathcal{M}_{q}}^{2}\!a\!_{\mathcal{M}_{q}}}\!\right]\!\!\cdot\! \mathbb{E}\!\!\left[e^{\!-\theta_{q}^{2}\!r_{\mathcal{M}_{q}}}\!\right]}
  \end{aligned}
\end{equation}
\end{small}

\vspace{-0.5em}

\par Combining (\ref{B2}) and (\ref{B4}), and the definition of convex functions, the convexity of kernel function $K_{\mathcal{M}_{q}}(\theta_{\mathcal{M}_{q}},-w_{q}^{\ast})$ can be inferred. So \emph{Theorem \ref{theo2}} is concluded.

\vspace{-1.2em}

\section{Proof of Theorem 4}
\setcounter{equation}{0}
\renewcommand\theequation{C-\arabic{equation}}

\vspace{-0.3em}

\par The convex optimization problem $\mathcal{P}4^{'}$ can be solved by exploiting Karush-Kuhn-Tucker (KKT) method. The Lagrange function of the $\mathcal{P}4^{'}$ can be constructed as
\vspace{-0.5em}
\begin{small}
\begin{small}
\begin{equation}\label{C1}
  \begin{aligned}
    &\mathcal{L}\!\left(t_{\mathcal{M}_{q}},\psi_{\mathcal{M}_{q}},\mu_{\zeta}\!\right)= \mathbb{E}_{\zeta}\bigg\{\!\sum_{q \in \mathcal{Q}}\sum_{\mathcal{M}_{q} \in \mathcal{H}_{q}}\!\!\! t_{\mathcal{M}_{q}}\!\!\bigg\} + \mathbb{E}_{\zeta}\!\bigg\{\!\mu_{\zeta}\bigg(\!\sum_{q \in \mathcal{Q}}\sum_{\mathcal{M}_{q} \in \mathcal{H}_{q}}\!\!\!\! t\!_{\mathcal{M}_{q}} \!-\! T \!\bigg) \!\!\bigg\} + \sum_{q \in \mathcal{Q}}\sum_{\mathcal{M}_{q} \in \mathcal{H}_{q}}\!\!\! \psi_{\mathcal{M}_{q}}\!\bigg( \!\! F\!\!\left(t_{\mathcal{M}_{q}}\!\right)\!-\! e^{-\theta_{\mathcal{M}_{q}}^{\ast}\!\left|\mathcal{R}_{\mathcal{M}_{q}}\!\right|\overline{\mathcal{EB}}_{\mathcal{M}_{q}}}\!\!\bigg),
  \end{aligned}
\end{equation}
\end{small}
\end{small}
where $\psi_{\mathcal{M}_{q}}, q \in \mathcal{Q}, \mathcal{M}_{q} \in \mathcal{H}_{q}$ and $\mu_{\zeta}$ are Lagrange multipliers of constraint (24a) and (24b), respectively, and $\overline{\mathcal{EB}}_{\mathcal{M}_{q}} = BT\mathcal{EB}\!_{\mathcal{M}_{q}}\!\big(\theta_{\mathcal{M}_{q}}^{\ast}|A_{\mathcal{M}_{q}}\!\big)$.

\par By taking the partial derivative of $\mathcal{L}$ w.r.t. $t_{\mathcal{M}_{q}}$ and setting its value to zero, yields:
\vspace{-0.4em}
\begin{equation}\label{C3}
  \frac{\partial \mathcal{L}}{\partial t_{\mathcal{M}_{q}}} \!=\!  \big(\! 1+\mu_{\zeta}-\psi_{\mathcal{M}_{q}}\Theta_{\mathcal{M}_{q}}^{\ast}\!\! (1+\zeta^{'})^{-\Theta_{\mathcal{M}_{q}}^{\ast}t_{\mathcal{M}_{q}}}\cdot\ln(1+\zeta^{'}\!) \big) \! \cdot \! f(\zeta,M) \d \zeta = 0.
\end{equation}

\vspace{-0.3em}
\par Note that if there exists an optimal solution $\boldsymbol{t}^{\star}$, then $\boldsymbol{t}^{\star}$ and its corresponding optimal Lagrange multipliers $\mu_{\zeta}^{\star}$ and $\boldsymbol{\psi}^{\star} \triangleq \big\{\psi_{\mathcal{M}_{q}}^{\star}\big\}_{q \in \mathcal{Q},\mathcal{M}_{q} \in \mathcal{H}_{q}}$ must satisfy KKT conditions \cite{boyd2004convex}, which are given as
\vspace{-0.5em}
\begin{small}
\begin{equation}\label{C2}
   \quad \begin{cases}
   \mu_{\zeta}^{\star}\big\{\!\big(\!\sum\limits_{q\in\mathcal{Q}}\sum\limits_{\mathcal{M}_{q}\in\mathcal{H}_{q}} \!\!\!t_{\mathcal{M}_{q}}^{\star} \!- \! T \big)\!\big\} = 0,\ \forall \zeta,\\
    \sum\limits_{q\in \mathcal{Q}}\sum\limits_{\mathcal{M}_{q}\in\mathcal{H}_{q}}\!\!\! \psi^{\star}_{\mathcal{M}_{q}}\!\!\left(\! F(t_{\mathcal{M}_{q}}^{\star})-e^{-\theta_{\mathcal{M}_{q}}^{\ast}\left|\mathcal{R}_{\mathcal{M}_{q}}\right|\overline{\mathcal{EB}}_{\mathcal{M}_{q}}} \!\!\right)= 0,\ \forall \zeta, \\
   \mu_{\zeta}^{\star} \ge 0, \psi_{\mathcal{M}_{q}}^{\star} \geq 0,\ \forall q, \mathcal{M}_{q}, \forall \zeta,\\
   \frac{\partial \mathcal{L}}{\partial t_{\mathcal{M}_{q}}}\bigg |_{t_{\mathcal{M}_{q}} = t_{\mathcal{M}_{q}}^{\star}} = 0, \ \forall q, \mathcal{M}_{q}, \forall \zeta.
 \end{cases}
\end{equation}
\end{small}
\par Substituting $t_{\mathcal{M}_{q}} \triangleq t_{\mathcal{M}_{q}}^{\star}$ into (\ref{C2}), we can finally obtain the optimal solution of $\boldsymbol{t}^{\star}$, which is given as
\vspace{-0.6em}
\begin{small}
\begin{equation}\label{C4}
   \begin{aligned}
  \!\!t_{\mathcal{M}_{q}}^{\star} \!=\! \left[\frac{\ln\left(\psi_{\mathcal{M}_{q}}^{\star}\Theta_{\mathcal{M}_{q}}^{\ast}\ln\big(1+\zeta^{'}\big)\right)-\ln\big(1+\mu_{\zeta}^{\star}\big)}{\Theta_{\mathcal{M}_{q}}^{\ast}\ln\big(1+\zeta^{'}\big)}\right]^{+}\!\!\!.
  \end{aligned}
\end{equation}
\end{small}
\par So the proof of \emph{Theorem 4} is concluded.

\vspace{-1em}

\section{Proof of Theorem 5}
\setcounter{equation}{0}
\renewcommand\theequation{D-\arabic{equation}}

\par Given $\zeta$ and $\psi_{\mathcal{M}_{q}}^{\star}$, $t_{\mathcal{M}_{q}}\left(\zeta,\psi_{\mathcal{M}_{q}}^{\star},\mu_{\zeta}\right)$ is a strictly monotonically decreasing function with respect to $\mu_{\zeta}$, since
\vspace{-0.4em}
\begin{equation}\label{D1}
  \frac{\partial t_{\mathcal{M}_{q}}}{\partial \mu_{\zeta}} = \min\bigg\{-\frac{1}{\left(1+\mu_{\zeta}\right)g\left(\theta_{q}^{\ast},\zeta\right)},0\bigg\} \leq 0,
  \vspace{-1em}
\end{equation}
which implies that $\lim\limits_{\mu_{\zeta} \rightarrow \infty}$ $t_{\mathcal{M}_{q}}\left(\zeta,\psi_{\mathcal{M}_{q}}^{\star},\mu_{\zeta}\right) = 0^{+}$. Hence, if the following inequality holds, i.e.,
\vspace{-0.8em}
\begin{equation}\label{D2}
  \lim_{\mu_{\zeta}\rightarrow 0}\sum\limits_{q\in\mathcal{Q}}\sum\limits_{\mathcal{M}_{q}\in\mathcal{H}_{q}}t_{\mathcal{M}_{q}}\left(\zeta,\psi_{\mathcal{M}_{q}}^{\star},\mu_{\zeta}\right) \geq T,
\end{equation}
\vspace{-0.3em}
there exists a unique solution $\mu_{\zeta}^{\star} > 0$, which satisfies
\vspace{-0.3em}
\begin{equation}\label{D3}
  \sum\limits_{q\in\mathcal{Q}}\sum\limits_{\mathcal{M}_{q}\in\mathcal{H}_{q}}t_{\mathcal{M}_{q}}\left(\zeta,\psi_{\mathcal{M}_{q}}^{\star},\mu_{\zeta}^{\star}\right) = T,
\end{equation}
\vspace{-1.5em}
\par Otherwise, if $\forall\mu_{\zeta}\geq 0$, it always has
\vspace{-1em}
\begin{equation}\label{D4}
  \sum\limits_{q\in\mathcal{Q}}\sum\limits_{\mathcal{M}_{q}\in\mathcal{H}_{q}}t_{\mathcal{M}_{q}}\left(\zeta,\psi_{\mathcal{M}_{q}}^{\star},\mu_{\zeta}\right) < T,
  \vspace{-1em}
\end{equation}
\vspace{-0.3em}
then, we can deduce $\mu_{\zeta}^{\star} = 0$ from the second equation of KKT conditions. Finally, according to (\ref{D1})-(\ref{D4}), the proof of \emph{Theorem \ref{theo5}} is concluded.

\vspace{-1.7em}

\section{Proof of Theorem 6}
\setcounter{equation}{0}
\renewcommand\theequation{E-\arabic{equation}}

\par In order to prove problem $\mathcal{P}4$ is convex, we need only to examine the convexity of the constraints (\ref{e31}) and (\ref{e33}) with respect to $\left(\boldsymbol{t}, \boldsymbol{\mathcal{J}}\right)$, which is not hard to prove by exploiting derivation rule. The Lagrangian of problem $\mathcal{P}6$ is formulated as follows:
\vspace{-0.3em}
\begin{small}
\begin{equation}\label{E1}
   \begin{aligned}
      & \mathcal{L} =\mathbb{E}_{\zeta}\bigg\{\!\!\sum_{q\in\mathcal{Q}} \sum_{\mathcal{M}_{q} \in \mathcal{H}_{q}} \!\!\!\!\! t_{\!\mathcal{M}_{q}} \!\!\bigg\} + \mathbb{E}_{\zeta}\bigg\{\! \mu_{\zeta}\bigg(\sum_{q\in\mathcal{Q}} \sum_{\mathcal{M}_{q} \in \mathcal{H}_{q}} \!\!\!\!\! t_{\!\mathcal{M},q} - T\! \bigg)\!\!\bigg\} \! + \!\sum\limits_{q \in \mathcal{Q}}\! \sum\limits_{\mathcal{M}_{q}\in\mathcal{H}_{q}}\!\!\!\!\! \psi_{\!\mathcal{M}_{q}}\bigg(\! \mathbb{E}_{\zeta}\bigg\{\! e^{-\Theta_{\mathcal{M}_{q}}^{\star}\big(\mathcal{J}_{\!\mathcal{M}_{q}} + t_{\!\mathcal{M}_{q}}\! \log\big(1 + \zeta^{'}\big) \big)}\!\bigg\}\!-\!U_{\!\mathcal{M}_{q}}\bigg) \\
      & \quad \quad \quad + \sum\limits_{q \in \mathcal{Q}}\! \sum\limits_{\mathcal{M}_{q}\in\mathcal{H}_{q}}\!\!\!\!\! \phi_{\!\mathcal{M}_{q}}\mathbb{E}_{\zeta}\bigg\{\!\big(1\!-\!\mathcal{Y}_{\!\mathcal{M}_{q}}^{th}\big)\mathcal{J}_{\!\mathcal{M}_{q}}\! - \! \mathcal{Y}_{\!\mathcal{M}_{q}}^{th}t_{\!\mathcal{M}_{q}}\!\log\big(1+\zeta^{'}\big)\!\bigg\},
   \end{aligned}
\end{equation}
\end{small}
where $\psi_{\mathcal{M}_{q}}$, $\mu_{\zeta}$, and $\phi_{\mathcal{M}_{q}}$ denote the Lagrangian multipliers of constraints (30), (12b), and (32), respectively. By exploiting KKT method, the optimal solution of $\left(\boldsymbol{t}^{\star},\boldsymbol{\mathcal{Y}}^{\star}\right)$ satisfies:
\begin{small}
\begin{equation}\label{E2}
  \left\{\!\!
    \begin{array}{ll}
      \!\!\! \frac{\partial \mathcal{L}}{\mathcal{J}_{\mathcal{M}_{q}}}\big|_{\mathcal{J}_{\mathcal{M}_{q}} = \mathcal{J}_{\mathcal{M}_{q}}^{\star}} \!\! = 0, \quad \quad \quad \quad \quad \quad \quad \quad \quad \quad \quad \quad\hbox{$\forall \zeta, q, \mathcal{M}_{q}$;} \\
      \!\!\! \frac{\partial \mathcal{L}}{t_{\mathcal{M}_{q}}}\big|_{t_{\mathcal{M}_{q}} = t_{\mathcal{M}_{q}}^{\star}} \!\! = 0, \quad \quad \quad \quad \quad \quad \quad \quad \quad \quad \quad \quad \ \ \hbox{$\forall \zeta, q, \mathcal{M}_{q}$;} \\
      \!\!\! \mu_{\zeta}^{\star}\big(\!\!\sum\limits_{q\in\mathcal{Q}} \sum\limits_{\mathcal{M}_{q} \in \mathcal{H}_{q}}\!\!\!\! t_{\mathcal{M}_{q}}^{\star}\!-\!T\big) = 0, \quad \quad \quad \quad \quad \quad \quad \quad \quad \hbox{$\forall \zeta$;} \\
      \!\!\! \psi_{\mathcal{M}_{q}}^{\star}\!\bigg(\! \mathbb{E}_{\zeta}\bigg\{\! e^{-\Theta_{\mathcal{M}_{q}}^{\ast}\big(\mathcal{J}_{\!\mathcal{M}_{q}}^{\star} + t_{\!\mathcal{M}_{q}}^{\star}\! \log\left(1 + \zeta\right) \big)}\!\bigg\}\!-\!U_{\!\mathcal{M}_{q}}\!\!\bigg)\! = \!0, \hbox{$\forall \zeta, q, \mathcal{M}_{q}$;} \\
      \!\!\! \phi_{\!\mathcal{M}_{q}}^{\star}\!\mathbb{E}_{\zeta}\bigg\{\!\!\big(1\!-\!\mathcal{Y}_{\!\mathcal{M}_{q}}^{th}\big)\mathcal{J}_{\!\mathcal{M}_{q}}^{\star}\!\! - \! \mathcal{Y}_{\!\mathcal{M}_{q}}^{th}t_{\!\mathcal{M}_{q}}^{\star}\!\log\big(1\!+\!\zeta\big)\!\bigg\} \!\!= 0, \hbox{$\forall \zeta, q, \mathcal{M}_{q}$.}
    \end{array}
  \right.
\end{equation}
\end{small}
Then, we substituted (\ref{E1}) into (\ref{E2}) to solve the equation (\ref{E2}), which yields (\ref{e34}) and (\ref{e35}), respectively. Then, we can optimize $\mu_{\zeta}^{\star}$, $\psi_{\mathcal{M}_{q}}^{\star}$, and $\phi_{\mathcal{M}_{q}}^{\star}$ in a similar way to \emph{Theorem 3} and \emph{Algorithm 3} in the previous subsection. So the proof of \emph{Theorem 6} is concluded.
\end{appendices}

\vspace{-0.6em}
\footnotesize
\bibliographystyle{IEEEtran}
\bibliography{IEEEabrv,ref}

\end{document}